\documentclass[11 pt, a4paper]{article}
\usepackage{array,epsfig}
\usepackage{amsthm}
\usepackage{amsmath}
\usepackage{amssymb}
\usepackage{color}
\usepackage{enumitem}
\usepackage{booktabs}
\usepackage{sidecap}
\usepackage{setspace}
\usepackage{comment}
\usepackage{url}
\usepackage{float}
\usepackage{apacite} 
\usepackage{parskip}
\usepackage{standalone}
\usepackage{tabu}
\usepackage{aligned-overset}
\usepackage[font=small,labelfont=bf]{caption}

\usepackage{bm, bbm, mdframed, mathtools, mathabx, multirow, subfigure, tikz, multirow, tcolorbox, soul, makecell}
\usepackage[left=2.5cm, right=2.5cm, top=3cm, bottom=3cm]{geometry}
\usepackage[english]{babel}
\usepackage[mathscr]{euscript}
\numberwithin{equation}{section} 
\usepackage{mathdots}
\usepackage{lscape}
\usepackage{pdflscape}

\usepackage[round, longnamesfirst]{natbib}
\usepackage[T1]{fontenc} 
\usepackage[autostyle]{csquotes}  
\usepackage[citecolor=black, linkcolor = black, colorlinks=true, allcolors=black]{hyperref}
\usepackage{glossaries}
\usepackage{appendix}
\usepackage{verbatim}
\usepackage{tcolorbox}
\usepackage{xcolor}
\usepackage{listings}
\usepackage{rotating}
\usepackage{comment}
\usepackage{graphicx}
\usepackage{algorithm2e}
\usepackage{bookmark}
\usepackage{setspace}
\usepackage{xr}
\usepackage{threeparttable}
\usepackage{etoolbox}
\usepackage{tabularx}
\usepackage{supertabular}
\usepackage{enumitem}
\usepackage{XCharter} 
\RestyleAlgo{ruled}
\SetKwComment{Comment}{/* }{ */}
\usepackage{tikz}
\usetikzlibrary{shapes.geometric, arrows, positioning, decorations.pathreplacing}
\tikzstyle{c_solid} = [circle, minimum width=1cm, minimum height=1cm,text centered,draw=black, fill=white]
\tikzstyle{c_dashed} = [circle, minimum width=1cm, minimum height=1cm,text centered,draw=black, dashed]
\tikzstyle{arrow} = [thick,->,>=stealth]
\tikzstyle{startstop} = [rectangle, rounded corners, minimum width=3cm, minimum height=1cm, draw=black, fill=gray!20]
\tikzstyle{barrow} = [thick,->,>=stealth, bend left=60]

\allowdisplaybreaks



\newtheorem{assumption}{Assumption}
\newtheorem{theorem}{Theorem}

\newtheorem{definition}{Definition}
\newtheorem{example}{Example}

\numberwithin{equation}{section}
\numberwithin{theorem}{section}
\numberwithin{assumption}{section}
\numberwithin{definition}{section}
\numberwithin{example}{section}

\newcommand{\E}{\mathbb{E}}
\newcommand{\ind}{\perp \!\!\!\! \perp}

\DeclareMathOperator{\Var}{Var}

\AtBeginEnvironment{tabular}{\small}
\AtBeginEnvironment{tabularx}{\small}

\title{\textbf{Evaluating Program Sequences with Double Machine Learning: An Application to Labor Market Policies}}
\author{
Fabian Muny\thanks{University of St.Gallen, Rosenbergstrasse 22, 9000 St.Gallen, CH, E-mail: \texttt{fabian.muny@unisg.ch}\\
Financial support from the Swiss National Science Foundation (SNSF) is gratefully acknowledged. The study is part of the project ``Chances and risks of data-driven decision making for labour market policy'' (grant number SNSF 407740\_187301) of the Swiss National Research programme ``Digital Transformation'' (NRP 77). This work used GPT-4o for editorial help. I thank Elena Bassoli, Sandro Heiniger, and Michael Lechner for valuable feedback and suggestions and Yuqian Zhang for sharing his code. All errors are my own.}}
\date{} 
\begin{document}
\begingroup
\let\newpage\relax
\maketitle
\endgroup

 \vspace{0.5cm}

\begin{center}
 \vspace{0.3cm}
  \textbf{Abstract}\vspace{0.3cm}   \\
\end{center}

\begin{minipage}{\textwidth}
    \small
Many programs evaluated in observational studies incorporate a sequential structure, where individuals may be assigned to various programs over time. While this complexity is often simplified by analyzing programs at single points in time, this paper reviews, explains, and applies methods for program evaluation within a sequential framework. It outlines the assumptions required for identification under dynamic confounding and demonstrates how extending sequential estimands to dynamic policies enables the construction of more realistic counterfactuals. Furthermore, the paper explores recently developed methods for estimating effects across multiple treatments and time periods, utilizing Double Machine Learning (DML), a flexible estimator that avoids parametric assumptions while preserving desirable statistical properties. Using Swiss administrative data, the methods are demonstrated through an empirical application assessing the participation of unemployed individuals in active labor market policies, where assignment decisions by caseworkers can be reconsidered between two periods. The analysis identifies a temporary wage subsidy as the most effective intervention, on average, even after adjusting for its extended duration compared to other programs. Overall, DML-based analysis of dynamic policies proves to be a useful approach within the program evaluation toolkit.\\[0.5cm]
\end{minipage}

\thispagestyle{empty}
\newpage

\newpage
\setcounter{page}{1}

\setcounter{page}{1} 

\section{Introduction}

Program evaluations based on observational data are used by economists to assess the impact of policy measures such as training programs, public transfer schemes, and healthcare interventions \citep{abadie2018econometric}.
The standard approach in the literature involves comparing outcomes between  treatment groups of program participants and a control group of non-participants.
If all confounding variables jointly influencing the outcome and program assignment are observed, the causal effect of the intervention can be identified. By collapsing program assignment, participation, and completion into a single treatment state, the described approach considerably simplifies the complexities prevalent in many non-experimental settings, limiting the ability to address many important questions. For example, program duration may vary between participants, which could have a strong impact on program efficiency. Furthermore, individuals  may participate not only in one but in several successive measures, which makes it difficult to match them to one particular program type. Importantly, a specific program sequence may result from reassignment following the initial placement. For instance, a program may be shortened if an individual is no longer eligible, or the program may be switched if the initial program is deemed unsuccessful. After all, a particular program sequence may be effective for one person but not for another, which cannot be distinguished when only analyzing population-level effects.

This paper addresses the previously mentioned challenges by adopting a framework to evaluate program sequences instead of single time-point interventions. The framework is based on ideas from biostatistics, originally formulated by \citet{robins1986new, robins1987addendum} and further developed in subsequent research, as reviewed in \citet{richardson2014causal}. While these methods have frequently been applied to the analysis of sequences of medical treatments \citep[e.g.][and many more]{hernan2002estimating, taubman2009intervening, young2011comparative}, they have so far seen limited adoption in the econometric program evaluation literature. Wider adoption has mainly been hindered by the stricter identification and estimation requirements in the sequential setting compared to one-time interventions. This paper demonstrates that these concerns can be addressed by (i) extending sequential estimands to dynamic policies that facilitate the credible identification of counterfactual treatment sequences using observational data and (ii) leveraging recently proposed machine learning-based estimation strategies to flexibly estimate these quantities. These innovations enhance the applicability of sequential analysis for assessing policy measures commonly studied in economics, ensuring a better reflection of their sequential nature.

The key challenge in analyzing program sequences from observational data is that the complete trajectory of treatments is typically not predetermined prior to the start of the sequence. This must be considered both when accounting for the program assignment mechanism underlying the observed data (\textit{dynamic confounding}) and when designing counterfactual scenarios that depend on time-varying covariates (\textit{dynamic policies}). While the former has been addressed in prior econometric research \citep[e.g.][]{lechner2009sequential}, the latter has thus far received little attention in evaluation studies. Instead, dynamic policies have primarily been used in the context of optimal dynamic treatment regimes \citep{murphy2003optimal, zhang2013robust, sakaguchi2024robust}. This body of literature develops procedures to select an optimal dynamic policy from a class of feasible policies, with the objective of maximizing the mean response in the population. While such approaches allow to provide individualized treatment recommendations, identifying the full set of policies in a given class requires strong assumptions that are often implausible in non-experimental contexts. In contrast, the present work adopts an alternative objective. Rather than optimizing over an extensive policy class, it is proposed to selectively isolate specific policies that closely align with the assumed underlying assignment process and are therefore convincingly identifiable from observable data. Accordingly, instead of focusing on individualized treatment recommendations, dynamic policies are leveraged as a framework to uncover aggregate and group-level effects that enable robust estimation and inference.

\defcitealias{chernozhukov2018double}{Chernozhukov, Chetverikov, Demirer, Duflo, Hansen, Newey, \& Robins,   2018;} 

A major challenge for statistical inference in sequential settings is that the number of possible program sequences expands exponentially with the number of periods, leading to data sparsity for individual sequences. For example, a setting with five treatments and five time periods already yields $5^{5}=3,125$ possible treatment sequences, and ensuring sufficient observations for each becomes increasingly difficult. Hence, while nonparametric identification of dynamic policies is achievable, conventional estimation approaches, such as structural nested mean models \citep{robins1989analysis, robins1994correcting} or marginal structural models \citep{robins2000marginal}, rely on structural assumptions to extrapolate into regions with limited data support.\footnote{For an overview of these methods, see Appendix \ref{ap:other_estimators_conv}}  
However, if a fair number of observations is available for each program sequence of interest, more flexible, machine learning-based estimators can be employed. Aggregating information across periods and/or treatments can enhance feasibility, especially when credible information about the functional form of the underlying data-generating process is lacking. Building on this approach, the present paper demonstrates the potential of recently proposed double machine learning (DML) methods \citetext{\citetalias{chernozhukov2018double} \citealp{bodory2022evaluating, bradic2021high}} for analyzing dynamic policies. The key advantages of DML methods are that they do not require parametric assumptions and that they can handle a potentially large covariate space while maintaining desirable statistical properties.

In the empirical application, the analysis focuses on identifying which implementation aspects are most beneficial for individuals selected to participate in active labor market policies (ALMP). ALMP are interventions designed to improve employment outcomes for unemployed individuals and are widely used by governments as policy tools.\footnote{For an overview, refer to reviews and meta-studies by \citet{card2010active,card2018works,crepon2016active,vooren2019effectiveness}} In Switzerland, which is the focus of this paper, two-thirds of unemployed are assigned to these measures within the first twelve months of their unemployment period, with approximately 60\% of them participating in more than one program. The application is the first to jointly address dynamic confounding and dynamic policies in evaluating ALMP sequences, whereas prior work \citep{lechner2013does, laffers2024locking} addressed only the former. Specifically, it is shown that introducing a dynamic policy based on intermediate outcomes allows to identify a more practically relevant estimand with only minor modifications to the identifying assumptions. Furthermore, the application innovates by applying flexible dynamic DML estimation in the context of ALMP, offering more reliable effect estimates compared to prior sequential studies that relied on parametric estimators. Leveraging DML in combination with a large dataset containing extensive intermediate details about the unemployed also allows to assess group-level effect heterogeneity in a sequential ALMP evaluation. The analysis shows that a particular temporary wage subsidy is most effective on average, even after aligning program duration across programs. Moreover, individuals with limited language skills profit more from this program in comparison to extended training courses. The findings emphasize the practical value of employing DML-based estimation for the empirical evaluation of programs sequences, allowing policymakers to develop more nuanced and targeted policies.

The remainder of the paper is structured as follows: The following Section~\ref{sec:dyneffect} introduces the notation, the conceptual framework, the estimands of interest and their identification. Section~\ref{sec:dynDML}, presents different estimators for DML under static and dynamic confounding and discusses their properties. Section~\ref{sec:application} applies these estimators to the empirical application, followed by the conclusion in Section~\ref{sec:conclusion}. Further derivations and additional results are provided in the Appendix.

\section{Identifying Effects of Sequential Policies} \label{sec:dyneffect}

\subsection{Notation}

This section introduces the framework for assessing sequential policies, drawing primarily on \citet{hernan2020causal}, while employing a notation more commonly used in econometrics. The framework is presented in a setting with $T=2$ time periods, as used in the application in Section \ref{sec:application}. For clarity, the presentation avoids using a general $T$, although the framework naturally extends to cases with $T>2$. When $T=1$, it coincides with the standard setting of a single time-point intervention. Let $\mathcal{D}_t = \left\{0,1,...,M_t\right\}$ denote the set of existing programs in period $t\in \left\{1,2\right\}$ and $D_{i,t}$ a discrete random variable indicating the observed program of individual $i \in \left\{1,...,N\right\}$ in period $t$. Prior to each program, a vector of covariates $X_{i,t-1} \in \mathcal{X}_{t-1}$ is observed, which may contain lagged outcomes $Y_{i,t-1} \in \mathcal{Y}_t$. The main outcome of interest is observed after the final treatment $Y_{i}:=Y_{i,T}=Y_{i,2}$, where the time subscript is dropped for the sake of readability. This variable may include outcome information from any period following the start of the final treatment. In general, capital letters (except for $T$, $M$ and $N$) denote random variables, lower case letters denote their realizations and boldface letters denote vectors of variable histories up to $t$, i.e. $\mathbf{X}_{i,t} = (X_{i,0}, ..., X_{i,t})$ and $\mathbf{d}_t = (d_1, ..., d_t)$ such that for example $\mathbf{d}_1=d_1$ and $\mathbf{d}_2=(d_1, d_2)$. To simplify the notation further, the individual identifier $i$ is omitted when not explicitly needed and all random variables for which realizations can be observed are collected in the vector $\mathbf{W}_2=(Y,\,\mathbf{D}_2,\,\mathbf{X}_1)$. 

The potential outcome framework of \citet{rubin1974estimating} is adopted to study causal effects, where $Y^{\mathbf{d}_2}$ denotes the hypothetical outcome under a particular treatment sequence $\mathbf{d}_2 \in \mathcal{D}_1 \times \mathcal{D}_2$. Similarly, potential values of the time-varying covariates are defined as $X_1^{d_1}$. The relationship between potential and observed variables follows the standard observation rule, where an observed variable is assumed to correspond to the potential variable associated with the assigned treatments.
\begin{assumption}
\label{as:sutva}
Stable Unit Treatment Value Assumption (SUTVA)
\begin{enumerate}[label={[\alph*]}]
    \item \label{as:sutvaY} $Y = \sum_{\mathbf{d}_2 \in \mathcal{D}_1 \times \mathcal{D}_2} \mathbbm{1}\left\{\mathbf{D}_2 = \mathbf{d}_2\right\} Y^{\mathbf{d}_2}$
    \item \label{as:sutvaX1} $X_1 = \sum_{d_1 \in \mathcal{D}_1} \mathbbm{1}\left\{D_1 = {d}_1\right\} X_1^{d_1}$
\end{enumerate}
\end{assumption}
This assumption implicitly requires that there are no unrepresented programs in the population of interest (everyone is assigned to a particular program sequence $\mathbf{d}_2$) and that there are no relevant interactions between individuals, meaning that the program of one individual does not affect the final outcome and intermediate covariates of another individual.

\subsection{Conceptual Framework}

The key challenge in analyzing program sequences from observational data is that the program assignment mechanism underlying the observed data might be affected by time-varying feedback between treatments, covariates and outcomes. Hence, to identify the causal effect, it is essential to consider that sequential treatment assignment is a decision process potentially influenced by dynamic confounding. 
\begin{definition}[Confounding] \label{def:dynamic_confounding}
    A sequential program assignment mechanism exhibits confounding if $\E[Y^{\mathbf{d}_2}] \neq \E[Y|\mathbf{D}_2 = \mathbf{d}_2]$. It is characterized by static confounding if $\E[Y^{\mathbf{d}_2}|X_0] = \E[Y|\mathbf{D}_2 = \mathbf{d}_2,X_0]$ and by dynamic confounding if $\E[Y^{\mathbf{d}_2}|X_0] \neq \E[Y|\mathbf{D}_2 = \mathbf{d}_2,X_0]$ \citep{hernan2020causal}.
\end{definition}
According to this definition, confounding is static if the expected outcome of those treated with $\mathbf{D}_2=\mathbf{d}_2$ equals the expected outcome if everyone were treated with $\mathbf{d}_2$, conditional on pre-treatment information. This implies that the entire treatment sequence is pre-determined given $X_0$. Conversely, if there is feedback between treatments, covariates, and outcomes across periods, this equality will generally not hold, and the confounding is considered dynamic. Under dynamic confounding, controlling only for pre-treatment information often proves inadequate when assessing the impacts of treatment sequences. Therefore, existing research has developed alternative identification strategies that rely on modified identification assumptions and also imply new challenges for effect estimation \citep{robins2009estimation}.

The causal diagram \citep{pearl1995causal} in Figure \ref{fig:mainDAG} illustrates confounding the two-period setup. Prior to the first period, the pre-treatment covariates $X_0$ are observed. These variables may have a causal effect (arrow) on any future treatment, intermediate covariate, or outcome. In period $t=1$, individuals are treated with $D_1$, which may impact their time-varying characteristics in the current and subsequent periods as well as any future treatment assignments. The same happens in period $t=2$. Under dynamic confounding, the initial treatment assignment $D_1$ induces changes in the covariates $X_1$, which subsequently influence the second treatment $D_2$ and the outcome $Y$, as highlighted by the blue bold arrows. In contrast, if any one of the bold blue arrows is absent, confounding is considered static, meaning that program assignment in all periods is predetermined conditional on pre-treatment information.\footnote{In absence of a causal pathway from $D_1$ to $X_1$, the variable $X_1$ can also considered to be a pre-treatment variable.}

\begin{figure}[ht]
    \centering
    \captionsetup{font=small, width=0.6\linewidth}
    \caption{Causal pathways in the sequential treatment effect model with two time periods}
    \begin{tikzpicture}[node distance=3cm]
    \node (D1) [c_solid, align=center] {$D_1$};
    \node (X1) [c_solid, below right of=D1, align=center, draw=blue] {\textcolor{blue}{$X_1$}};
    \node (D2) [c_solid,  above right of=X1, align=center] {$D_2$};
    \node (X0) [c_solid, below left of=D1, align=center] {$X_0$};
    \node (Y2) [c_solid, below right of=D2, align=center] {$Y$};
    \draw [arrow] (X0) -- (D1);
    \draw [arrow] (X0) -- (D2);
    \draw [arrow, bend right=20] (X0) to (Y2);
    \draw [arrow] (X0) -- (X1);
    \draw [arrow] (D1) -- (D2);
    \draw [arrow, color=blue, line width=1.5] (D1) -- (X1);
    \draw [arrow] (D1) -- (Y2);
    \draw [arrow, color=blue, line width=1.5] (X1) -- (Y2);
    \draw [arrow, color=blue, line width=1.5] (X1) -- (D2);
    \draw [arrow] (D2) -- (Y2);
\end{tikzpicture}
    \caption*{\scriptsize \textit{Notes:} The arrows in the diagram illustrate the allowed causal pathways between the variables in the sequential model with dynamic confounding. At any point in time $t$, covariates $X_t$ and treatments $D_t$ may influence any future treatment, covariate, or outcome. If one of the blue bold arrows is deleted from the figure, confounding is static, meaning that the treatment sequence is predetermined conditional on covariates.}
    \label{fig:mainDAG}
\end{figure}

Besides addressing dynamic confounding, dynamics also need to be considered in the design of counterfactual scenarios. While confounding is determined by the observational setting, counterfactuals can be freely specified to meet the research objectives. However, as discussed below, aligning counterfactuals with the assumed confounding structure can substantially facilitate identification and estimation. In the sequential setting, static counterfactuals that do not depend on time-varying covariates, such as ``two consecutive periods of a training course,'' are often of limited relevance because they fail to account for the possibility of dynamic decision-making \citep{wager2024causal}. For instance, the mean potential outcome for such a sequence reflects a scenario in which all individuals remain in the program for the entire sequence. This implies continued participation even in cases where individuals would no longer be eligible. Dynamic decision-making can be formalized through the concept of dynamic policies\footnote{Policies are also called treatment rules or regimes. Following \citet{wager2024causal}, here they are denoted as policies as this is the standard terminology in the econometrics literature.} \citep{murphy2003optimal}. 

\begin{definition}[Dynamic policy] \label{def:dynamic_policy}    
    Let $\mathcal{R}_t \subseteq \mathcal{D}_t$ denote a subset of the possible treatments in period~$t \in \{1,2\}$. Let $V_{t-1}$, with realizations $v_{t-1} \in \mathcal{V}_{t-1}$, be a vector of decision variables consisting of a subset of the covariates $X_{t-1}$. Denote by $V_1^{d_1}$ the potential decision variables at time 1 if the first‐period treatment were $d_1$. Then, a dynamic policy is defined as a pair of deterministic functions $\mathbf{g}_2 = (g_1, g_2)$, where each $g_t: \mathcal{V}_0 \times ... \times \mathcal{V}_{t-1} \rightarrow \mathcal{R}_t$
    assigns a treatment $g_t(\mathbf{v}_{t-1}) \in \mathcal{R}_t$ based on the history of decision variables up to time $t-1$. The potential decision variables and potential outcomes associated with a policy $\mathbf{g}_2$ are defined as
    \begin{align*}
      V_1^{g_1} &:= \sum_{d_1 \in \mathcal{D}_1 }\mathbbm{1}\left\{{g}_1({V}_0) = {d}_1\right\} V_1^{{d}_1}\quad \text{and}\\
      Y^{\mathbf{g}_2} &:= \sum_{\mathbf{d}_2 \in \mathcal{D}_1 \times \mathcal{D}_2} \mathbbm{1}\left\{\mathbf{g}_2(\mathbf{V}_1^{d_1}) = \mathbf{d}_2\right\} Y^{\mathbf{d}_2} \quad \text{with} \quad \mathbf{V}_1^{d_1} = (V_0, V_1^{d_1})\text{.}
\end{align*}
\end{definition}

The policy $\mathbf{g}_2$ is defined as a mapping from the (potential) decision variables to a subset of the possible treatment states. In the most flexible setting, decision variables can include all covariates, i.e. $\mathbf{V}^{d_1}_1=\mathbf{X}^{d_1}_1$, and the policy can return any possible program, i.e. $\mathcal{R}_t = \mathcal{D}_t$. Often, however, it is useful to align $\mathbf{V}^{d_1}_1$ and $\mathcal{R}_t$ with the requirements of the counterfactual scenario under consideration. For instance, the subsequent application will define $\mathbf{V}^{d_1}_1$ in a way to ensure that policies can depend on the evolution of the outcome variable. Importantly, while dynamic policies may be restricted to depend on selected decision variables, such restrictions do not limit the nature of confounding present in the data.

\begin{example} \label{ex:1}
    A dynamic policy of interest might be specified as: ``Assign to program $d$ in the first period. Continue program $d$ in the second period if the potential intermediate outcome $Y_1^{d}$ equals zero, otherwise assign program $d'$.'' This policy can be formalized as
    \begin{align*}
    g_1(V_0) = d \quad \text{and} \quad g_2(\mathbf{V}_1^{d}) = d \cdot \mathbbm{1}\left\{Y_1^{d} = 0\right\} + d' \cdot \mathbbm{1}\left\{Y_1^{d} \neq 0\right\}\text{,}
\end{align*}
where the potential intermediate outcome is the only decision variable and the range of the policy in the first and second period equals $\mathcal{R}_1 = \left\{d\right\}$ and $\mathcal{R}_2 = \left\{d, d'\right\}$, respectively.
\end{example}

A key property of dynamic policies is that two individuals following the same policy $\mathbf{g}_2$ may follow different program sequences $\mathbf{d}_2$ depending on their covariates. For instance, if $Y^d_1$ in Example~\ref{ex:1} represents a binary employment indicator, then (i) individuals who remain unemployed after receiving treatment $d$ in the first period under the sequence $(d,d)$, and (ii) individuals who become employed and follow sequence $(d,d')$, would both comply with the policy. A sequence $\mathbf{d}_2$ constitutes a specific instance of a policy, where $\mathbf{g}_2$ is expressed as a constant function. This type of policy is referred to as a \textit{static policy}. In what follows, $\mathbf{d}_2$ is used to represent static policies exclusively, whereas $\mathbf{g}_2$ is used in contexts that encompass both static and dynamic policies. An overview of possible scenarios involving static and dynamic confounding and policies, along with accompanying examples, is given in Table \ref{tab:summary_dynamics}.

\begin{table}[ht]
\caption{The possible dynamic scenarios with accompanying examples. }
\label{tab:summary_dynamics}
\centering
\renewcommand{\arraystretch}{1.5}
\begin{tabular}{m{0.15\textwidth}  m{0.04\textwidth} |>{\centering}m{0.35\textwidth} | >{\centering\arraybackslash}m{0.35\textwidth} |}
\cline{3-4}
&&\multicolumn{2}{c|}{\makecell[l]{\\[-0.2cm]
$\quad\quad\quad\quad$Treatment assignment based on time-varying covariates in\\
\textbf{$\quad\quad\quad\quad\quad\quad\;\;$assignment mechanism underlying the data?}}}\\
&&\textit{NO}&\multicolumn{1}{c|}{\textit{YES}}\\ \cline{1-4}
\multicolumn{1}{|m{0.15\textwidth}}{\multirow[c]{8}{=}{\makecell[l]{Treatment\\ assignment\\based on\\ time-varying\\covariates in\\\textbf{counterfactual}\\\textbf{of interest?} \\} }} & 
\multirow[c]{3}{=}{\textit{NO}}& 
{\footnotesize \textbf{Static policy under static confounding:}}& 
{\footnotesize \textbf{Static policy under dynamic confounding:}}\\[-.2cm] 
\multicolumn{1}{|m{0.15\textwidth}}{}&&
{\footnotesize \textit{E.g.,} effect of two-period training course estimated from data with fixed initial treatment assignment.}&
{\footnotesize \textit{E.g.,} effect of two-period training course estimated from data with time-varying treatment assignment.}\\
\cline{2-4}
\multicolumn{1}{|m{0.15\textwidth}}{}& 
\multirow[c]{4}{=}{\textit{YES}}&
{\footnotesize \textbf{Dynamic policy under static confounding:}}&
{\footnotesize \textbf{Dynamic policy under dynamic confounding:}}\\[-.2cm] 
\multicolumn{1}{|m{0.15\textwidth}}{}&&
{\footnotesize \textit{E.g.,} effect of training course, extending to period 2 only if eligible throughout period 1, estimated from data with fixed initial treatment assignment.}&
{\footnotesize \textit{E.g.,} effect of training course, extending to period 2 only if eligible throughout period 1, estimated from data with time-varying treatment assignment.}\\
\cline{1-4}
\end{tabular}
\caption*{{\scriptsize\textit{Note:} Overview of the possible dynamic scenarios. Depending on the assumptions about the underlying assignment mechanism and the counterfactual of interest, different estimands may be analyzed.}}
\end{table}

\subsection{Estimation Targets}

The analysis aims at evaluating effects of sequential policies at different levels of granularity. At the broadest level of the entire population, the primary target parameter is the average potential outcome (APO) under a particular policy, defined as
\begin{align*}
   \theta^{\mathbf{g}_2} := \E[Y^{\mathbf{g}_2}]\text{.}
\end{align*}
This parameter represents the mean outcome if all individuals were assigned according to the policy $\mathbf{g}_2$ (or according to the sequence $\mathbf{d}_2$ if the policy is static). To assess heterogeneity, the focus can be switched to a specific subgroup of interest, for example to individuals that participated in a labor market program during a previous unemployment spell. This can be studied by examining the group average potential outcome (GAPO)
\begin{align*}
    \theta^{\mathbf{g}_2}(z_0) := \E[Y^{\mathbf{g}_2}| Z_0 = z_0]\text{,}
\end{align*}
where $Z_0$ is a column or deterministic function of $X_0$ with low cardinality.\footnote{Subgroups are defined based on pre-treatment covariates, since estimands involving time-varying covariates are not identifiable under the assumptions stated below.} The following discussion of identification and estimation will focus on the parameters $\theta^{\mathbf{g}_2}$ and $\theta^{\mathbf{g}_2}(z_0)$, respectively. The average treatment effect (ATE) between implementing policy $\mathbf{g}_2$ and alternative policy $\mathbf{g}'_2$ is obtained subsequently by taking the difference between two average potential outcomes, i.e. 
\begin{align*}
    \tau^{\mathbf{g}_2, \mathbf{g}'_2} := \E[Y^{\mathbf{g}_2}-Y^{\mathbf{g}'_2}] = \theta^{\mathbf{g}_2} - \theta^{\mathbf{g}'_2}\text{,}
\end{align*}
which directly follows from the linearity of the expectation operator. Subgroup-specific average treatment effects (GATE) are similarly obtained as
\begin{align*}
    \tau^{\mathbf{g}_2, \mathbf{g}'_2}(z_0) = \E[Y^{\mathbf{g}_2}-Y^{\mathbf{g}'_2}|Z_0=z_0] 
    = \theta^{\mathbf{g}_2}(z_0) - \theta^{\mathbf{g}'_2}(z_0) \text{.}
\end{align*}
Heterogeneous treatment effects of this type are also referred to as conditional average treatment effects (CATE) in the literature. Following \citet{knaus2021machine}, they are denoted as GATE to emphasize the focus on large, discrete subgroups of the population rather than granular individualized effects.\footnote{More granular individualized effects are not considered because DML-based estimators of such effects lack statistical guarantees and perform poorly in finite samples under confounding, as shown for example by \citet{lechner2024comprehensive} for single time-point interventions. The development of robust estimation methods for individualized effects in the sequential setting remains an open question for future research beyond the scope of this study.} 

\subsection{Identification of Static Policies}

The prior section defined the estimands of interest using potential outcomes. However, since each individual is observed in only one particular program sequence, the remaining potential outcomes are unobservable, necessitating additional assumptions for their identification. The assumptions required vary based on the type of confounding and the nature of the policy of interest. Before introducing dynamic policies, the identification assumptions for static policies are revisited.

\subsubsection{Static Confounding}
In the presence of static confounding, the APO of a static policy $\mathbf{g}_2(\mathbf{V}^{g_1}_1) = \mathbf{d}_2$ can be identified if the following conditional independence assumption (CIA) and overlap assumption are satisfied.

\begin{assumption}[Identification assumptions for static policies under static confounding]$ $
\label{as:identification_static}
\begin{enumerate}[label={[\alph*]}]
    \item \label{as:CIA_static}
    \makebox[1.3cm][l]{CIA}
    \makebox[1.5cm][l]{($t=1$):}
    \makebox[1.8cm][l]{$\forall x_0 \in \mathcal{X}_0:$}
    \makebox[1.8cm][r]{$Y^{\mathbf{d}_2}$}
    \makebox[4cm][l]{$\ind D_1 | X_{0}=x_0 $}\\
    \makebox[1.3cm][l]{}
    \makebox[1.5cm][l]{($t=2$):}
    \makebox[1.8cm][l]{$\forall x_0 \in \mathcal{X}_0:$}
    \makebox[1.8cm][r]{$Y^{\mathbf{d}_2}$}
    \makebox[4cm][l]{$\ind D_2 | X_{0}=x_0, D_1=d_1$}
    \item \label{as:CS_static}
    \makebox[1.3cm][l]{Overlap}
    \makebox[1.5cm][l]{($t=1$):}
    \makebox[1.8cm][l]{$\forall x_0 \in \mathcal{X}_0:$}
    \makebox[1.7cm][r]{$p_{d_1}(x_0)$}
    \makebox[5.7cm][l]{$:=\Pr(D_1=d_1|X_0=x_0)$}
    \makebox[0.5cm][l]{$>0$}\\
    \makebox[1.3cm][l]{}
    \makebox[1.5cm][l]{($t=2$):}
    \makebox[1.8cm][l]{$\forall x_0 \in \mathcal{X}_0:$}
    \makebox[1.7cm][r]{$p_{d_2}(x_0)$}
    \makebox[5.7cm][l]{$:=\Pr(D_2=d_2|X_0=x_0, D_1=d_1)$}
    \makebox[0.5cm][l]{$>0$}
\end{enumerate}
\end{assumption}

Assumption \ref{as:identification_static}\ref{as:CIA_static} is called full conditional independence assumption in \citet{lechner2001potential}. It requires that potential outcomes are independent of program assignment in the first period for given values of the pre-treatment covariates. Thus, $X_0$ must include all variables that influence both the first period program assignment and the outcomes simultaneously. Additionally, the assumption asserts that allocation to a program in the second period is random, conditional on pre-treatment covariates and prior program participation. This implies that intermediate characteristics $X_1$ are either unaffected by previous programs or do not simultaneously influence both program assignment in the second period and potential outcomes. Furthermore, Assumption \ref{as:identification_static}\ref{as:CS_static} requires that for a given history of previous program participation and any realization of pre-treatment characteristics, it must be possible to observe individuals with a program $D_t = d_t$. Otherwise it would be impossible to construct appropriate counterfactuals. \citet{lechner2001potential} show that Assumption \ref{as:identification_static} can be re-written using basic probability theory:
\begin{assumption}[Identification assumptions for static policies under static confounding (alternative formulation)]$ $
\label{as:identification_static2}
\begin{enumerate}[label={[\alph*]}]
    \item \label{as:CIA_static2}
    \makebox[2cm][l]{CIA:}
    \makebox[1.8cm][l]{$\forall x_0 \in \mathcal{X}_0:$}
    \makebox[1.8cm][r]{$Y^{\mathbf{d}_2}$}
    \makebox[4cm][l]{$\ind \mathbf{D}_2 | X_{0}=x_0 $}
    \item \label{as:CS_static2}
    \makebox[2cm][l]{Overlap:}
    \makebox[1.8cm][l]{$\forall x_0 \in \mathcal{X}_0:$}
    \makebox[1.7cm][r]{$p_{\mathbf{d}_2}(x_0)$}
    \makebox[3.7cm][l]{$:=\Pr(\mathbf{D}_2=\mathbf{d}_2|X_0=x_0)>0$}
\end{enumerate}

\end{assumption}
This formulation shows that the sequential framework with static confounding is equivalent to the static single-period setting for multiple treatments \citep{Imbens:2000,lechner2001identification}, where sequence $\mathbf{d}_2$ is considered a single treatment state, and identification is achieved by conditioning on pre-treatment covariates only. It is simple to show that APO and GAPO of a static policy $\mathbf{g}_2(\mathbf{V}_1) = \mathbf{d}_2$ can be identified under this assumption. The proof follows the standard identification argument for single time-point interventions as seen e.g. in \citet{rosenbaum1983central}.\footnote{While the focus is on constant static policies defined as a fixed program sequence $\mathbf{d}_2$, the results can be generalized to static policies $h_t: \mathcal{V}_0 \rightarrow \mathcal{D}_t$ that depend on pre-treatment decision variables $V_0$ but do not depend on time-varying information. Specifically, $E[Y^{\mathbf{h}_2}]$ is identified under the condition that Assumption \ref{as:identification_static} or \ref{as:identification_static2} holds for all $\mathbf{d}_2$ within the range of policies $\mathbf{h}_2(V_0)$. Similar identification results are commonly applied in optimal policy learning in the single-period setting \citep[e.g.][]{Athey:2021}.} 

\begin{theorem}[Identification of static policies under static confounding]$ $\label{th:identification_static}\\
Define $\mu_{\mathbf{d}_2}(X_0) := \E[Y|X_0, \mathbf{D}_2=\mathbf{d}_2]$. Assume that either Assumption \ref{as:identification_static} or Assumption \ref{as:identification_static2} holds. Then $\mu_{\mathbf{d}_2}(X_0) = \E[Y^{\mathbf{d}_2}|X_0]$ and the estimands are identified as 
    \begin{align*}
        \theta^{\mathbf{d}_2} = \E_{X_0}[\mu_{\mathbf{d}_2}(X_0)] \quad \text{and} \quad \theta^{\mathbf{d}_2}(z_0) = \E_{X_0}[\mu_{\mathbf{d}_2}(X_0)|Z_0=z_0]\text{.}
    \end{align*}
\end{theorem}

\subsubsection{Dynamic Confounding}
In observational settings, static confounding may be implausible if the assignment process underlying the data depends on time-varying information. Causal effects under dynamic confounding have first been studied in epidemiology and biostatistics starting with the seminal work by \citet{robins1986new, robins1987addendum}. He demonstrated that under feedback effects between treatments and covariates, standard approaches do not allow for a causal comparison of treatment sequences, even when all pre-treatment confounding factors are controlled for. Instead, using a graphical model, \citet{robins1986new} came up with the idea of what is called today a sequential randomized experiment and the sequential randomization assumption \citep{richardson2014causal}. For static policies, these ideas have been introduced to econometrics by \citet{lechner2009sequential} and \citet{lechner2010identification} who termed the requirement for identification as weak dynamic conditional independence:

\begin{assumption}[Identification assumptions for static policies under dynamic confounding]$ $
\label{as:identification}
\begin{enumerate}[label={[\alph*]}]
    \item \label{as:CIA}
    \makebox[1.3cm][l]{CIA}
    \makebox[1.5cm][l]{($t=1$):}
    \makebox[2.8cm][l]{$\forall x_0 \in \mathcal{X}_0:$}
    \makebox[1.8cm][r]{$Y^{\mathbf{d}_2}$}
    \makebox[4cm][l]{$\ind D_1 | X_0=x_0$}\\
    \makebox[1.3cm][l]{}
    \makebox[1.5cm][l]{($t=2$):}
    \makebox[2.8cm][l]{$\forall \mathbf{x}_1 \in \mathcal{X}_0 \times \mathcal{X}_1:$}
    \makebox[1.8cm][r]{$Y^{\mathbf{d}_2} $}
    \makebox[4cm][l]{$\ind D_2 | \mathbf{X}_{1}=\mathbf{x}_1, D_1=d_1$}
    \item \label{as:CS}
    \makebox[1.3cm][l]{Overlap}
    \makebox[1.5cm][l]{($t=1$):}
    \makebox[2.8cm][l]{$\forall x_0 \in \mathcal{X}_0:$}
    \makebox[1.7cm][r]{$p_{d_1}(x_0)$}
    \makebox[5.7cm][l]{$:=\Pr(D_1=d_1|X_0=x_0)$}
    \makebox[0.5cm][l]{$>0$}\\
    \makebox[1.3cm][l]{}
    \makebox[1.5cm][l]{($t=2$):}
    \makebox[2.8cm][l]{$\forall \mathbf{x}_1 \in \mathcal{X}_0 \times \mathcal{X}_1:$}
    \makebox[1.7cm][r]{$p_{d_2}(d_1, \mathbf{x}_1)$}
    \makebox[5.7cm][l]{$:=\Pr(D_2=d_2|\mathbf{X}_1 = \mathbf{x}_1, D_1=d_1)$}
    \makebox[0.5cm][l]{$>0$}
\end{enumerate}
\end{assumption}

For $t=1$, Assumption \ref{as:identification} is equivalent to Assumption \ref{as:identification_static}. However, in the second period, conditioning on pre-treatment characteristics $X_0$ and the treatment history no longer suffices to establish independence between potential outcomes and treatment state $D_2$. Instead, Assumption \ref{as:identification}\ref{as:CIA} requires conditioning on the whole covariate history $\mathbf{X}_{1}$, which may be influenced by previous treatments. Hence, additional information about time-varying covariates that are expected to determine dynamic program selection is necessary. Importantly, these covariates may not be influenced by programs in future periods in a way that is related to the outcome variable.\footnote{This exogeneity requirement can be explicitly stated as $\forall \mathbf{d}_2, \mathbf{d}'_2 \in \mathcal{D}_1 \times \mathcal{D}_2: X_0 ^{\mathbf{d}_2} = X_0^{\mathbf{d}'_2}$ and $X_1^{{d}_2} = X_1^{{d}'_2}$, where $X_t^{\mathbf{d}_2}$ denotes the potential covariates in period $t$ under treatment sequence $\mathbf{d}_2$.} The overlap assumption \ref{as:identification}\ref{as:CS} also requires additional conditioning on time-varying covariates for the treatment probability in the second period. Under Assumption \ref{as:identification}, the APO and GAPO are identified, as stated in the following theorem:

\begin{theorem}[Identification of static policies under dynamic confounding]$ $\label{th:identification}\\Define $\mu_{\mathbf{d}_2}(\mathbf{X}_1) := \E[Y|\mathbf{X}_1, \mathbf{D}_2=\mathbf{d}_2]$ and $\nu_{\mathbf{d}_2}(X_0) := \E_{X_1} \big[ \mu_{\mathbf{d}_2}(\mathbf{X}_1) \big| X_0, D_1=d_1\big]$. Assume that Assumption \ref{as:identification} holds. Then $\nu_{\mathbf{d}_2}(X_0) = \E[Y^{\mathbf{d}_2}|X_0]$ and the estimands are identified as 
    \begin{align*}
        \theta^{\mathbf{d}_2} = \E_{X_0}[\nu_{\mathbf{d}_2}(X_0)] \quad \text{and} \quad \theta^{\mathbf{d}_2}(z_0) = \E_{X_0}[\nu_{\mathbf{d}_2}(X_0)|Z_0=z_0]\text{.}
    \end{align*}
\end{theorem}

Hence, the identification result for dynamic confounding is analogous to the case of static confounding, with $\mu_{\mathbf{d}_2}(X_0)$ replaced by $\nu_{\mathbf{d}_2}(X_0)$.\footnote{Assumption \ref{as:identification} allows to identify additional parameters such as the average potential outcome within a particular treatment group in the first period $\E[Y^{\mathbf{d}_2}|D_1=d'_1] \; \forall\; d'_1 \in \mathcal{D}_1$. However, identification for subgroups defined by later periods, i.e. $\E[Y^{\mathbf{d}_2}|\mathbf{D}_2=\mathbf{d}'_2]$, is not possible under Assumption \ref{as:identification} if $d_1 \neq d_1'$, as noted by \citet{lechner2010identification}.} The proof of Theorem \ref{th:identification} first establishes that $\mu_{\mathbf{d}_2}(\mathbf{X}_1) = \E[Y^{\mathbf{d}_2}|\mathbf{X}_1, D_1=d_1]$. However, due to potential feedback between the treatments through $X_1$, the conditional outcome $\mu_{\mathbf{d}_2}(\mathbf{X}_1)$ cannot simply be averaged over the population to determine the APO. Instead, an extra averaging step over the  conditional distribution of covariates $X_1$ is required. Identification through $\nu_{\mathbf{d}_2}(X_0)$ is known in the literature as the $g$-formula or iterated conditional expectations \citep{robins1986new, robins1987addendum, hernan2020causal}.

\subsection{Identification of Dynamic Policies} 

As previously discussed, prior econometric applications have mainly focused on static policies while program assignment in practice often involves dynamic decision-making. When considering dynamic policies, it is necessary to adjust identification assumptions, as they are based on intermediate \textit{potential} decision variables. As a result, the counterfactual scenarios become more complex, since the policy can deterministically depend on these intermediate variables. For instance, the policy defined in Example \ref{ex:1} assigns an individual to program $d_2$ or $d'_2$ depending on whether the potential decision variable in the first period equals one or not. These potential intermediate values need to be considered in the conditional independence assumption.

\begin{assumption}[Identification assumptions for dynamic policies with dynamic confounding]$ $
\label{as:identification_dyn}
\begin{enumerate}[label={[\alph*]}]
    \item \label{as:CIA_dyn}
    \makebox[1.3cm][l]{CIA}
    \makebox[1.3cm][l]{($t=1$):}
    \makebox[5.8cm][l]{$\forall x_0 \in \mathcal{X}_0, \;\forall\mathbf{d}_2 \in \mathcal{R}_1 \times \mathcal{R}_2:$}
    \makebox[1.5cm][r]{$\big\{Y^{\mathbf{d}_2}, V_1^{d_1} \big\}$}
    \makebox[2.6cm][l]{$\ind D_1 | X_0=x_0 $}\\
    \makebox[1.3cm][l]{}
    \makebox[1.3cm][l]{($t=2$):}
    \makebox[5.8cm][l]{$\forall \mathbf{x}_1 \in \mathcal{X}_0 \times \mathcal{X}_1,\;\forall \mathbf{d}_2 \in \mathcal{R}_1 \times \mathcal{R}_2:$}
    \makebox[1.5cm][r]{$Y^{\mathbf{d}_2}$}
    \makebox[2.6cm][l]{$\ind D_2 | \mathbf{X}_{1}=\mathbf{x}_1, D_1=g_1(v_0)$}
    \item \label{as:CS_dyn}
    \makebox[1.3cm][l]{Overlap}
    \makebox[1.3cm][l]{($t=1$):}
    \makebox[2.6cm][l]{$\forall x_0 \in \mathcal{X}_0:$}
    \makebox[1.7cm][r]{$p_{g_1}(x_0)$}
    \makebox[7cm][l]{$:=\Pr(D_1=g_1(v_0)|X_0=x_0)$}
    \makebox[0.5cm][l]{$>0$}\\
    \makebox[1.3cm][l]{}
    \makebox[1.3cm][l]{($t=2$):}
    \makebox[2.6cm][l]{$\forall \mathbf{x}_1 \in \mathcal{X}_0 \times \mathcal{X}_1:$}
    \makebox[1.7cm][r]{$p_{g_2}(g_1, \mathbf{x}_1)$}
    \makebox[7cm][l]{$:=\Pr(D_2=g_2(\mathbf{v}_1)|\mathbf{X}_1 = \mathbf{x}_1, D_1=g_1(v_0))$}
    \makebox[0.5cm][l]{$>0$}
\end{enumerate}
\end{assumption}

Assumption \ref{as:identification_dyn} extends the notion of `full' conditional exchangeability in \citet{robins2009estimation} by employing $\mathbf{V}_1^{d_1}$ instead of $\mathbf{X}_1^{d_1}$ and $\mathcal{R}_t$ instead of $\mathcal{D}_t$. Compared to Assumption \ref{as:identification} for static policies, Assumption \ref{as:identification_dyn} introduces two key differences. First, conditional independence and overlap must hold for all possible treatment sequences that the dynamic policy could generate, rather than just for one prespecified sequence. Second, given the pre-treatment characteristics, the assignment in the first period must be independent not only of any potential final outcomes but also of the potential intermediate variables that govern the second-period treatment. This additional requirement ensures that no unmeasured confounding affects the relationship between the first-period treatment and these intermediate variables. For example, consider an unmeasured confounder that influences both $D_1$ and $V^{{g}_1}_1$ but does not affect $Y^{\mathbf{g}_2}$. In this case, $\E[Y^{\mathbf{g}_2}]$ is not identifiable as $\mathbf{g}_2$ is a function of $V^{{g}_1}_1$ \citep{robins1986new, robins2009estimation}.

Importantly, in contrast to the existing literature that defines policies as functions of the entire covariate vector, the present framework introduces a distinction between confounders and decision variables. Consequently, even though all covariates $\mathbf{X}_1$ might influence the underlying dynamic selection, conditional independence of the treatment in $t=1$ needs to hold only with respect to the potential $V_1^{d_1}$ that are part of the dynamic policy (and $Y^{\mathbf{d}_2}$) but not with respect to all $X_1^{d_1}$. Therefore, the choice of the structure of the dynamic policy critically influences the increased restrictiveness of Assumption \ref{as:identification_dyn}. For instance, in Example~\ref{ex:1}, the dynamic policy depends only on potential intermediate outcomes, making the additional requirements less demanding. This is because, in many practical contexts, the CIA between the first-period treatment and the final outcome naturally extends to intermediate outcomes as well. Furthermore, if the chosen dynamic policy $\mathbf{g}_2$ more closely aligns with the assignment process underlying the observed data than a static policy $\mathbf{d}_2$, the overlap condition in Assumption \ref{as:identification_dyn}\ref{as:CS_dyn} becomes more credible than its static counterpart \ref{as:identification}\ref{as:CS}. An example of such a setting is provided below in the empirical application.

\begin{theorem}[Identification of dynamic policies under dynamic confounding]$ $\label{th:identification_dyn}\\Define $\mu_{\mathbf{g}_2}(\mathbf{X}_1) := \E[Y|\mathbf{X}_1, \mathbf{D}_2=\mathbf{g}_2(\mathbf{V}_1)]$ and $\nu_{\mathbf{g}_2}(X_0) := \E_{X_1} \big[ \mu_{\mathbf{g}_2}(\mathbf{X}_1) \big| X_0, D_1=g_1(V_0)\big]$. Assume Assumption \ref{as:identification_dyn} holds. Then $\nu_{\mathbf{g}_2}(X_0) = \E[Y^{\mathbf{g}_2}|X_0]$ and the estimands are identified as 
    \begin{align*}
        \theta^{\mathbf{g}_2} = \E_{X_0}[\nu_{\mathbf{g}_2}(X_0)] \quad \text{and} \quad \theta^{\mathbf{g}_2}(z_0) = \E_{X_0}[\nu_{\mathbf{g}_2}(X_0)|Z_0=z_0]\text{.}
    \end{align*}
\end{theorem}

Under Assumption \ref{as:identification_dyn}, identification is achieved in a similar way as for static policies, as shown in Theorem \ref{th:identification_dyn}. Hence, despite the need for additional identification conditions under dynamic policies, the logic of identification mirrors that of static policies once these are met. As shown, the average outcome associated with following policy $\mathbf{g}_2(\mathbf{V}_1^{g_1})$ can be inferred from individuals whose observed treatments and covariates align with following strategy $\mathbf{g}_2(\mathbf{V}_1)$.\footnote{Thus far, the case of dynamic policies under static confounding has not been considered. When relying on dynamic policies, it is often desirable to construct counterfactuals that closely reflect observed practice. Hence, in observational studies, dynamic policies usually should be accompanied by dynamic confounding in the underlying data. Dynamic policies under static confounding might become relevant in experimental settings, for example when using stratified-on-$X_0$ randomization. For such cases, Assumption \ref{as:identification_dyn} can be weakened by conditioning only on $X_0=x_0$ instead of $\mathbf{X}_1 = \mathbf{x}_1$ in the second period. An explicit statement of this assumption is omitted for the sake of brevity.}

\subsection{Identification using Augmented Inverse Probability Weighting}\label{sec:identification_dr}

In addition to the identification results using the $g$-formula discussed in the previous subsections, alternative identification approaches based on Assumptions \ref{as:sutva}-\ref{as:identification_dyn} offer favorable robustness properties through additional reweighting by propensity scores. Let
\begin{align}
    \Theta^{st}_{\mathbf{d}_2}(\mathbf{W}_2) &:= \mu_{\mathbf{d}_2}(X_0) + \frac{\left(Y  -  \mu_{\mathbf{d}_2}(X_0)  \right) \mathbbm{1}\left\{\mathbf{D}_2=\mathbf{d}_2\right\} }{p_{\mathbf{d}_2}(X_0)}\quad\text{and} \label{eq:DR_static}\\
    \Theta^{dy}_{\mathbf{g}_2}(\mathbf{W}_2) &:= \nu_{\mathbf{g}_2}(X_0) + \frac{\left(\mu_{\mathbf{g}_2}(\mathbf{X}_1)  -  \nu_{\mathbf{g}_2}(X_0)  \right) \mathbbm{1}\left\{D_1=g_1(V_0)\right\} }{p_{g_1}(X_0)} +
    \frac{\left(Y - \mu_{\mathbf{g}_2}(\mathbf{X}_1) \right) \mathbbm{1}\left\{\mathbf{D}_2=\mathbf{g}_2(\mathbf{V}_1)\right\}}{p_{g_2}(\mathbf{X}_1, g_1)p_{g_1}(X_0)}  \label{eq:DR}
\end{align}
denote so-called score functions for the settings under $st$atic and {$dy$}namic confounding, respectively. It can be shown that the estimands of interest are identified as expectations of these scores:

\begin{theorem}[Identification using Augmented Inverse Probability Weighting]$ $\label{th:identification_dr}\\Define $\Theta^{j}_{\mathbf{g}_2}(\mathbf{W}_2)$ for $j \in \left\{st, dy\right\}$ as in \ref{eq:DR_static} and \ref{eq:DR}. The estimands of interest are identified as 
    \begin{align*}
        \theta^{\mathbf{g}_2} =  \E [ \Theta^{j}_{\mathbf{g}_2}(\mathbf{W}_2)] \quad\quad\quad\text{and}\quad\quad\quad
    \theta^{\mathbf{g}_2}(z_0) =  \E [ \Theta^{j}_{\mathbf{g}_2}(\mathbf{W}_2)|Z_0=z_0]\text{,}
    \end{align*}
for $j=st$ and a static policy under Assumptions \ref{as:identification_static} or \ref{as:identification_static2}, for $j=dy$ and a static policy under Assumption \ref{as:identification} and for $j=dy$ and a dynamic policy under Assumption \ref{as:identification_dyn}.
\end{theorem}

These identification results go back to \citet{robins1994estimation} for static confounding and  \citet{robins2000robust} for dynamic confounding. The former possesses the well-known double robustness property in the sense that it even identifies the APO if either $\mu_{\mathbf{d}_2}(\cdot)$ or $p_{\mathbf{d}_2}(\cdot)$ is replaced by any alternative function. Under dynamic confounding, this result extends to multiple robustness, which ensures identification if at least one component from each of the two periods (i.e. $\nu_{\mathbf{g}_2}(\cdot)$ or $p_{g_1}(\cdot)$ and $\mu_{\mathbf{d}_2}(\cdot)$ or $p_{g_2}(\cdot)$) is correctly specified. The score functions (\ref{eq:DR_static}) and (\ref{eq:DR}) will become important for the estimation procedures discussed in the following section.


\section{Evaluating Sequential Policies with Double Machine Learning} \label{sec:dynDML}
\subsection{Causal Machine Learning Based on Neyman Orthogonal Scores} 

The previous section introduced several estimands of interest and showed that, under identification assumptions, aggregates of their unobserved components can be expressed in terms of random variables, from which observations can be sampled. Given a set of such random samples, different techniques proposed in the literature can be used to estimate the parameters of interest and to conduct inference. This paper considers flexible machine learning estimators of $\theta^{\mathbf{g}_2}$ and $\theta^{\mathbf{g}_2}(z_0)$ that do not require functional form assumptions about the underlying data generating process and allow to use high-dimensional covariates, enhancing the credibility of the identification arguments in observational settings.\footnote{The focus here is on machine learning-based estimators; for an overview of conventional parametric methods in the sequential setting see Appendix~\ref{ap:other_estimators_conv} or the textbook introduction in \citet{hernan2020causal}.} However, machine learning estimators trade-off bias against variance to obtain a low mean squared error. Hence, naïve plug-in estimators that solely rely on machine learning estimates of $\mu_{\mathbf{d}_2}(X_0)$ or $\nu_{\mathbf{g}_2}(X_0)$ are biased due to regularization. To address this issue, it has been shown that plug-in estimators can be de-biased using an orthogonal moment condition, which adds an influence function adjustment to the target parameter \citep{chernozhukov2018double}. Estimators based on the influence function have desirable properties, such as $\sqrt{N}$-convergence and asymptotic normality. Furthermore, an estimator needs to solve the influence function in order to be asymptotically efficient \citep{robins1994estimation}.

The orthogonal moment condition of the the average potential outcome $\theta^{\mathbf{g}_2}$ is given by
\begin{align}\label{eq:EIF}
    \E[\Theta^{j}_{\mathbf{g}_2}(\mathbf{W}_2, \eta^j) - \theta^{\mathbf{g}_2}] = 0 \quad\text{for} \; j \in \left\{st, dy\right\}\text{,}
\end{align}
where $\eta^{st} = (\mu_{\mathbf{d}_2}(X_0), p_{\mathbf{d}_2}(X_0))$ and $\eta^{dy} = (\nu_{\mathbf{g}_2}(X_0), \mu_{\mathbf{g}_2}(\mathbf{X}_1), p_{d_2}(\mathbf{X}_1, g_1), p_{g_1}(X_0))$ denote the vectors of nuisance functions under static and dynamic confounding, respectively. For true $\theta^{\mathbf{g}_2}$ and $\eta^j$ the condition is satisfied, as shown by the identification result in the previous section. This suggests to construct an estimator solving an empirical equivalent of it,
\begin{align} \label{eq:emp_EIF}
    \hat\theta^{\mathbf{g}_2} = \frac{1}{N} \sum_{i=1}^N  \Theta^{j}_{\mathbf{g}_2}(\mathbf{W}_{i2}, \hat\eta^j)\text{,}
\end{align}
where $\hat\eta^j$ refers to the estimated nuisance functions. In addition, for the group average potential outcome, $\theta^{\mathbf{g}_2}(z_0)$, the moment condition
\begin{align*}
    \E[\Theta^{j}_{\mathbf{g}_2}(\mathbf{W}_2, \eta^j) - \theta^{\mathbf{g}_2}(z_0)| Z_0=z_0] = \frac{1}{\Pr(Z_0=z_0)} \E [ \Theta^{j}_{\mathbf{g}_2}(\mathbf{W}_{2}, \eta^j) \mathbbm{1}\left\{Z_0=z_0\right\}] - \theta^{\mathbf{g}_2}(z_0) = 0
\end{align*}
suggests to use the estimator
\begin{align*}
    \hat\theta^{\mathbf{g}_2}(z_0) = \frac{1}{\frac{1}{N} \sum_{i=1}^N \mathbbm{1}\left\{Z_{0i}=z_0\right\}} \sum_{i=1}^N \Theta^{j}_{\mathbf{g}_2}(\mathbf{W}_{i2}, \hat\eta^j) \mathbbm{1}\left\{Z_{0i}=z_0\right\}\text{.}
\end{align*}
Hence, estimation of both parameters of interest can be based on the same scores \citep{chernozhukov2024applied}.

Under static confounding ($j=st$), estimation is implemented by separately predicting $\hat\mu_{\mathbf{d}_2}(X_0)$ and $\hat p_{\mathbf{d}_2}(X_0)$ by machine learning methods, plugging them into Equation (\ref{eq:DR_static}), i.e.
\begin{align} \label{eq:DML_emp_static}
{\Theta}^{st}_{\mathbf{d}_2}(\mathbf{W}_{2}, \hat\eta^{st}) = \hat{\mu}_{\mathbf{d}_2}(X_{0})
    + \frac{\left(Y -  \hat{\mu}_{\mathbf{d}_2}(X_{0})\right) \cdot \mathbbm{1}\left\{\mathbf{D}_{2}=\mathbf{d}_2\right\}}{\hat{p}_{\mathbf{d}_2}(X_{0})}\text{,}
\end{align}
and finally averaging this score over the population of interest. For statistical inference, the variance of the scores can be used to construct a standard $t$-test statistic. The structure of the score ${\Theta}^{st}_{\mathbf{d}_2}(\mathbf{W}_{2}, \hat\eta^{st})$ demonstrates that the regularization bias in the estimation of $\hat{\mu}_{\mathbf{d}_2}(X_{0})$ is corrected by adding an adjustment term, consisting of the conditional outcome residuals, re-weighted by the inverse treatment probability. Hence, the adjustment increases as the prediction deviates further from the observed outcome and as the conditional treatment probability decreases.

DML in the single-period setting has been proposed by \citet{chernozhukov2018double}. In their seminal paper, the authors demonstrate the importance of addressing regularization bias and overfitting in the estimates of the nuisance functions when employing the machine learning-based plug-in approach. This is achieved by (i) using an orthogonal moment function, and (ii) by using a cross-fitting procedure. The orthogonal moment condition (i) allows to obtain $\sqrt{N}$-consistency even when using machine learning estimators that typically converge at relatively slow rates. This is because it satisfies a certain `Neyman'-orthogonality property that makes it locally insensitive to small biases in the nuisance estimates. In analogy to the concept of identification double robustness discussed in Section \ref{sec:identification_dr}, this property is also referred to as rate double robustness in the literature \citep{knaus2022double}.
The cross-fitting procedure (ii) avoids overfitting by ensuring that observations are not used to predict their own nuisance functions. Therefore, the set of observations $\mathcal{W} = \left\{1, ..., N\right\}$ is split into $K$ equally sized subsamples $\mathcal{W}_k$. For each $k=1,...,K$ the nuisance parameters are first trained on the complementing subset $\mathcal{W}_{-k}$, then predicted in the subsample $\mathcal{W}_k$ and plugged into the score function. Finally, the DML estimate is obtained by taking the mean of the cross-fitted scores. In line with earlier reasoning, this procedure designed for the single-period setting, can be directly applied to sequential estimation under static confounding by treating each program sequence $\mathbf{d}_2$ as a distinct treatment state.

\subsection{Sequential Double Machine Learning under Dynamic Confounding}

Several recent contributions \citep{bodory2022evaluating, bradic2021high, singh2021finite, chernozhukov2022automaticdyn} propose extensions of single-period DML to the sequential setting under dynamic confounding. All these extensions are introduced within a framework of static policies. However, once identification is established, DML-based estimation can proceed similarly for both static and dynamic policies, as detailed below. Accordingly, the procedures are presented directly using the more general notation for dynamic policies. 

All mentioned contributions are based on the orthogonal moment condition (\ref{eq:EIF}) with score
\begin{align} \label{eq:DML_emp}
{\Theta}^{dy}_{\mathbf{g}_2}(\mathbf{W}_{2}, \hat\eta^{dy}) = \hat{\nu}_{\mathbf{g}_2}(X_{0})
    &+ \frac{\left(\hat{\mu}_{\mathbf{g}_2}(\mathbf{X}_{1})  -  \hat{\nu}_{\mathbf{g}_2}(X_{0})  \right) \cdot \mathbbm{1}\left\{D_{1}=g_1(V_{0})\right\} }{\hat{p}_{g_1}(X_{0})}\nonumber\\
    &+ \frac{\left(Y -  \hat{\mu}_{\mathbf{g}_2}(\mathbf{X}_{1})    \right) \cdot \mathbbm{1}\left\{\mathbf{D}_{2}=(g_1(V_{0}), g_2(\mathbf{V}_{1}))\right\}}{\hat{p}_{g_2}(\mathbf{X}_{1},  g_{1})\hat{p}_{g_1}(X_{0})}\text{,}
\end{align}
which now consists of two re-weighted outcome residuals, one for each period. This corresponds to the efficient influence function for longitudinal treatment effects \citep{robins2000robust, bang2005doubly}. The authors demonstrate that this score satisfies the Neyman orthogonality condition, indicating its suitability for the DML approach. However, the extension to dynamic confounding is complicated by the fact that $\nu_{\mathbf{g}_2}(X_0)$ cannot be directly represented by observable variables as it nests the conditional mean outcome of the second period $\mu_{\mathbf{g}_2}(\mathbf{X}_{1})$. Hence, estimation of $\nu_{\mathbf{g}_2}(X_0)$ requires estimates $\hat{\mu}_{\mathbf{g}_2}(\mathbf{X}_{1})$ as inputs, which can be implemented in various ways.

An initial approach proposed by \citet{bodory2022evaluating} is to estimate $\hat\nu_{\mathbf{g}_2}(X_0)$ by regression of $\hat{\mu}_{\mathbf{g}_2}(\mathbf{X}_{1})$ on $X_0$ for observations following $D_1 = g_1(V_0)$, i.e.
\begin{align} \label{eq:nu_bodory}
   \hat{\nu}^{\text{BHL22}}_{\mathbf{g}_2}(X_0) &= \hat{\E}[\hat{\mu}_{\mathbf{g}_2}(\mathbf{X}_{1})|X_0, D_1=g_1(V_0)]\text{,}
\end{align}
where $\hat{\E}[A|B, C=c]$ denotes predictions from a regression of $A$ on $B$ for observations with $C=c$. This requires an additional split of the subsamples $\mathcal{W}_{-k}$ to avoid overfitting, such that $\hat\nu_{\mathbf{g}_2}(X_0)$ and $\hat{\mu}_{\mathbf{g}_2}(\mathbf{X}_{1})$ are not learned from the same sample. In a setting with more than two time periods, the number of splits would increase even further. 

\citet{bradic2021high} propose another method for implementing DML under dynamic confounding, introducing an additional bias correction term for estimating the nested conditional outcome. In particular, they estimate $\hat\nu_{\mathbf{g}_2}(X_0)$ as
\begin{align} \label{eq:nu_bradic}
   \hat{\nu}^{\text{BJZ24}}_{\mathbf{g}_2}(X_0) &= \hat{\E}\left[\hat{\mu}_{\mathbf{g}_2}(\mathbf{X}_{1}) +  \frac{\mathbbm{1}\left\{D_{2}=g_2(\mathbf{V}_1)\right\}\left(Y-\hat{\mu}_{\mathbf{g}_2}(\mathbf{X}_{1})\right)}{\hat{p}_{g_2}(\mathbf{X}_1, g_1)} \middle|X_0, D_1=g_1(V_0)\right]\text{,}
\end{align}
where the pseudo-outcome that is regressed on $X_0$ in the subsample $D_1=g_1(V_0)$ corresponds to the doubly robust score of the treatment effect of the second period. Again, this requires second-order sample splitting, but the authors propose to regain full sample size efficiency by cross-fitting. In a first step, $\hat{\mu}_{\mathbf{g}_2}(\mathbf{X}_{1})$ is estimated using the first subsample of $\mathcal{W}_{-k}$, and predictions are made on the second subsample. These predictions are then used to estimate $\hat{\nu}^{\text{BJZ24}}_{\mathbf{g}_2}(X_0)$. The process is repeated by reversing the roles of the subsamples, resulting in two estimates of $\hat{\nu}^{\text{BJZ24}}_{\mathbf{g}_2}(X_0)$. Subsequently, the observations from fold $\mathcal{W}_{k}$ are applied to both estimated functions, and the predictions are averaged for each observation. While the original paper is only formulated for a binary treatment setting, it is extended here to the case of multiple treatments. The exact algorithm and a comparison to the method of \citet{bodory2022evaluating} is provided in Appendix \ref{sec:appendix_algo}.

For both approaches, the properties of $\sqrt{N}$-consistency and asymptotic normality extend from single-period DML to the sequential setting. Besides standard regularity conditions, the theoretical guarantees rely on four (instead of two in the single-period framework) consistent nuisance parameter predictions $\hat{\mu}_{\mathbf{g}_2}(\mathbf{X}_{1})$, $\hat{\nu}_{\mathbf{g}_2}(X_0)$, $\hat{p}_{g_1}(X_0)$ and $\hat{p}_{g_2}(\mathbf{X}_1,  g_1)$. In addition, \citet{bodory2022evaluating} require three product rate conditions: The two within-period products of the rates of convergence between $\hat{\nu}_{\mathbf{g}_2}(X_0)$ and $\hat{p}_{g_1}(X_0)$, and between $\hat{\mu}_{\mathbf{g}_2}(\mathbf{X}_{1})$ and $\hat{p}_{g_2}(\mathbf{X}_1, g_1)$, as well as the cross-period product rate between $\hat{\mu}_{\mathbf{g}_2}(\mathbf{X}_{1})$ and $\hat{p}_{g_1}(X_0)$ need to be at least as fast as $N^{1/2}$ (see Assumption 4(d) therein). In \citet{bradic2021high}, the additional doubly robust step reduces the number of required product rate conditions to just two. Specifically, only the within-period products need to be considered, making the cross-period product between $\hat{\mu}_{\mathbf{g}_2}(\mathbf{X}_{1})$ and $\hat{p}_{g_1}(X_0)$ no longer necessary. As will be demonstrated later, the practical significance of the weakened assumption appears minimal within the context of the following empirical application.

Both methods were originally developed for static policies under dynamic confounding and are extended here to accommodate dynamic policies. While the identification assumptions differ between static and dynamic policies as discussed in Section \ref{sec:dyneffect}, both identification results have the same functional form, with $\mathbf{d}_2$ replaced by $\mathbf{g}_2(\mathbf{V}_1)$. DML is an estimator plugging nuisance estimates directly into the empirical representation of the orthogonal moment condition. These nuisance functions can be estimated in the same way for both static and dynamic policies. Moreover, dynamic policies are deterministic functions of the same covariates already appearing in the estimator for static policies under dynamic confounding. Hence, the estimators naturally extend to dynamic policies, allowing estimation to proceed in the same manner regardless of whether the interest lies in $\theta^{\mathbf{d}_2}$ or $\theta^{\mathbf{g}_2}$. For completeness, Appendix \ref{sec:app_neyman} provides a proof of Neyman orthogonality for the setting with dynamic policies. A recent exposition of sequential DML in the context of dynamic policies is provided by \citet[][Chapter 14]{wager2024causal}.

The presented estimators offer ready-to-implement algorithms based on robust theoretical foundations under standard assumptions. In addition, several alternative machine learning estimators based on orthogonal scores have been proposed in the literature, including \citet{singh2021finite, chernozhukov2022automaticdyn, van2012targeted, lewis2020double}. However, they rely on stronger assumptions or lack practical implementability in the present setting. A detailed overview of these methods is provided in Appendix \ref{ap:other_estimators_EIF}.

\section{Application: Evaluation of ALMP Sequences in Switzerland}\label{sec:application}

\subsection{Introduction and Related Literature}

Active labor market policies (ALMP) are government programs provided to unemployed individuals with the objective of facilitating their return to the labor market. While much of the existing literature has focused on the effects of the first or the longest program, this study investigates two practical aspects among participants: the duration and the order of the programs. 

A limited number of prior studies have analyzed ALMP sequences under dynamic confounding. \citet{lechner2013does} offers the first comprehensive evaluation using Austrian data and inverse probability weighting. The authors find that jobs search assistance is more effective after a qualification program compared to the reverse order, and that two consecutive qualification programs outperform a single one. Adopting the same approach, \citet{dengler2015effectiveness,dengler2019effectiveness} find similar results for public employment programs and classroom training in Germany. More recently, \citet{laffers2024locking} analyze ALMP sequences for young adults in Slovakia using the DML estimator of \citet{bodory2022evaluating}, also reporting positive effects from combining programs. However, all mentioned studies are restricted to static policies, which makes the resulting treatment effects hard to identify and difficult to interpret. For example, \citet{laffers2024locking} define unemployed non-participants and employed non-participants as two distinct control groups, which imposes strong assumptions: the former requires fixing unemployment duration a priori, while the latter assumes that exiting unemployment is conditionally independent of potential outcomes. These limitations are addressed in the present paper by introducing dynamic policies.

Complementary work by \citet{vikstrom2017dynamic} examines ALMP sequences in Sweden using a survival time framework, where the outcome of interest is the probability of remaining unemployed up to a specific period, which differs from this paper's emphasis on medium-term labor market outcomes. In contrast to earlier findings, this study suggests that participating in two consecutive programs offers no substantial advantage over a single program. Finally, another strand of literature considers dynamic assignment to ALMP, accounting for potential non-randomness of program start dates \citep[e.g.][]{sianesi2004evaluation, crepon2009active, van2019long, kastoryano2022dynamic}. While these papers adapt their identification and estimation procedures to dynamic confounding, they remain within a framework that compares single ALMP but does not allow the evaluation of program sequences.

\subsection{Institutional Background}

In Switzerland, ALMP are nationally regulated but implemented by regional employment offices (REOs). Unemployed individuals who have worked at least 12 months in the previous two years can register at a REO to receive income maintenance based on their past salary for up to 24 months. To receive these benefits, individuals must actively search for a job and participate in assigned ALMP. The programs can be categorized into five groups:
\begin{itemize}
    \item \textit{Job search assistance (JA)}: Orientation measures and courses for basic job acquisition skills.
    \item \textit{Training courses (TC)}: Language, IT, and sector-specific vocational training courses.
    \item \textit{Employment program (EP)}: Unpaid employment outside the regular labor market (not in competition with other firms) providing a meaningful activity and daily routine.
    \item \textit{Temporary wage subsidy (WS)}: Monetary compensation incentivizing individuals to accept temporary jobs paying lower wage than the unemployment benefit.
    \item \textit{Other programs (OP):} Small programs not included in the previous four categories, such as training grants, vocational placements, and internships.
\end{itemize}
Two thirds of individuals take part in one of these programs within the first twelve months of their unemployment spell, with approximately 60\% of them participating in more than one program. Programs are assigned by caseworkers using information from the registration process and consultation sessions. They have the authority to sanction individuals who decline to participate.

\subsection{Data and Panel Design}

The analysis is based on Swiss administrative records for the period 2004 to 2018. The population under study is defined as individuals aged 25-55\footnote{Younger and older individuals are excluded to avoid dealing with educational and (early) retirement choices.} who became eligible for programs and received unemployment benefits between April 2011 and January 2015,\footnote{The time frame is restricted by a major revision of Swiss unemployment insurance in early 2011 and the need for a follow-up period of at least three years after program start to measure outcomes.} following a minimum of three months of prior unsubsidized employment. Among these individuals, all individuals with a program lasting at least five business days and starting within 12 months of the beginning of their unemployment spell are selected.\footnote{These restrictions ensure that assessments conducted before the allocation to a program are excluded and that there is sufficient time left to participate in a program within the entitlement period.} For the final sample of 191,619 individuals, monthly information on employment status, program participation and covariates is observed. For details on the institutional background and data, see \citet{mascolo2024heterogeneous}, who base their study on the same public records.

The information is aggregated to a panel of two three-month periods, starting from the month of the first program. This structure has been chosen to meet two key requirements: Firstly, it should capture as good as possible the true assignment process. Secondly, it should ensure a sufficient number of observations per program sequence to enable a meaningful econometric analysis.\footnote{Since aggregation inherently introduces inaccuracies, maintaining the dataset at the highest possible level of granularity would be ideal. When opting against aggregation, an alternative approach involves using estimators relying on parametric assumptions, such as marginal structural models \citep{robins2000marginal}, which enable extrapolation into regions beyond observed data support. However, given limited knowledge about the true data-generating process in this application, an aggregation approach was selected instead, prioritizing flexible estimation.} Defining the panel’s reference point as the start month of the first program, rather than the start month of the unemployment spell, ensures that all individuals receive a treatment in the first period. This approach substantially increases the number of individuals with identical program sequences.\footnote{For example, consider two individuals who are unemployed for ten months. Individual 1 is assigned to a six-week training course in the first month of the unemployment spell. Individual 2 is assigned to the same program in the fifth month. Despite the different start times of the program, they both exhibit the same program sequence, ``training course - no program''.} The lack of sequences beginning with non-participation is of minor concern, as the analysis targets implementation details among participants, while addressing dynamic confounding.\footnote{Alternative designs might be preferable if the focus of the evaluation is when to start programs \citep[timing-to-treatment framework, e.g.][]{nie2021learning} or program duration without considering dynamics \citep[dose response framework, e.g.][]{Imbens:2000}.} The time between the start of unemployment and the first program is controlled for as a pre-treatment covariate.

Period length is determined in a data-driven way. The sample reveals a median interval of 3 months between program starts and a median program duration of 45 days. Opting for three-month periods maximizes the number of program starts in the second period, compared to two-month or four-month intervals. With a three-month period length, individuals have on average $2.1$ appointments with their caseworker in the first period, which seems reasonable since the assignment decision is expected to be reconsidered less frequently than at every single meeting. In addition, the proportion of individuals with changes in covariates $X_1$ relative to $X_0$ increases considerably when moving from 2-month to 3-month intervals, but shows a much smaller increase when comparing 3-month to 4-month intervals. This suggests that a three-month period is sufficient for time-varying selection to occur.

Finally, the number of periods is another key factor influencing the range of possible program sequences. With each additional period, the number of sequences for a given set of programs increases exponentially. This makes it increasingly difficult to find individuals with the same sequence as the number of periods rises. In the available data, sample sizes for most sequences proved insufficient when considering $T > 2$ periods. Therefore, the analysis is restricted to two periods, consistent with the formal notation introduced earlier.

\subsection{Definition of Treatments and Outcomes}

To evaluate the effectiveness of different program sequences, the outcome is defined as the sum of months employed over 30 months starting from the second period. This measure captures the medium-term impact of program sequences on employment trajectories, which aligns with the Swiss government’s policy objective of promoting sustained employment through ALMP.

In each of the two periods of the panel, each individual is assigned to a treatment state corresponding to one of the four programs \textit{JA}, \textit{TC}, \textit{EP}, or \textit{WS} introduced above. For a program to be considered a treatment state, it must last at least five business days within the period, regardless of its start date. For instance, if an individual participates in an employment program for five months, the treatment state in the second period is \textit{EP}, even if the program started in the first period. Individuals participating in multiple different programs within the same period are assigned to the longest program.\footnote{Hence, within-period dynamics in program assignment are disregarded, i.e. it is assumed that assignment to a second program within a period does not depend on the first program within the same period. This seems reasonable given the limited duration of the periods.} In the second period, two additional treatment states are introduced. Individuals who remain without program participation throughout the period are categorized as \textit{No program (NP)}. This group includes both unemployed individuals who are no longer assigned to a program and individuals who are not assigned because they have exited unemployment. This additional treatment state allows analyzing sequences with a program in the first period but no program in the second period. Individuals who are assigned to \textit{OP} in the second period remain in the sample for modeling treatment assignment in the first period. However, they are not considered in the analysis due to the small number of participants and special admission requirements for these programs, which prevent credible identification.

\begin{figure}[htbp!]
    \centering
    \captionsetup{font=small, width=0.8\linewidth} 
    \caption{Alluvial plot of program sequences considered in the analysis}
    \includegraphics[width=0.8\linewidth]{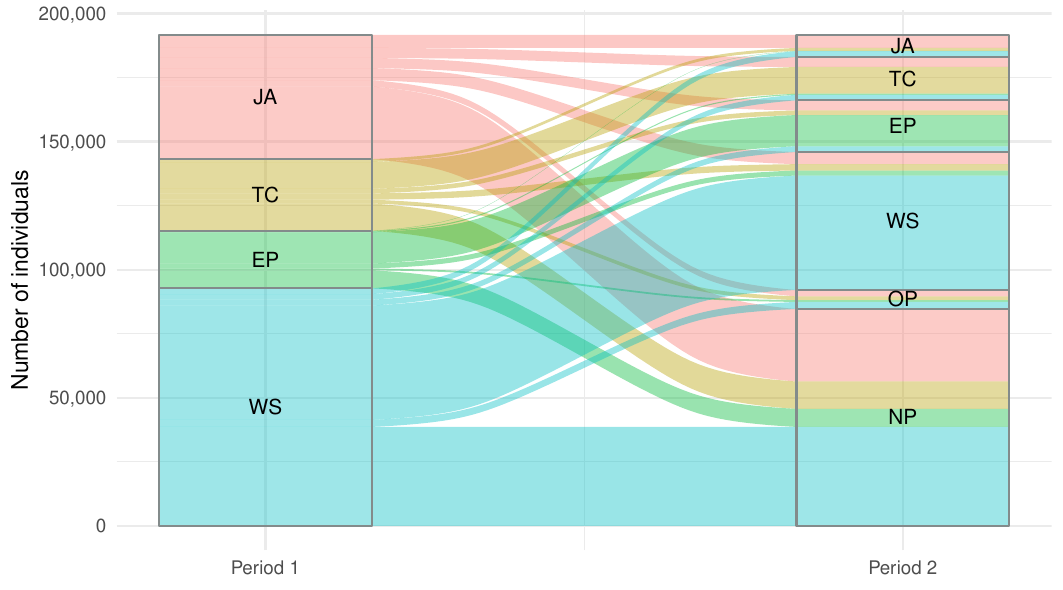}
    \caption*{\scriptsize \textit{Notes:} Program frequencies and transitions between first and second period. \textit{JA}: Job-search assistance, \textit{TC}: Training course, \textit{EP}: Employment program,  \textit{WS}: Wage subsidy, \textit{OP}: Other programs, \textit{NP}: No program. Appendix Figure~\ref{fig:alluvial_trim} shows the same plot after applying the trimming procedure described in Section~\ref{sec:implement}.} 
    \label{fig:alluvial}
\end{figure}

Figure \ref{fig:alluvial} illustrates the distribution of treatment states across the two periods, revealing insights about program size and duration. In the first period, temporary wage subsidies comprise nearly half of the beneficiaries, followed by job-search assistance. Most recipients of temporary wage subsidies in the first period continue in the second period, while recipients of job-search assistance often transition to other program states. Overall, more than half of the individuals remain in a program in the second period. The plot highlights the large variety of transitions, emphasizing the importance of sequential analysis.

When evaluating ALMP over time, program assignments critically depend on the evolution of employment status. For instance, if an individual exits unemployment during the first period, this directly influences program allocation decisions in the second period. Consequently, evaluating counterfactuals that impose a fixed sequence of two program periods $(d_1, d_2)$ for all individuals, without accounting for changes in employment status, may lack meaningful insights. The present analysis solves this issue using dynamic policies. Specifically, as previewed earlier in Example \ref{ex:1}, the decision variable $V^{d_1}_1$, which determines program allocation in the second period, is defined as the potential intermediate outcome $Y_1^{d_1}$. Here, $Y_1^{d_1}$ is a binary indicator equal to one if an individual treated with program $d_1$ exits unemployment during the first period, and zero otherwise. Using this decision variable, 20 dynamic policies of interest are defined as
\begin{align}\label{eq:g_application}
    \mathbf{g}_2(Y_1^{d_1}) &= \left(d_1,  \mathbbm{1}\left\{Y_1^{d_1}=0\right\} \cdot d_2 +\mathbbm{1}\left\{Y_1^{d_1}=1\right\} \cdot \text{NP}\right)\text{,} 
\end{align}
with $d_1, d_2 \in  \left\{\text{JA}, \text{TC}, \text{EP}, \text{WS}\right\} \times  \left\{\text{JA}, \text{TC}, \text{EP}, \text{WS}, \text{NP}\right\}$.
The policies are dynamic only in the second period, while they assign a fixed program $d_1$ in the first period. This allows to construct counterfactuals in which every individual is initially assigned to program $d_1$ during the first period, with program $d_2$ assigned in the second period only to those who remain unemployed throughout the first period. The treatment $d_2$ may also be defined as \textit{NP}, in which case the policy is fully static. Policy (\ref{eq:g_application}) is defined in terms of potential intermediate outcomes $Y_1^{d_1}$ as the assignment to the second program should depend on the employment status after the first counterfactual program $d_1$ and not on the employment status following the observed program $D_1$.

\subsection{Identification}

For single time-point interventions, the literature widely acknowledges that the effects of ALMP can be plausibly identified using observational data, provided a comprehensive set of control variables is available \citep{caliendo:2017, lechner2013sensitivity}. Since individuals are randomly assigned to caseworkers, who then determine program assignments, the primary challenge lies in accounting for all factors considered by caseworkers during this process. In the current setting, the relevant covariates identified in prior research are available. These include, among others, socio-demographic characteristics, employment and earnings histories spanning the past seven years, details about the last job and prior unemployment spells, regional labor market conditions, and caseworkers' assessments of individuals' job search efforts prior to program start. Appendix \ref{sec:appendix_descriptives} provides an overview of all control variables used.

For the analysis of sequential treatments, the identification assumptions must be extended to address static or dynamic confounding, as well as static or dynamic policies, as discussed in Section \ref{sec:dyneffect}. Under static confounding, Assumption \ref{as:identification_static2} requires the entire program path to be predetermined, given the available information before the start of the sequence, which seems unlikely, as caseworkers may adjust their assignments over time. This concern is supported by an analysis of covariate means by assigned program type, which reveals substantial differences across program sequences for both baseline and time-varying covariates (see Tables \ref{tab:balance_x0_2nd} and \ref{tab:balance_x1_2nd} in the Appendix). Therefore, this analysis focuses on the setting under dynamic confounding, which permits a more flexible program assignment process in the underlying data.

For static policies under dynamic confounding, Assumption \ref{as:identification}\ref{as:CIA} consists of two parts, one for every period. In the first period, the identification argument aligns with that of a standard non-sequential evaluation and should therefore be reasonable given the available information. The additional difficulty of dynamic confounding is addressed by the second part of the assumption, which requires controlling for all factors that influence both potential outcomes and program assignment in the second period, conditional on program participation in the first period. To address this assumption, five types of covariates are considered, that may change during the first period and are expected to determine dynamic selection into the second program. Firstly, it is essential to control for the intermediate outcome, employment status, as only unemployed individuals are eligible for assignment to a program in the second period. Secondly, financial aspects are taken into account by controlling for the amount of unemployment benefits and other state subsidies, as well as intermediate earnings during the first period. Thirdly, the selection likely hinges on an individual's job prospects. These are assessed through variables such as the number of applications written and caseworkers' assessments of individual's job search efforts, employability, and qualification needs. Fourthly, caseworkers' decisions might be influenced by the cooperativeness of individuals. This is measured through the number of scheduled, postponed, and canceled appointments at the REO, as well as incidents resulting in sanction days. Lastly, changes in the personal situation such as relocation, pregnancy, and the number of sickness days, are also accounted for. 

A key difficulty in identifying static program sequences is that program participation depends on remaining unemployed. Consequently, both unemployment status and program assignment must be accounted for when addressing selection bias. This essentially requires that, conditional on covariates and prior program participation, not only program assignment but also the chance of remaining unemployed are random, which makes credible identification significantly more demanding. This issue can be addressed by redefining the estimand in terms of dynamic policies, as defined in equation (\ref{eq:g_application}). These policies account for the possibility of leaving unemployment in the counterfactual scenario, such that assignment to the treatment group is no longer conditional on employment status. However, this advantage comes at the cost of the stricter identification requirements stated in Assumption \ref{as:identification_dyn}. In particular, conditional on pre-treatment information, assignment to the first program needs to be independent of final \textit{and intermediate} potential outcomes. Given the context of this application, the additional restrictiveness of the assumption appears to be unproblematic. For static policies, the conditional independence assumption applies to the final potential outcome, i.e. the joint distribution of potential employment status over the 30 months starting from the second period. The additional complexity introduced by the dynamic strategy necessitates independence with respect to employment status during the three months of the first period as well. This restriction aligns with the assumption made in any single-period program evaluation that measures outcomes from the start of the first treatment.

Finally, the overlap assumptions \ref{as:identification}\ref{as:CS} and \ref{as:identification_dyn}\ref{as:CS_dyn} require that, in both periods, there is a non-zero probability of following the (static or dynamic) policy of interest, conditional on pre-period information. This assumption is expected to hold among unemployed individuals, as all the programs considered are accessible to them, irrespective of their covariate history. Thus, although treatment probabilities may vary based on previous characteristics, any unemployed individual should, in principle, be eligible for admission to any program. Again, the fact that individuals might exit unemployment during the first period poses a challenge to the credibility of the assumption for static policies. For example, individuals who leave unemployment in the first period are only eligible for treatment in the second period if they re-enter unemployment within that three-month window, resulting in a very low treatment probability. This issue does not arise for dynamic strategies, as such situations are excluded under any treatment strategy of interest. Specifically, for any realization of the function $g_2(Y_1^{d_1})$, individuals employed in the first period are not assigned to a program in the second period. Hence, carefully designing the structure of the dynamic policy allows to obtain counterfactuals that are more credibly identified.

\subsection{Implementation}\label{sec:implement}

To estimate the effects of interest, the two dynamic estimation procedures by \citet{bodory2022evaluating}\footnote{Note that the implementation of the procedure by \citet{bodory2022evaluating} used in this paper differs from that in the original \texttt{causalweight} \textbf{\textsf{R}}-package. In particular, \citet{bodory2022evaluating} fit $\hat\mu$ stratified by programs but include $D_1$ as a covariate when estimating $\hat p_{d_2}$. In contrast, the approach adopted here stratifies both $\hat\mu$ and $\hat p_{d_2}$. This adjustment is expected to reduce bias in the estimates of $\hat p_{d_2}$, but may increase variance if there are few individuals following a particular program $d_1$.} and \citet{bradic2021high} described in Section \ref{sec:dynDML} are applied with 5-fold cross-fitting. The nuisance functions are estimated by random forests using the Python package \texttt{scikit-learn} \citep{pedregosa2011scikit}. In line with standard practice, predictors with low variance or high collinearity are removed prior to the analysis. Following \citet{bach2024hyperparameter}, the hyperparameters of the random forest are tuned on the full sample using \texttt{FLAML} \citep{wang2021flaml} with a maximum time budget of 10 minutes per nuisance estimation. The minimum number of trees used in a forest is set to 500.\footnote{Python code implementing the procedures is available at \href{https://github.com/fmuny/dynamicDML}{https://github.com/fmuny/dynamicDML}}

Estimators based on propensity score reweighting, such as DML, are known to produce unstable estimates when estimated propensity scores assign large weights to certain observations \citep{khan2010irregular}. Hence, even if overlap holds in the population, it may not be satisfied in the sample if the treatment probabilities for certain programs are very low for specific individuals. To address this issue, observations with extreme propensity scores are dropped from the sample, extending the minmax trimming procedure proposed in \citet{lechner2019practical} to the sequential setting. In particular, for the first period propensity scores $\hat p_{d_1}(X_0)$, first the minimum value among individuals within the subgroups $D_1 = d_1$ and $D_1 \neq d_1$ are identified, respectively. Then, the largest of these two values is selected, and all individuals with a $\hat p_{d_1}(X_0)$ smaller than this threshold are removed from the sample. Analogously, all units with $\hat p_{d_1}(X_0)$ larger than the smallest maximum are also removed. For the second period propensity scores $\hat p_{g_2}(\mathbf{X}_1, d_1)$ the same procedure is applied for individuals with $(D_1 = d_1$ and $D_2 = g_2(Y_1))$ vs. $(D_1 \neq d_1$ or $D_2 \neq g_2(Y_1))$, separately for the subgroups with $g_2(Y_1) = d_2$ and $g_2(Y_1) = \text{NP}$, respectively. After repeating the procedure for all 20 dynamic policies of interest, 7\% of the observations are dropped, resulting in a final sample size of 177,856 observations. A comparison of pre-treatment covariate means shows that trimmed units are more likely to come from the German-speaking region of Switzerland, have longer prior unemployment durations, and have received greater previous government support. Nonetheless, overall differences between trimmed and retained observations appear moderate, as shown in Appendix Table \ref{tab:balance_x0_trim}. 
Plots of the propensity score densities after trimming for dynamic policies are presented in Figures \ref{fig:app_pscores_dynpolJATC} and \ref{fig:app_pscores_dynpolEPWS}.

\subsection{Results}\label{sec:results}

\subsubsection{Program Duration}

Table \ref{tab:DATE_first+duration_em332_short} presents the main results of the analysis. It illustrates average treatment effects for different comparisons of policies and estimation techniques. Each column refers to a comparison of programs in the first period while in the panels below, several scenarios of second period policies are compared.\footnote{While Table \ref{tab:DATE_first+duration_em332_short} only reports ATEs and standard errors, Table \ref{tab:DATE_first+duration_em332} in the Appendix provides an extended version including the trimmed number of observations for each policy and results for both dynamic DML estimators discussed in Section \ref{sec:dynDML}.} Panel A provides a setting where the second period programs are unrestricted. These are the results one would obtain from a standard single-period evaluation analyzing the first program only. The discussion first focuses on these baseline results before highlighting the added value of the dynamic evaluation methods. The results indicate that \textit{WS} is the most effective program on average, leading to significantly more months in employment in the medium term compared to all other programs. The effect sizes range from 2.65 months more employment compared to \textit{JA} to 1.94 months more employment compared to \textit{TC} in the 30 months starting from the second period. \textit{TC} is identified as the second most effective program, yielding 0.71 and 0.59 months of increased employment compared to \textit{JA} and \textit{EP}, respectively. No statistically significant difference is observed between the programs \textit{JA} and \textit{EP}.

\begin{table}[htbp!]
\centering
\begin{threeparttable}
\caption{Average treatment effects for program duration.}
\label{tab:DATE_first+duration_em332_short}
\begin{tabularx}{\textwidth}{rXrrrrrrr}
\toprule
 &  &  & \makecell{$d_1 =$ JA\\ $d'_1 =$ TC} & \makecell{$d_1 =$ JA\\ $d'_1 =$ EP} & \makecell{$d_1 =$ JA\\ $d'_1 =$ WS} & \makecell{$d_1 =$ TC\\ $d'_1 =$ EP} & \makecell{$d_1 =$ TC\\ $d'_1 =$ WS} & \makecell{$d_1 =$ EP\\ $d'_1 =$ WS} \\
\midrule
\multicolumn{9}{l}{\textit{Panel A: Second period program unrestricted (single-period intervention):}}\\
\midrule\ & \multirow[c]{2}{*}{ATE (static conf.)} &  & -0.71***\phantom{(} & -0.12\phantom{***(} & -2.65***\phantom{(} & 0.59***\phantom{(} & -1.94***\phantom{(} & -2.53***\phantom{(} \\
 &  &   & (0.11)\phantom{***} & (0.16)\phantom{***} & (0.09)\phantom{***} & (0.15)\phantom{***} & (0.09)\phantom{***} & (0.14)\phantom{***} \\
\midrule
\multicolumn{9}{l}{\textit{Panel B: Second period without program (static policy):} $g_2(Y_1^{d_1}) =$ NP, $g'_2(Y_1^{d'_1}) =$ NP}\\
\midrule\ & \multirow[c]{2}{*}{ATE (dyn. conf.)} &  & -1.43***\phantom{(} & -1.47***\phantom{(} & -4.19***\phantom{(} & -0.03\phantom{***(} & -2.75***\phantom{(} & -2.72***\phantom{(} \\
 &  &   & (0.21)\phantom{***} & (0.33)\phantom{***} & (0.15)\phantom{***} & (0.34)\phantom{***} & (0.17)\phantom{***} & (0.31)\phantom{***} \\
 & \multirow[c]{2}{*}{ATE (static conf.)} &  & -1.61***\phantom{(} & -2.74***\phantom{(} & -4.67***\phantom{(} & -1.13***\phantom{(} & -3.06***\phantom{(} & -1.92***\phantom{(} \\
 &  &   & (0.23)\phantom{***} & (0.37)\phantom{***} & (0.19)\phantom{***} & (0.35)\phantom{***} & (0.15)\phantom{***} & (0.33)\phantom{***} \\
\midrule
\multicolumn{9}{l}{\textit{Panel C: Same program for at least two periods (static policy):} $g_2(Y_1^{d_1}) = d_1$, $g'_2(Y_1^{d'_1}) = d'_1$}\\
\midrule\ & \multirow[c]{2}{*}{ATE (dyn. conf.)} &  & -0.94***\phantom{(} & -0.38\phantom{***(} & -2.86***\phantom{(} & 0.57*\phantom{**(} & -1.91***\phantom{(} & -2.48***\phantom{(} \\
 &  &   & (0.34)\phantom{***} & (0.35)\phantom{***} & (0.26)\phantom{***} & (0.34)\phantom{***} & (0.24)\phantom{***} & (0.26)\phantom{***} \\
 & \multirow[c]{2}{*}{ATE (static conf.)} &  & -1.37**\phantom{*(} & -0.86\phantom{***(} & -2.87***\phantom{(} & 0.51***\phantom{(} & -1.50***\phantom{(} & -2.01***\phantom{(} \\
 &  &   & (0.59)\phantom{***} & (0.59)\phantom{***} & (0.58)\phantom{***} & (0.19)\phantom{***} & (0.14)\phantom{***} & (0.15)\phantom{***} \\
\midrule
\multicolumn{9}{l}{\textit{Panel D: Same program for at least two periods if not employed in first period (dynamic policy):}}\\
\multicolumn{9}{l}{\phantom{\textit{Panel D: }}$g_2(Y_1^{d_1}) = \mathbbm{1}\{Y_1^{d_1}=0\}d_1 +\mathbbm{1}\{Y_1^{d_1}=1\}\text{NP}$ and $g'_2(Y_1^{d'_1}) = \mathbbm{1}\{Y_1^{d'_1}=0\}d'_1 +\mathbbm{1}\{Y_1^{d'_1}=1\}\text{NP}$}\\
\midrule\ & \multirow[c]{2}{*}{ATE (dyn. conf.)} &  & -1.24***\phantom{(} & -0.61**\phantom{*(} & -3.51***\phantom{(} & 0.63***\phantom{(} & -2.27***\phantom{(} & -2.90***\phantom{(} \\
 &  &   & (0.21)\phantom{***} & (0.24)\phantom{***} & (0.16)\phantom{***} & (0.24)\phantom{***} & (0.15)\phantom{***} & (0.19)\phantom{***} \\
\bottomrule
\end{tabularx}
\caption*{\scriptsize \textit{Note:} This table reports ATEs with standard errors for various comparisons of policies. *, **, *** indicate p-values below 10\%, 5\%, and 1\%. $d_1$ and $d'_1$ represent first-period programs in the treatment and control states, respectively, while $g_2$ and $g'_2$ denote second-period policies dependent on the potential intermediate outcomes (1 if an individual exits unemployment in the first period). JA: Job-search assistance, TC: Training course, EP: Employment program, WS: Temporary wage subsidy, NP: No program. Outcome: Cumulative months in employment in the 30 months from start of the second period. ATE (dyn. conf.) [ATE (static conf.)] indicates that dynamic [static] confounding is assumed and the estimator by \citet{bodory2022evaluating}  [\citet{chernozhukov2018double}] is used for estimation. Additional details can be found in Table \ref{tab:DATE_first+duration_em332}.}
\end{threeparttable}
\end{table}

The median duration of the four programs, measured by the number of business days, are 23 for \textit{JA}, 39 for \textit{TC}, 80 for \textit{EP}, and 65 for \textit{WS}. A natural question arising from these differences is whether program duration affects effectiveness, i.e., whether some programs are superior to others due to their varying lengths. One approach to address this question is to use the sequential framework, comparing sequences as if individuals only participated during the first period.\footnote{An alternative approach to account for program duration would be to estimate a continuous treatment effect \citep{Imbens:2000}. This framework, however, requires program duration to be determined before treatment start, while here it is allowed that program duration can be updated once in-between. Repetitions of the same program type are also considered.} Policies of this type are not dynamic as \textit{NP} can occur in the second period regardless of whether an individual is unemployed or not. The results are presented in Panel B of Table \ref{tab:DATE_first+duration_em332_short}. The first row of the panel presents the ATE estimated using the dynamic DML method of \citet{bodory2022evaluating}, while the second row shows estimates ignoring dynamic confounding.

The results show that when restricting program duration to a maximum of three months, the ranking of the programs slightly changes. While \textit{WS} remains the most beneficial and \textit{JA} the least beneficial program, there is no longer a significant difference between \textit{TC} and \textit{EP}, when considering dynamic confounding. Furthermore, aligning program duration enhances the advantage of \textit{WS} relative to the other programs, for instance, by 1.5 months compared to \textit{JA} and 0.8 months compared to \textit{TC}. 
When comparing dynamic to static confounding, a systematic over-estimation of ATEs is observed for programs with a larger average duration. For example, when comparing the shortest program (\textit{JA}) to the longest program (\textit{EP}), the relative effectiveness of the latter almost doubles when dynamic confounding is ignored (1.47 vs. 2.74). Consequently, disregarding the feedback between treatments and intermediate outcomes results in an overly optimistic assessment of longer programs. Furthermore, Appendix Table~\ref{tab:DATE_first+duration_em332} shows that results from the methods of \citet{bodory2022evaluating} and \citet{bradic2021high} yield very similar estimates, indicating no clear advantage of either method in this application.

Instead of restricting programs to the first period, another approach to align program duration is to require at least two consecutive periods of the same program. Panel C of Table \ref{tab:DATE_first+duration_em332_short} presents estimates for the static scenario in which all individuals are assigned to the same program in both periods. Instead, Panel D presents results for dynamic policies in which individuals are reassigned to the same program only if they remain unemployed
throughout the first period. Overall, the results show patterns similar to those already observed for shorter or unrestricted sequences. 
Compared to programs lasting only one period, longer programs tend to slightly narrow the differences in effectiveness. However, these reductions are modest, suggesting that program duration is unlikely to be the primary driver of the observed differences between programs. The comparison between static and dynamic policies under dynamic confounding reveals that effect sizes increase for all comparison when considering dynamic policies. For example, under static policies, two periods of \textit{WS} result in 2.86 months more employment than \textit{JA}, whereas under dynamic policies, the difference increases to 3.51 months. This underscores that the choice of counterfactual can have economically significant implications.

\subsubsection{Program Order}

Besides duration or multiple participation, the sequential treatment effect framework can be exploited to obtain insights on the effective ordering of programs. As seen in Figure \ref{fig:alluvial}, a considerable number of individuals is assigned to different programs across the two periods. In such cases, it could be interesting to assess whether reversing the order of two programs leads to better outcomes. Table \ref{tab:DATE_order_em332_short} presents the results of such an analysis. Panel A and B present the average treatment effects implementing a specific pair of programs, compared to implementing the same pair in reverse order, assuming static and dynamic policies, respectively. 

The results for static policies indicate that when \textit{WS} is combined with \textit{TC} or \textit{EP}, it should be implemented as the second program rather than the first. This suggests that quitting a subsidized job to join an alternative program is less effective than first participating in the alternative program and then transitioning to the subsidized job. However, under dynamic policies, all effects become smaller and statistically insignificant. Hence, the previous conclusion holds only if individuals remain unemployed for at least two periods, which is unknown a priori. Instead, when allowing for the possibility of reemployment during the first period in the counterfactual, no clear superiority of any combination is observed. This suggests that conclusions drawn from ALMP evaluations based solely on static policies may not be reliable.\footnote{A challenge with analyzing the order of programs is that the second program may extend into subsequent periods, resulting again in comparisons of programs with differing durations. While a three-period setup with a final \textit{NP} period, as in \citet{lechner2013does}, could address this issue, it is not used here due to limited sample sizes.}

\begin{table}[htbp!]
\centering
\begin{threeparttable}
\caption{Average treatment effects for program order.}
\label{tab:DATE_order_em332_short}
\begin{tabularx}{\textwidth}{rXrrrrrrr}
\toprule
 &  &  & \makecell{$d_1 =$ JA\\ $d'_1 =$ TC} & \makecell{$d_1 =$ JA\\ $d'_1 =$ EP} & \makecell{$d_1 =$ JA\\ $d'_1 =$ WS} & \makecell{$d_1 =$ TC\\ $d'_1 =$ EP} & \makecell{$d_1 =$ TC\\ $d'_1 =$ WS} & \makecell{$d_1 =$ EP\\ $d'_1 =$ WS} \\
\midrule
\multicolumn{9}{l}{\textit{Panel A: Program order (static policy): $g_2(Y_1^{d_1}) = d'_1$, $g'_2(Y_1^{d'_1}) = d_1$}} \\
\midrule\ & \multirow[c]{2}{*}{ATE (dyn. conf.)} &  & -0.01\phantom{***(} & 5.30\phantom{***(} & 0.52\phantom{***(} & -0.51\phantom{***(} & 1.65***\phantom{(} & 1.89***\phantom{(} \\
 &  &   & (0.59)\phantom{***} & (5.54)\phantom{***} & (0.48)\phantom{***} & (0.62)\phantom{***} & (0.59)\phantom{***} & (0.59)\phantom{***} \\
 & \multirow[c]{2}{*}{ATE (static conf.)} &  & -0.43\phantom{***(} & -0.25\phantom{***(} & 1.03***\phantom{(} & -0.17\phantom{***(} & 1.09***\phantom{(} & 1.29***\phantom{(} \\
 &  &   & (0.31)\phantom{***} & (1.05)\phantom{***} & (0.25)\phantom{***} & (0.53)\phantom{***} & (0.28)\phantom{***} & (0.39)\phantom{***} \\
\midrule
\multicolumn{9}{l}{\textit{Panel B: Program order (dynamic policy):}}\\
\multicolumn{9}{l}{\phantom{\textit{Panel B: }}$g_2(Y_1^{d_1}) = \mathbbm{1}\{Y_1^{d_1}=0\}d'_1 +\mathbbm{1}\{Y_1^{d_1}=1\}\text{NP}$ and $g'_2(Y_1^{d'_1}) = \mathbbm{1}\{Y_1^{d'_1}=0\}d_1 +\mathbbm{1}\{Y_1^{d'_1}=1\}\text{NP}$}\\
\midrule\ & \multirow[c]{2}{*}{ATE (dyn. conf.)} &  & -0.58\phantom{***(} & -0.40\phantom{***(} & 0.22\phantom{***(} & -0.13\phantom{***(} & 0.19\phantom{***(} & 0.13\phantom{***(} \\
 &  &   & (0.41)\phantom{***} & (0.46)\phantom{***} & (0.31)\phantom{***} & (0.66)\phantom{***} & (0.32)\phantom{***} & (0.88)\phantom{***} \\
\bottomrule
\end{tabularx}
\caption*{\scriptsize \textit{Note:} This table reports ATEs with standard errors for various comparisons of policies. *, **, *** indicate p-values below 10\%, 5\%, and 1\%. See the note to Table \ref{tab:DATE_first+duration_em332_short} for further context; additional details are provided in Table \ref{tab:DATE_order_em332}.}
\end{threeparttable}
\end{table}

\subsubsection{Effect Heterogeneity}

A key advantage of causal machine learning methods is that they allow to flexibly analyze effect heterogeneities. In the DML setting, heterogeneous effects can be obtained by aggregating the estimated scores $\hat{\Theta}^{dy}_{\mathbf{g}_2}$ for specific sub-groups of interest. This analysis focuses on two types of heterogeneities: local language skill level and prior program participation in previous unemployment spells. Language skills, a proxy for migration background, are included since significant effect heterogeneity has been observed for this variable in multiple previous studies \citep[e.g.][]{Cockx:2020}. Previous program participation is included to enable an even more detailed analysis of program combinations. Of course, the procedure could be applied to any other discrete pre-treatment characteristic, provided there is a sufficiently large sample size. 

\begin{figure}[htbp!]
    \centering
    \captionsetup{font=small, width=0.9\linewidth} 
    \caption{GATE-ATE by local language knowledge and previous program participation}
    \includegraphics[width=0.9\linewidth]{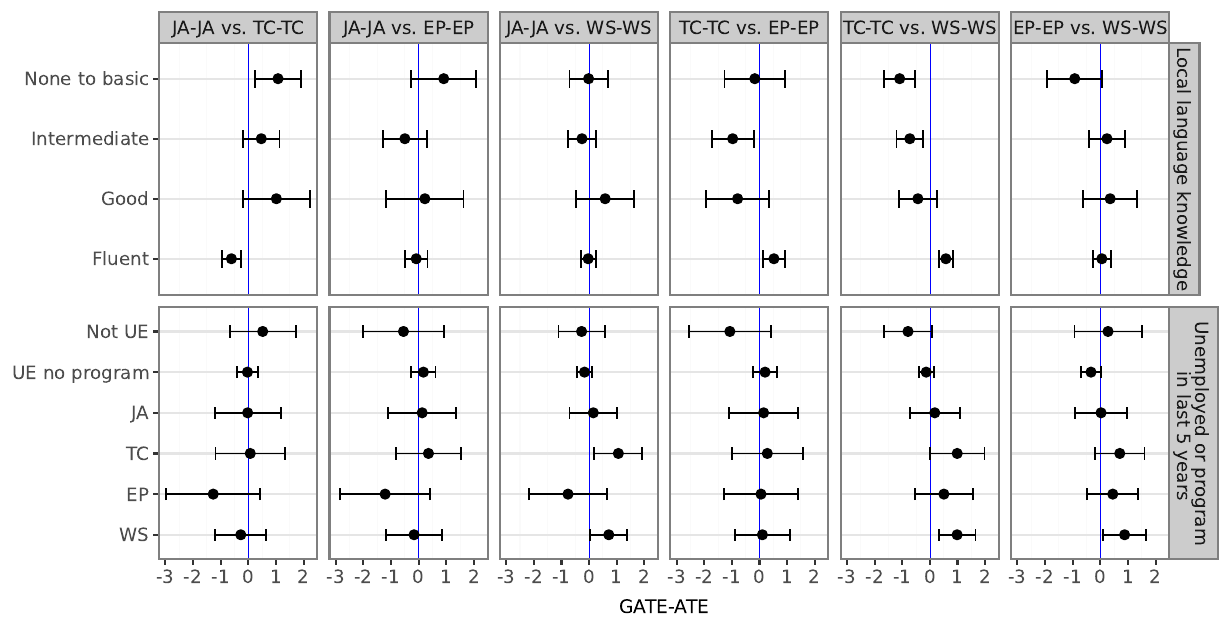}
    \caption*{\scriptsize \textit{Notes:} This figure shows GATE-ATE with 95\% confidence intervals by local language knowledge and previous program participation. Each column represents the comparison of two dynamic policies, where the first period program is continued in the second period if the individual remains unemployed in the first period (no program otherwise). The effects are estimated using the method proposed in \citet{bodory2022evaluating}. JA: Job-search assistance, TC: Training course, EP: Employment program, WS: Temporary wage subsidy. Outcome: Cumulative months in employment in the 30 months from start of the second period. Tabular results are available in Appendix Table \ref{tab:GATEmATE_combined_em332}.}
    \label{fig:gate_combined}
\end{figure}

The first row of panels in Figure \ref{fig:gate_combined} presents the results for heterogeneities based on local language knowledge. Specifically, estimates of the difference GATE-ATE are shown for dynamic policies (\ref{eq:g_application}) with $d_2=d_1$, estimated by the method of \citet{bodory2022evaluating}. The effects are presented as the difference between GATE and ATE, where a significant deviation from zero indicates significant heterogeneity relative to the average.\footnote{As derived in Appendix \ref{ap:gate_ate_var}, standard errors for this difference can be computed as the square root of $\Var(\hat \theta (z_0)-\hat \theta) = \Var(\hat \theta (z_0))+\Var(\hat \theta) - \frac{2N_{z_0}}{N} \Var(\hat \theta (z_0))$, where $N_{z_0}$ denotes the number of observations with $Z_0=z_0$.} The results show that for this type of long programs,  \textit{TC} is particularly effective for individuals fluent in the local language, whereas those with limited language skills benefit significantly less than the average. This finding holds true when comparing \textit{TC} to each of the other programs. The result is surprising, given that \textit{TC} includes language courses alongside various other training programs. In addition, individuals with none to intermediate proficiency in the local language benefit significantly more than average from \textit{WS} compared to \textit{TC}, suggesting that extended programs incorporating work experience may be more effective than extended training courses for those with limited language skills.

The second row of Figure \ref{fig:gate_combined} presents the same analysis using information about unemployment spells in the five years prior to program start. The results are mostly close to zero and insignificant with a few exceptions. In particular, individuals who participated in \textit{WS} during a previous unemployment spell benefit above-average from programs other than \textit{WS}. This suggests that repeated participation in \textit{WS} across multiple unemployment spells leads to reduced program effectiveness. In addition, individuals who have not been unemployed in Switzerland before, profit less from \textit{TC} compared to \textit{WS} (significant at 10\% level). This aligns with the findings for local language knowledge, as recent immigrants are overrepresented among the first-time unemployed.

\section{Conclusion} \label{sec:conclusion}
This paper reviewed, explained, and applied methods for the evaluation of program sequences and introduced the concept of dynamic policies to the econometric program evaluation literature. By summarizing the identification process under static and dynamic confounding, it was demonstrated that assessing dynamic policies allows for the construction of more realistic counterfactuals, requiring only minor adjustments to identification assumptions. In addition, the analysis illustrated how dynamic DML can be employed to flexibly estimate the effects of dynamic policies. The presented methods provide a foundation for more effective policy design in settings where program assignments depend on time-varying characteristics.

The empirical application evaluated sequences of Swiss ALMP across two consecutive periods, starting at the beginning of the first program. An initial descriptive analysis revealed a significant variety of program sequences, highlighting the necessity of a sequential analysis. Using standard results from single time point interventions as a benchmark, it was demonstrated how an assessment of sequential policies can provide additional insights into implementation details of ALMP. This revealed that Temporary Wage Subsidies are most effective on average, even after adjusting program duration across different program types. 
In particular, individuals with limited proficiency of the local language profit more from programs related to obtaining work experience in comparison to extended training courses. 
In the application, disregarding dynamic confounding resulted in an overly optimistic assessment of longer programs. Moreover, the choice between static and dynamic policies was shown to result in economically and statistically significant differences in estimated effect sizes.

Overall, DML-based estimation of effects of dynamic policies turns out to be a valuable addition to the standard program evaluation toolkit. A key limitation of any sequential analysis is the need for sufficient sample sizes for all sequences of interest to ensure robust estimation. This challenge is particularly pronounced for flexible machine learning methods, which avoid parametric structural assumptions but require large datasets to capture non-linear relationships. In the application, this constraint necessitated the aggregation of program categories into four major groups. This aggregation potentially masks distinct effects of individual sub-programs, limiting the detail of the findings. For the same reason, only a few relatively long periods could be used in the panel, which might not perfectly capture the true dynamic selection process. Although sequential methods are data-intensive, their potential is expected to increase as larger datasets become available. 

\newpage 


\bibliography{literature.bib} 
\bibliographystyle{apacite} 

\newpage
\begin{appendices}
\renewcommand\thetable{\thesection.\arabic{table}} 
\counterwithin{table}{section}
\renewcommand\thefigure{\thesection.\arabic{figure}} 
\counterwithin{figure}{section}

\section{Estimators for Effects of Sequential Treatments}\label{ap:other_estimators}

\subsection{Overview of Conventional Estimators} \label{ap:other_estimators_conv}

To situate the machine learning-based estimators used in this paper in the broader methodological context, this section briefly overviews conventional estimation techniques commonly applied in the analysis of sequential treatments. For a comprehensive textbook introduction to these methods see \citet{hernan2020causal}.

The early literature proposed estimation of dynamic treatment effects based on the so-called $g$-formula. The $g$-formula is equivalent to identification result discussed in Theorem \ref{th:identification} above. It is called $g$-formula as it is designed for the estimation of $g$eneralized treatment effects beyond the static case. The associated estimation method, referred to as $g$-computation \citep{robins1986new}, requires estimation of the mean outcomes for all treatment sequences of interest, conditional on covariates, as well as the estimation of conditional densities of all time-varying covariates. 
Originally, this has been implemented using a parametric plug-in procedure in which, for example, the conditional outcomes and (discrete) covariate distributions were estimated by linear and logistic regressions, respectively. 
Later it has been acknowledged that the latter can be avoided by exploiting a representation in terms of nested conditional expectations, which can be used as updated outcomes in an iterative estimation procedure \citep{robins2000robust, murphy2003optimal}. Based on this idea, \citet{lechner2010identification} propose propensity score matching, where instead of conditioning on the covariates, it is conditioned on the propensity scores.

Alternatively, causal effects of time-varying treatments can also be estimated using inverse probability weighting (IPW) \citep{robins2000marginal}, which goes back to \citet{horvitz1952generalization}. Conceptually, while the $g$-formula mimics a case in which everyone receives the same treatment, the idea behind IPW is to re-establish a setting in which there is no confounding and everyone receives treatment at random \citep{robins2009estimation}. Therefore, each observation in the sample is re-weighted by the inverse probability of receiving the treatment it actually got, given covariates. In the dynamic setting, the standard procedure is complicated by the occurrence of multiple treatments and their dependence on previous treatments and covariates. This leads to a product of propensity scores in the denominator of the weights, which might amplify the problem of extreme weights already known from the static setup. Note that in observational studies the propensity scores need to be estimated, see \citet{lechner2009sequential} for an application in the ALMP context. 

For cases where the number of program sequences is large compared to the sample size, literature has drawn on structural mean models, which model the relationship between program assignment and the mean potential outcome. For example, a popular estimation approach based on \citet{robins2000marginal} are marginal structural models (MSM), which are specifically designed for cases with many periods and many levels of treatments. In its simplest form, an MSM estimates the expected potential outcome using a least-squares regression of the outcome on the cumulative number of treated periods in the pseudo-population reweighted by inverse probability weights. Another approach is the $g$-estimation of structural nested mean models \citep{robins1989analysis, robins1994correcting}. At each period, this approach models the effect of changing the treatment in that period, conditional on treatment and covariate history. Then, starting in the last period, the models are solved using a backward induction algorithm that recursively iterates outward and applies the sequential randomization assumption in each step. To simplify computation, a linear specification of the equations is typically assumed.

A further option to estimate dynamic treatment effects are methods based on augmented inverse probability weighting (AIPW) that combine the $g$-formula with IPW. When using parametric models for nuisance parameter estimation, AIPW methods are useful as they remain consistent even if one of the parametric models is misspecified. AIPW estimation goes back to \citet{robins1994estimation} and \citet{scharfstein1999adjusting} who showed that AIPW is doubly robust. \citet{bang2005doubly} extended the procedure to the longitudinal setting and proposed to implement AIPW estimation parametrically using inverse propensity scores as ``clever covariates'' in the nested outcome models. This procedure can be seen as a precursor of targeted minimum loss-based estimation, which is discussed in the following section. The main body of this paper focuses on similar doubly-robust estimators that, however, do not rely on parametric models for the estimation of the nuisance functions.

\subsection{Alternative Machine Learning Estimators based on Influence Functions} \label{ap:other_estimators_EIF}

Besides the DML-estimators discussed in Section \ref{sec:dynDML}, alternative influence function-based estimators have been proposed in the literature. \citet{singh2021finite} provide a very general framework, which applies to several longitudinal parameters such as sequential and mediated treatment effects or long-term effects using surrogates. Their paper can be seen as a sequential extension of \citet{chernozhukov2023simple}, and covers both ATEs and GATEs as special cases. Their theory provides a finite sample Gaussian approximation under regularity conditions, as well as $\sqrt{N}$-consistency and asymptotic normality under assumptions on the learning rate of the nuisance estimates, similar to \citet{bodory2022evaluating}. The authors do not commit to a specific estimator for the nested conditional outcome $\nu_{\mathbf{g}_2}(X_0)$, as long as it converges at a fast enough rate. However, they provide estimation theory for an adversarial nested instrumental variable regression procedure that avoids using $\hat{\mu}_{\mathbf{g}_2}(\mathbf{X}_{1})$ as a pseudo-outcome. Hence, no additional sample splits within the cross-fitting folds are needed, which becomes increasingly relevant as the number of time periods increases. The application in the main text does not consider their adversarial estimator, given that it remains within a setting limited to a maximum of two time periods and does not require instrumental variables.

\citet{chernozhukov2022automaticdyn} propose to estimate the ATE under dynamic confounding using automatic DML (Auto-DML). Instead of directly using the orthogonal score function (\ref{eq:DML_emp}), Auto-DML is based on the more general function
\begin{align*}
    \Theta_{\mathbf{g}_2}^{\text{AUTO}}(\mathbf{W}_{2}) :=&\; \nu_{\mathbf{g}_2}(X_0) + a_1(D_1, X_0)\left(\mu_{\mathbf{g}_2}(\mathbf{X}_1)  -  \nu_{\mathbf{g}_2}(X_0)  \right)  +a_2(\mathbf{D}_2, \mathbf{X}_1) \left(Y -  \mu_{\mathbf{g}_2}(\mathbf{X}_1)    \right)  \text{,}
\end{align*}
where the weights $a_1$ and $a_2$ denote the recursive Riesz representer functions. This representation is guaranteed to exist if treatments are discrete and the overlap assumption holds \citep{chernozhukov2022automatic}. While the previously proposed estimators exploit closed-form solutions of $a_1$ and $a_2$, and plug-in estimates of the propensity scores $\hat p_{g_1}$ and $\hat p_{g_2}$, Auto-DML avoids  using the analytical form and learns the Riesz representers $\hat a_t$ directly from numerical optimization. As the authors argue, this leads to improved behavior of the estimated weights even in settings where the functional form is known, since plugging-in estimated probabilities in the denominators of the weights can be avoided. Similar to the nested outcomes, a particular difficulty is that only the Riesz representer of the final period is directly identified from the data. For earlier periods, $a_t$ must be learned based on the estimated representer from the previous period, which complicates the extension of the procedures developed for single time periods. While \citet{chernozhukov2022automaticdyn} provide promising theoretical results, unsolved questions remain regarding the practical implementation of Auto-DML in the sequential setting. For example, it is unclear which estimators for the nested conditional outcomes and nested Riesz representers satisfy the theoretical requirements. In addition, the performance of the approach in finite samples is not yet well understood. Finally, the literature is still lacking empirical applications of Auto-DML even in single-period setting.

A method closely related to DML is targeted minimum loss-based estimation (TMLE), originally proposed in \citet{van2006targeted}. In the basic setting with static policies and static confounding, TMLE updates the initial (biased) prediction of the conditional outcome $\hat\mu_{\mathbf{d}_2}(X_0)$ using the adjustment
\begin{align*}
    \Tilde{\mu}_{\mathbf{d}_2}(X_{0}) &= \hat{\mu}_{\mathbf{d}_2}(X_{0}) + \underbrace{\frac{\frac{1}{N}\sum_{j=1}^N \hat\alpha(X_{0j}, \mathbf{D}_{2j}) \left(Y_j -  \hat{\mu}_{\mathbf{d}_2}(X_{0j})\right) }{\frac{1}{N}\sum_{j=1}^N \hat\alpha(X_{0j}, \mathbf{D}_{2j})^2}}_{=: \hat\epsilon} \hat\alpha(X_{0}, \mathbf{d}_2)\text{,}
\end{align*}
with $\hat\alpha(X_{0}, \mathbf{D}_{2}) = \mathbbm{1}\left\{\mathbf{D}_{2}=\mathbf{d}_2\right\}/\hat{p}_{\mathbf{d}_2}(X_{0})$. Hence, $\Tilde{\mu}_{\mathbf{d}_2}(X_{0})$ corresponds to the conditional outcome $\hat{\mu}_{\mathbf{d}_2}(X_{0})$ plus the predicted value of a regression of the residual $Y-\hat{\mu}_{\mathbf{d}_2}(X_{0})$ on the ``clever covariate'' $\hat \alpha(X_{0}, \mathbf{D}_{2})$ with regression coefficient $\hat\epsilon$, evaluated at $\mathbf{D}_{2}=\mathbf{d}_2$. Conceptually, the key distinction between DML and TMLE lies in their approach to obtaining an estimator where the orthogonal moment condition equals zero. DML imposes this condition upfront and derives the de-biased estimator by solving the condition. In contrast, TMLE starts with the original prediction of the conditional outcome and fluctuates it until the moment function is zero and no further de-biasing is needed. This can be seen from plugging $\Tilde{\mu}_{\mathbf{d}_2}(X_{0i})$ for $\hat{\mu}_{\mathbf{d}_2}(X_{0i})$ into the condition (\ref{eq:emp_EIF}), which results in a zero de-biasing term, i.e. 
\begin{align*}
     \Tilde\theta^{\mathbf{g}_2} = \frac{1}{N} \sum_{i=1}^N  \Tilde{\mu}_{\mathbf{d}_2}(X_{0i})
    + \frac{\left(Y_i -  \Tilde{\mu}_{\mathbf{d}_2}(X_{0i})\right) \cdot \mathbbm{1}\left\{\mathbf{D}_{2i}=\mathbf{d}_2\right\}}{\hat{p}_{\mathbf{d}_2}(X_{0i})}
      = \frac{1}{N} \sum_{i=1}^N  \Tilde\mu_{\mathbf{d}_2}(X_{0i})\text{.}
\end{align*}
Hence, the target parameter estimate can be directly obtained by averaging $\Tilde{\mu}_{\mathbf{d}_2}(X_{0i})$ over the population of interest. DML and cross-fitted TMLE share the same statistical properties and are asymptotically equivalent, as shown in \cite{chernozhukov2022automatic}. However, TMLE might be more stable than DML in finite samples since the working model on the ``clever covariate'' can be exploited to impose global constraints, for example if the outcome $Y$ is bounded in some interval \citep{kennedy2022semiparametric}.

TMLE has been extended to dynamic policies and dynamic confounding, referred to as longitudinal TMLE (LTMLE). Similar to DML, the estimated (and targeted) conditional mean outcome of the second period, $\Tilde{\mu}_{\mathbf{g}_2}(\mathbf{X}_{1})$, is employed as a pseudo-outcome in a regression on $X_0$ within the subset $D_1=g_1(V_0)$. This produces an estimate $\hat{\nu}_{\mathbf{g}_2}(X_{0})$, which is subsequently targeted again to yield $\Tilde{\nu}_{\mathbf{g}_2}(X_{0})$. In direct extension of the previous exposition, the targeted estimates $\Tilde{\mu}_{\mathbf{g}_2}(\mathbf{X}_{1})$ and $\Tilde{\nu}_{\mathbf{g}_2}(X_{0})$ lead to a zero de-biasing term when plugged into the score function (\ref{eq:emp_EIF}). \citet{tran2019double} show that LTMLE outperforms competing estimators of sequential effects in a simulation study. However, to the best of the author's knowledge, no cross-fitted version of LTMLE currently exists, and the optimal approach for implementing sample splitting in the longitudinal context remains unclear. Hence, the procedure is prone to overfitting, unless restrictive complexity constraints on the nuisance estimators are met, which exclude commonly applied machine learning estimators such as lasso or random forests \citep{kennedy2022semiparametric}. For this reason, the method is not adopted in the current study.

Finally, \citet{lewis2020double} introduce DML for effect estimation of sequential treatments as an extension of the classical $g$-estimation of structural nested mean models \citep{robins1989analysis, robins1994correcting}. Unlike the procedures discussed in the previous section, their approach is not directly based on the identification result (\ref{eq:DR}) but instead relies on an expansion of the potential outcome, 
\begin{align*}
    Y^{\mathbf{g}_2} = Y^{g_1, g_2} = Y + \Big(Y^{g_2} -Y\Big) + \Big(Y^{g_1, {g}_2} - Y^{g_2}\Big)\text{,}
\end{align*}
with $Y^{{g}_2} := \sum_{\mathbf{d}_2 \in \mathcal{D}_1 \times \mathcal{D}_2} \mathbbm{1}\left\{D_1 = d_1, {g}_2(\mathbf{V}_1) = {d}_2\right\} Y^{\mathbf{d}_2}$. Hence, the target potential outcome is obtained from the observed outcome by adding the `blip effect' of changing only the last program plus the `blip effect' of switching the first program after having already switched the second program. This observation motivates a backward induction algorithm, which recursively estimates the `blip effects' in a Neyman orthogonal manner using cross-fitting. Starting in the final period, the `blip effect' is estimated and subtracted from the outcome to derive a new adjusted outcome that reflects the effect of the counterfactual program rather than the observed program in the final period. The adjusted outcome is subsequently used to estimate the `blip effect' of changing programs in the previous period.
While the `blip effects' are non-parametrically identified by applying Assumption \ref{as:identification} at each stage, a linear parametric form is imposed on them for estimation.
This eliminates the need to weight by inverse products of estimated propensity scores while ensuring $\sqrt{N}$ convergence and asymptotic normality, even when nuisance functions are estimated using machine learning methods.
Their method is not used in this study due to the restrictive nature of the parametric assumptions. Nevertheless, the additional structure provided by their approach may be beneficial in scenarios involving many or continuous treatments and many time periods, where non-parametric approaches become infeasible. 

\newpage
\section{Proofs}\label{ap:identification}

\subsection{Identification of Static Policies}

\paragraph{Proof of Theorem \ref{th:identification_static}:} The conditional APO is identified as
\begin{align}\label{eq:mu_stat}
    \E[Y^{\mathbf{d}_2}|X_0] = \E[Y^{\mathbf{d}_2}|X_0, \mathbf{D}_2=\mathbf{d}_2] = \E[Y|X_0, \mathbf{D}_2=\mathbf{d}_2] =: \mu_{\mathbf{d}_2}(X_0)\text{,}
\end{align}
where the first equality uses Assumption \ref{as:identification_static} or \ref{as:identification_static2} and the second equality uses Assumption \ref{as:sutva}. Given this result, the parameters of interest are derived using the law of iterated expectations as
\begin{align}\label{eq:identification_static_result}
    \theta^{\mathbf{d}_2} &= \E[Y^{\mathbf{d}_2}] = \E_{X_0}[\E[Y^{\mathbf{d}_2}|X_0]] = \E_{X_0}[\mu_{\mathbf{d}_2}(X_0)] \;\;\; \text{and}\nonumber\\
    \theta^{\mathbf{d}_2}(z_0) &= \E[Y^{\mathbf{d}_2}|Z_0=z_0] = \E_{X_0}[\E[Y^{\mathbf{d}_2}|X_0]|Z_0=z_0] = \E_{X_0}[\mu_{\mathbf{d}_2}(X_0)|Z_0=z_0]\text{.}
\end{align}

\paragraph{Proof of Theorem \ref{th:identification}:} The average potential outcome for a static program sequence, conditional on all information known at the end of the first period, is identified as
\begin{align}\label{eq:mu}
    \E[Y^{\mathbf{d}_2}|\mathbf{X}_1, D_1=d_1] = \E[Y^{\mathbf{d}_2}|\mathbf{X}_1, \mathbf{D}_2=\mathbf{d}_2] = \E[Y|\mathbf{X}_1, \mathbf{D}_2=\mathbf{d}_2] =: \mu_{\mathbf{d}_2}(\mathbf{X}_1)\text{,}
\end{align}
where the first equality uses Assumption \ref{as:identification} and the second equality uses Assumption \ref{as:sutva}. From here, the conditional APO is obtained as
\begin{align}\label{eq:nu}
    \E[Y^{\mathbf{d}_2}|X_0] &= \E[Y^{\mathbf{d}_2}|X_0, D_1=d_1] \nonumber\\ &= \E_{X_1} \big[ \E[Y^{\mathbf{d}_2}|\mathbf{X}_1, D_1=d_1] \big| X_0, D_1=d_1\big] \nonumber\\
    &= \E_{X_1} \big[ \mu_{\mathbf{d}_2}(\mathbf{X}_1) \big| X_0, D_1=d_1\big]\nonumber\\
    &=: \nu_{\mathbf{d}_2}(X_0)\text{,}
\end{align}
where the first equality again uses Assumption \ref{as:identification}, the second equality is obtained using the law of iterated expectations and the third equality plugs in equation (\ref{eq:mu}). Based on result (\ref{eq:nu}), the parameters of interest are identified using the law of iterated expectations as
\begin{align}\label{eq:identification_dynamic_result}
    \theta^{\mathbf{d}_2} &= \E[Y^{\mathbf{d}_2}] = \E_{X_0}[\E[Y^{\mathbf{d}_2}|X_0]] = \E_{X_0}[\nu_{\mathbf{d}_2}(X_0)] \;\;\;\text{and}\nonumber\\
    \theta^{\mathbf{d}_2}(z_0) &= \E[Y^{\mathbf{d}_2}|Z_0=z_0] = \E_{X_0}[\E[Y^{\mathbf{d}_2}|X_0]|Z_0=z_0] = \E_{X_0}[\nu_{\mathbf{d}_2}(X_0)|Z_0=z_0]\text{.}
\end{align}

\subsection{Identification of Dynamic Policies}

\paragraph{Proof of Theorem \ref{th:identification_dyn}:}

The average potential outcome for a dynamic policy, conditional on all information known at the end of the first period, is identified as
\begin{align}\label{eq:mu_dyn}
    \E[Y^{\mathbf{g}_2}|\mathbf{X}_1, D_1=g_1(V_0)] 
    &= \E\left[\sum_{\mathbf{d}_2 \in \mathcal{D}_1 \times \mathcal{D}_2} \mathbbm{1}\left\{g_1(V_0) = d_1, g_2(\mathbf{V}_1^{d_1}) = d_2\right\} Y^{\mathbf{d}_2}\middle|\mathbf{X}_1, D_1=g_1(V_0) \right]\nonumber\\
    &= \E\left[\sum_{\mathbf{d}_2 \in \mathcal{D}_1 \times \mathcal{D}_2} \mathbbm{1}\left\{D_1 = d_1, g_2(\mathbf{V}_1) = d_2\right\} Y^{\mathbf{d}_2}\middle|\mathbf{X}_1, D_1=g_1(V_0) \right]\nonumber\\
    &= \E\left[\sum_{\mathbf{d}_2 \in \mathcal{D}_1 \times \mathcal{D}_2} \mathbbm{1}\left\{D_1 = d_1, D_2 = d_2\right\} Y^{\mathbf{d}_2}\middle|\mathbf{X}_1, D_1=g_1(V_0), D_2=g_2(\mathbf{V}_1) \right]\nonumber\\
    &= \E[Y|\mathbf{X}_1, D_1=g_1(V_0), {D}_2=g_2(\mathbf{V}_1)]\nonumber\\ &=: \mu_{\mathbf{g}_2}(\mathbf{X}_1)\text{,}
\end{align}
where the first equality plugs in the definition of $Y^{\mathbf{g}_2}$, the second equality uses Assumption \ref{as:sutva}\ref{as:sutvaX1} for $V_1^{d_1}$, the third equality uses \ref{as:identification_dyn} for $t=2$, and the fourth equality applies Assumption \ref{as:sutva}\ref{as:sutvaY}. Using this result, it follows that
\begin{align*}
    \E&[Y^{\mathbf{g}_2}|X_0]\\
    &= \E\left[\sum_{\mathbf{d}_2 \in \mathcal{D}_1 \times \mathcal{D}_2} \mathbbm{1}\left\{g_1(V_0) = d_1, g_2(\mathbf{V}_1^{d_1}) = d_2\right\} Y^{\mathbf{d}_2}\middle|X_0 \right]\nonumber\\
    &= \E\left[\sum_{\mathbf{d}_2 \in \mathcal{D}_1 \times \mathcal{D}_2} \mathbbm{1}\left\{g_1(V_0) = d_1, g_2(\mathbf{V}_1^{d_1}) = d_2\right\} Y^{\mathbf{d}_2}\middle|X_0, D_1=g_1(V_0) \right]\nonumber\\
    &= \E_{X_1}\left[\E\left[\sum_{\mathbf{d}_2 \in \mathcal{D}_1 \times \mathcal{D}_2} \mathbbm{1}\left\{g_1(V_0) = d_1, g_2(\mathbf{V}_1^{d_1}) = d_2\right\} Y^{\mathbf{d}_2}\middle|\mathbf{X}_1, D_1=g_1(V_0) \right]\middle|X_0, D_1=g_1(V_0)\right]\nonumber\\
    &= \E_{X_1} \big[ \mu_{\mathbf{g}_2}(\mathbf{X}_1) \big| X_0, D_1=g_1(V_0)\big]\nonumber\\
    &=: \nu_{\mathbf{g}_2}(X_0)
\end{align*}
where the first equality again applies the definition of $Y^{\mathbf{g}_2}$, the second equality follows from Assumption \ref{as:identification_dyn} for $t=1$, the third equality applies the law of iterated expectations and the fourth equality plugs in result (\ref{eq:mu_dyn}). From here, $\theta^{\mathbf{g}_2}$ and $\theta^{\mathbf{g}_2}(z_0)$ are obtained identically to (\ref{eq:identification_dynamic_result}) by integrating over $X_0$.

\subsection{Identification using Augmented Inverse Probability Weighting}
\label{sec:DRproof}
This section proves identification of the target parameters $\theta^{\mathbf{g}_2}$ and $\theta^{\mathbf{g}_2}(z_0)$ by the score $\Theta_{\mathbf{g}_2}^{dy}(\mathbf{W}_{2})$ and shows that it fulfills the multiple robustness property. For a proof of identification double robustness under static confounding see e.g. \citet{knaus2022double}. \citet[][Lemma 1]{bradic2021high} provide a proof of multiple robustness of static policies under dynamic confounding. Here, their proof is extended to dynamic policies.

\paragraph{Proof of Theorem \ref{th:identification_dr}:} In what follows it is shown that
\begin{align*}
    \theta^{\mathbf{g}_2} &= \E[Y^{\mathbf{g}_2}] = \E_{X_0}[\E[Y^{\mathbf{g}_2}|X_0]] = \E_{X_0}[\E[\Theta_{\mathbf{g}_2}^{dy}(\mathbf{W}_{2})|X_0]] = \E[\Theta_{\mathbf{g}_2}^{dy}(\mathbf{W}_{2})]  \\
    \theta^{\mathbf{g}_2}(z_0) &= \E[Y^{\mathbf{g}_2}|Z_0=z_0] = \E_{X_0}[\E[Y^{\mathbf{g}_2}|X_0]|Z_0=z_0] = \E_{X_0}[\E[\Theta_{\mathbf{g}_2}^{dy}(\mathbf{W}_{2})|X_0]|Z_0=z_0] \\&= \E[\Theta_{\mathbf{g}_2}^{dy}(\mathbf{W}_{2})|Z_0=z_0]\text{.}
\end{align*}
Both statements hold true if $\E[Y^{\mathbf{g}_2}|X_0] = \E[\Theta_{\mathbf{g}_2}^{dy}(\mathbf{W}_{2})|X_0=x_0]$. One can write
\begin{align*}
    \E[\Theta_{\mathbf{g}_2}^{dy}(\mathbf{W}_{2})|X_0=x_0] &= \nu_{\mathbf{g}_2}(x_0)\\
    &+ \E \left[ \frac{\left(Y -  \mu_{\mathbf{g}_2}(\mathbf{X}_1)    \right) \mathbbm{1}\left\{\mathbf{D}_2=\mathbf{g}_2(\mathbf{V}_1)\right\}}{p_{g_2}(\mathbf{X}_1,  g_1)p_{g_1}(X_0)}\middle| X_0=x_0\right]\\
    &+ \E \left[ \frac{\left(\mu_{\mathbf{g}_2}(\mathbf{X}_1)  -  \nu_{\mathbf{g}_2}(X_0)  \right) \mathbbm{1}\left\{D_1=g_1(V_0)\right\} }{p_{g_1}(X_0)}\middle| X_0=x_0\right]\text{.}
\end{align*}
From statement \ref{eq:nu} it is known that $\nu_{\mathbf{g}_2}(x_0) = \E[Y^{\mathbf{g}_2}|X_0]$. Hence, to prove identification it suffices to show that the second and third term are zero. For the second term one can show
\begin{align}
    \E& \left[ \frac{\left(Y -  \mu_{\mathbf{g}_2}(\mathbf{X}_1)    \right) \mathbbm{1}\left\{\mathbf{D}_2=\mathbf{g}_2(\mathbf{V}_1)\right\}}{p_{g_2}(\mathbf{X}_1,  g_1)p_{g_1}(X_0)}\middle| X_0=x_0\right]\nonumber \\
    &= \E_{D_1} \left[ \E \left[ \frac{\left(Y -  \mu_{\mathbf{g}_2}(\mathbf{X}_1)    \right) \mathbbm{1}\left\{\mathbf{D}_2=\mathbf{g}_2(\mathbf{V}_1)\right\}}{p_{g_2}(\mathbf{X}_1,  g_1)p_{g_1}(X_0)}\middle| X_0=x_0, D_1\right]\middle| X_0=x_0 \right]  \nonumber\\
    &= \Pr(D_1=g_1(V_0)|X_0=x_0) \E \left[ \frac{\left(Y -  \mu_{\mathbf{g}_2}(\mathbf{X}_1)    \right) \mathbbm{1}\left\{D_2=g_2(\mathbf{V}_1)\right\}}{p_{g_2}(\mathbf{X}_1,  g_1)p_{g_1}(X_0)}\middle| X_0=x_0, D_1=g_1(V_0)\right] \nonumber\\
    &= \E \left[ \frac{\left(Y -  \mu_{\mathbf{g}_2}(\mathbf{X}_1)    \right) \mathbbm{1}\left\{D_2=g_2(\mathbf{V}_1)\right\}}{p_{g_2}(\mathbf{X}_1,  g_1)}\middle| X_0=x_0, D_1=g_1(V_0)\right] \nonumber\\
    &= \E_{X_1} \left[\E \left[ \frac{\left(Y -  \mu_{\mathbf{g}_2}(\mathbf{X}_1)    \right) \mathbbm{1}\left\{D_2=g_2(\mathbf{V}_1)\right\}}{p_{g_2}(\mathbf{X}_1,  g_1)} \middle| \mathbf{X}_1, D_1=g_1(V_0)\right]\middle|X_0=x_0, D_1=g_1(V_0)\right] \nonumber\\
    &= \E_{X_1} \left[ \Pr(D_2=g_2(\mathbf{V}_1)|\mathbf{X}_1, D_1=g_1(V_0)) \E \left[ \frac{Y -  \mu_{\mathbf{g}_2}(\mathbf{X}_1)}{p_{g_2}(\mathbf{X}_1,  g_1)}\middle| \mathbf{X}_1, \mathbf{D}_2=\mathbf{g}_2(\mathbf{V}_1)\right]\middle|X_0=x_0, D_1=g_1(V_0)\right] \nonumber\\
    &=  \E_{X_1} \Big[ \E \left[Y -  \mu_{\mathbf{g}_2}(\mathbf{X}_1) \middle| \mathbf{X}_1, \mathbf{D}_2=\mathbf{g}_2(\mathbf{V}_1)\right]\Big|X_0=x_0, D_1=g_1(V_0)\Big] \nonumber\\
    &=  \E_{X_1} \Big[ \E \left[Y \middle| \mathbf{X}_1, \mathbf{D}_2=\mathbf{g}_2(\mathbf{V}_1)\right] -  \mu_{\mathbf{g}_2}(\mathbf{X}_1) \Big|X_0=x_0, D_1=g_1(V_0)\Big] \nonumber\\
    &= 0 \text{.}\label{eq:DML_ymu}
\end{align}
For the third term one can show
\begin{align}   
    \E& \left[ \frac{\left(\mu_{\mathbf{g}_2}(\mathbf{X}_1)  -  \nu_{ \mathbf{g}_2}(X_0) \right) \mathbbm{1}\left\{D_1=g_1(V_0)\right\} }{p_{g_1}(X_0)}\middle| X_0=x_0\right]\nonumber\\
    &= \E_{D_1} \left[ \E_{X_1} \left[ \frac{\left(\mu_{\mathbf{g}_2}(\mathbf{X}_1)  -  \nu_{ \mathbf{g}_2}(X_0) \right) \mathbbm{1}\left\{D_1=g_1(V_0)\right\} }{p_{g_1}(X_0)}\middle| X_0=x_0, D_1\right]\middle| X_0=x_0 \right]  \nonumber\\
    &= \Pr(D_1=g_1(V_0)|X_0=x_0) \E_{X_1} \left[ \frac{\mu_{\mathbf{g}_2}(\mathbf{X}_1)  -  \nu_{ \mathbf{g}_2}(X_0)}{p_{g_1}(X_0)}\middle| X_0=x_0, D_1=g_1(V_0)\right] \nonumber\\
    &= \E_{X_1} \left[ \mu_{\mathbf{g}_2}(\mathbf{X}_1)  -  \nu_{ \mathbf{g}_2}(X_0)\middle| X_0=x_0, D_1=g_1(V_0)\right] \nonumber\\
    &= \E_{X_1} \left[ \mu_{\mathbf{g}_2}(\mathbf{X}_1)  \middle| X_0=x_0, D_1=g_1(V_0)\right] -  \nu_{\mathbf{g}_2}(x_0) \nonumber\\&=0\text{.} \nonumber
\end{align}
Hence, $\E[Y^{\mathbf{g}_2}|X_0] = \E[\Theta_{\mathbf{g}_2}^{dy}(\mathbf{W}_{2})|X_0=x_0]$ which completes the proof.

\paragraph{Proof of multiple robustness property:} One can also show that the score $\Theta_{\mathbf{g}_2}^{dy}(\mathbf{W}_{2})$ possesses so-called multiple robustness properties. Therefore,  replace the true functions $\mu_{\mathbf{g}_2}(\mathbf{X}_1)$, $\nu_{\mathbf{g}_2}(X_0)$, $p_{g_1}(X_0)$ and $p_{g_2}(\mathbf{X}_1,  g_1)$ in the second and third term by arbitrary functions $\Tilde{\mu}_{\mathbf{g}_2}(\mathbf{X}_1)$, $\Tilde{\nu}_{\mathbf{g}_2}(X_0)$, $\Tilde{p}_{g_1}(X_0)$ and $\Tilde{p}_{g_2}(\mathbf{X}_1, g_1)$. For the second term one can show
\begin{align}
    \E& \left[ \frac{\left(Y - \Tilde{\mu}_{\mathbf{g}_2}(\mathbf{X}_1)    \right) \mathbbm{1}\left\{\mathbf{D}_2=\mathbf{g}_2(\mathbf{V}_1)\right\}}{\Tilde{p}_{g_2}(\mathbf{X}_1,  g_1)\Tilde{p}_{g_1}(X_0)}\middle| X_0=x_0\right]\nonumber \\
    &= \E_{D_1} \left[ \E \left[ \frac{\left(Y - \Tilde{\mu}_{\mathbf{g}_2}(\mathbf{X}_1)    \right) \mathbbm{1}\left\{\mathbf{D}_2=\mathbf{g}_2(\mathbf{V}_1)\right\}}{\Tilde{p}_{g_2}(\mathbf{X}_1,  g_1)\Tilde{p}_{g_1}(X_0)}\middle| X_0=x_0, D_1\right]\middle| X_0=x_0 \right]  \nonumber\\
    &= \Pr(D_1=g_1(V_0)|X_0=x_0) \E \left[ \frac{\left(Y - \Tilde{\mu}_{\mathbf{g}_2}(\mathbf{X}_1)    \right) \mathbbm{1}\left\{D_2=g_2(\mathbf{V}_1)\right\}}{\Tilde{p}_{g_2}(\mathbf{X}_1,  g_1)\Tilde{p}_{g_1}(X_0)}\middle| X_0=x_0, D_1=g_1(V_0)\right] \nonumber\\
    &= \frac{p_{g_1}(x_0)}{\Tilde{p}_{g_1}(x_0)} \E \left[ \frac{\left(Y - \Tilde{\mu}_{\mathbf{g}_2}(\mathbf{X}_1)    \right) \mathbbm{1}\left\{D_2=g_2(\mathbf{V}_1)\right\}}{\Tilde{p}_{g_2}(\mathbf{X}_1,  g_1)}\middle| X_0=x_0, D_1=g_1(V_0)\right] \nonumber\\
    &= \frac{p_{g_1}(x_0)}{\Tilde{p}_{g_1}(x_0)}  \E_{X_1} \left[\E \left[ \frac{\left(Y - \Tilde{\mu}_{\mathbf{g}_2}(\mathbf{X}_1)    \right) \mathbbm{1}\left\{D_2=g_2(\mathbf{V}_1)\right\}}{\Tilde{p}_{g_2}(\mathbf{X}_1,  g_1)} \middle| \mathbf{X}_1, D_1=g_1(V_0)\right]\middle|X_0=x_0, D_1=g_1(V_0)\right] \nonumber\\
    &= \frac{p_{g_1}(x_0)}{\Tilde{p}_{g_1}(x_0)} \E_{X_1} \left[\frac{p_{g_2}(\mathbf{X}_1, g_1)}{\Tilde{p}_{g_2}(\mathbf{X}_1,  g_1)} \left( \E \left[Y\middle| \mathbf{X}_1, \mathbf{D}_2=\mathbf{g}_2(\mathbf{V}_1)\right]  - \Tilde{\mu}_{\mathbf{g}_2}(\mathbf{X}_1)\right)\middle|X_0=x_0, D_1=g_1(V_0)\right] \nonumber\\
    &= \frac{p_{g_1}(x_0)}{\Tilde{p}_{g_1}(x_0)} \E_{X_1} \left[\frac{p_{g_2}(\mathbf{X}_1, g_1)}{\Tilde{p}_{g_2}(\mathbf{X}_1,  g_1)}  \left( \mu_{\mathbf{g}_2}(\mathbf{X}_1)  - \Tilde{\mu}_{\mathbf{g}_2}(\mathbf{X}_1) \right)\middle|X_0=x_0, D_1=g_1(V_0)\right] \nonumber 
\end{align}
and for the third term one can show
\begin{align} 
    \E& \left[ \frac{\left(\Tilde{\mu}_{\mathbf{g}_2}(\mathbf{X}_1)  -  \Tilde{\nu}_{\mathbf{g}_2}(X_0) \right) \mathbbm{1}\left\{D_1=g_1(V_0)\right\} }{\Tilde{p}_{g_1}(X_0)}\middle| X_0=x_0\right]\nonumber\\
    &= \E_{D_1} \left[ \E_{X_1} \left[ \frac{\left(\Tilde{\mu}_{\mathbf{g}_2}(\mathbf{X}_1)  - \Tilde{\nu}_{\mathbf{g}_2}(X_0) \right) \mathbbm{1}\left\{D_1=g_1(V_0)\right\} }{\Tilde{p}_{g_1}(X_0)}\middle| X_0=x_0, D_1\right]\middle| X_0=x_0 \right]  \nonumber\\
    &= \Pr(D_1=g_1(V_0)|X_0=x_0) \E_{X_1} \left[ \frac{\Tilde{\mu}_{\mathbf{g}_2}(\mathbf{X}_1)  -  \Tilde{\nu}_{\mathbf{g}_2}(X_0)}{\Tilde{p}_{g_1}(X_0)}\middle| X_0=x_0, D_1=g_1(V_0)\right] \nonumber\\
    &= \frac{p_{g_1}(x_0)}{\Tilde{p}_{g_1}(x_0)} \E_{X_1} \left[\Tilde{\mu}_{\mathbf{g}_2}(\mathbf{X}_1) -  \Tilde{\nu}_{\mathbf{g}_2}(x_0) \middle| X_0=x_0, D_1=g_1(V_0)\right]\text{.} \nonumber
\end{align}
Hence, the conditional-on-$X_0$ expected score can be rewritten as
\begin{align*}
    \E[\Theta_{\mathbf{g}_2}^{dy}(\mathbf{W}_{2})|X_0=x_0]
    &= \Tilde{\nu}_{\mathbf{g}_2}(x_0)\\
    &+ \frac{p_{g_1}(x_0)}{\Tilde{p}_{g_1}(x_0)} \E_{X_1} \left[\frac{p_{g_2}(\mathbf{X}_1, g_1)}{\Tilde{p}_{g_2}(\mathbf{X}_1,  g_1)}  \left( \mu_{\mathbf{g}_2}(\mathbf{X}_1)  - \Tilde{\mu}_{\mathbf{g}_2}(\mathbf{X}_1) \right)\middle|X_0=x_0, D_1=g_1(V_0)\right]\\
    &+ \frac{p_{g_1}(x_0)}{\Tilde{p}_{g_1}(x_0)} \E_{X_1} \left[\Tilde{\mu}_{\mathbf{g}_2}(\mathbf{X}_1) -  \Tilde{\nu}_{\mathbf{g}_2}(x_0) \middle| X_0=x_0, D_1=g_1(V_0)\right]\text{.}
\end{align*}
Plugging in the telescoping sums $\Tilde{\mu}_{\mathbf{g}_2}(\mathbf{X}_1) -  \Tilde{\nu}_{\mathbf{g}_2}(x_0) = \Tilde{\mu}_{\mathbf{g}_2}(\mathbf{X}_1) - \mu_{\mathbf{g}_2}(\mathbf{X}_1) + \mu_{\mathbf{g}_2}(\mathbf{X}_1) -  \Tilde{\nu}_{\mathbf{g}_2}(x_0)$ and $\Tilde{\nu}_{\mathbf{g}_2}(x_0) = \nu_{\mathbf{g}_2}(x_0) - (\nu_{\mathbf{g}_2}(x_0) -\Tilde{\nu}_{\mathbf{g}_2}(x_0))$ and rearranging one gets
\begin{align*}
    \E[\Theta_{\mathbf{g}_2}^{dy}(\mathbf{W}_{2})|X_0=x_0]
    &= \nu_{\mathbf{g}_2}(x_0) - \left(\nu_{\mathbf{g}_2}(x_0) -\Tilde{\nu}_{\mathbf{g}_2}(x_0)\right)\\
    &+ \frac{p_{g_1}(x_0)}{\Tilde{p}_{g_1}(x_0)} \E_{X_1} \left[\frac{p_{g_2}(\mathbf{X}_1, g_1)}{\Tilde{p}_{g_2}(\mathbf{X}_1,  g_1)}  \left( \mu_{\mathbf{g}_2}(\mathbf{X}_1)  - \Tilde{\mu}_{\mathbf{g}_2}(\mathbf{X}_1) \right)\middle|X_0=x_0, D_1=g_1(V_0)\right]\\
    &+ \frac{p_{g_1}(x_0)}{\Tilde{p}_{g_1}(x_0)} \E_{X_1} \left[\Tilde{\mu}_{\mathbf{g}_2}(\mathbf{X}_1) - \mu_{\mathbf{g}_2}(\mathbf{X}_1) \middle| X_0=x_0, D_1=g_1(V_0)\right]\\
    &+ \frac{p_{g_1}(x_0)}{\Tilde{p}_{g_1}(x_0)} \E_{X_1} \left[\mu_{\mathbf{g}_2}(\mathbf{X}_1) -  \Tilde{\nu}_{\mathbf{g}_2}(x_0) \middle| X_0=x_0, D_1=g_1(V_0)\right]\\
    &= \nu_{\mathbf{g}_2}(x_0) - \left(\nu_{\mathbf{g}_2}(x_0) -\Tilde{\nu}_{\mathbf{g}_2}(x_0)\right)\\
    &+ \frac{p_{g_1}(x_0)}{\Tilde{p}_{g_1}(x_0)} \E_{X_1} \left[\left(\frac{p_{g_2}(\mathbf{X}_1, g_1)}{\Tilde{p}_{g_2}(\mathbf{X}_1,  g_1)}-1\right)  \left( \mu_{\mathbf{g}_2}(\mathbf{X}_1)  - \Tilde{\mu}_{\mathbf{g}_2}(\mathbf{X}_1) \right)\middle|X_0=x_0, D_1=g_1(V_0)\right]\\
    &+ \frac{p_{g_1}(x_0)}{\Tilde{p}_{g_1}(x_0)} \left( {\nu}_{\mathbf{g}_2}(x_0) -  \Tilde{\nu}_{\mathbf{g}_2}(x_0)\right)\\
    &= \nu_{\mathbf{g}_2}(x_0)\\
    &+ \frac{p_{g_1}(x_0)}{\Tilde{p}_{g_1}(x_0)} \E_{X_1} \left[\left(\frac{p_{g_2}(\mathbf{X}_1, g_1)}{\Tilde{p}_{g_2}(\mathbf{X}_1,  g_1)}-1\right)  \left( \mu_{\mathbf{g}_2}(\mathbf{X}_1)  - \Tilde{\mu}_{\mathbf{g}_2}(\mathbf{X}_1) \right)\middle|X_0=x_0, D_1=g_1(V_0)\right]\\
    &+ \left(\frac{p_{g_1}(x_0)}{\Tilde{p}_{g_1}(x_0)}-1\right) \left( {\nu}_{\mathbf{g}_2}(x_0) -  \Tilde{\nu}_{\mathbf{g}_2}(x_0)\right)\text{.}
\end{align*}

From this expression one obtains $\E[\Theta_{\mathbf{g}_2}^{dy}(\mathbf{W}_{2})|X_0=x_0]=\nu_{\mathbf{g}_2}(x_0)$ if
\begin{itemize}
    \item either $\Tilde{p}_{g_2}(\mathbf{X}_1,  g_1)=p_{g_2}(\mathbf{X}_1,  g_1)$ or $\Tilde{\mu}_{\mathbf{g}_2}(\mathbf{X}_1)=\mu_{\mathbf{g}_2}(\mathbf{X}_1)$ such that the second term equals zero
\end{itemize}
and
\begin{itemize}
    \item either $\Tilde{p}_{g_1}(X_0)=p_{g_1}(X_0)$ or $\Tilde{\nu}_{\mathbf{g}_2}(X_0)=\nu_{\mathbf{g}_2}(x_0)$ such that the third term equals zero.
\end{itemize}

\subsection{Neyman orthogonality} \label{sec:app_neyman}
The extension of the sequential DML estimator from static to dynamic policies is established by verifying Neyman orthogonality of $\E\left[ \Theta^{dy}_{\mathbf{g}_2}(\mathbf{W}_2, \eta^{dy}) - \theta^{\mathbf{g}_2} \right]$, where the policy $\mathbf{g}_2(\mathbf{V}_1)$ depends on time-varying covariates. The proof follows the structure outlined in \citet{knaus2022double} and \citet{bodory2022evaluating}. Assume the regularity conditions on the nuisance functions as stated in \citet{bodory2022evaluating} are satisfied. Recall the following definitions
\begin{align*}
    \theta^{\mathbf{g}_2} &= \E[Y^{\mathbf{g}_2}] = \E [ \Theta^{dy}_{\mathbf{g}_2}(\mathbf{W}_2, \eta^{dy})]\\
    \Theta^{dy}_{\mathbf{g}_2}(\mathbf{W}_2, \eta^{dy}) &= \nu_{\mathbf{g}_2}(X_0) + \frac{\left(\mu_{\mathbf{g}_2}(\mathbf{X}_1)  -  \nu_{\mathbf{g}_2}(X_0)  \right) \mathbbm{1}\left\{D_1=g_1(V_0)\right\} }{p_{g_1}(X_0)} +
    \frac{\left(Y - \mu_{\mathbf{g}_2}(\mathbf{X}_1) \right) \mathbbm{1}\left\{\mathbf{D}_2=\mathbf{g}_2(\mathbf{V}_1)\right\}}{p_{g_2}(\mathbf{X}_1, g_1)p_{g_1}(X_0)}\\
    \eta^{dy} &= (\nu_{\mathbf{g}_2}(X_0), \mu_{\mathbf{g}_2}(\mathbf{X}_1), p_{d_2}(\mathbf{X}_1, g_1), p_{g_1}(X_0))
\end{align*}
First introduce perturbations $\eta^{dy} + r(\Tilde \eta^{dy} - \eta^{dy})$ around the true nuisance parameters such that
\begin{align*}
    \E [ \Theta^{dy}_{\mathbf{g}_2}(\mathbf{W}_2, \eta^{dy} + r(\Tilde \eta^{dy} - \eta^{dy}))- \theta^{\mathbf{g}_2}] &= \E[A + B + C + D - \theta^{\mathbf{g}_2}]
\end{align*}
with
\begin{align*}
    A &= \nu_{\mathbf{g}_2}(X_0) + r\left(\Tilde\nu_{\mathbf{g}_2}(X_0) - \nu_{\mathbf{g}_2}(X_0)\right)\\
    B &= \frac{\big(\mu_{\mathbf{g}_2}(\mathbf{X}_1) + r(\Tilde\mu_{\mathbf{g}_2}(\mathbf{X}_1) - \mu_{\mathbf{g}_2}(\mathbf{X}_1))  \big) \mathbbm{1}\left\{D_1=g_1(V_0)\right\} }{p_{g_1}(X_0) + r(\Tilde p_{g_1}(X_0) - p_{g_1}(X_0))}\\
    C &= -\frac{\big(\nu_{\mathbf{g}_2}(X_0) + r(\Tilde\nu_{\mathbf{g}_2}(X_0) - \nu_{\mathbf{g}_2}(X_0))  \big) \mathbbm{1}\left\{D_1=g_1(V_0)\right\} }{p_{g_1}(X_0) + r(\Tilde p_{g_1}(X_0) - p_{g_1}(X_0))}\\
    D &= \frac{\big(Y - \mu_{\mathbf{g}_2}(\mathbf{X}_1) - r(\Tilde\mu_{\mathbf{g}_2}(\mathbf{X}_1) - \mu_{\mathbf{g}_2}(\mathbf{X}_1)) \big) \mathbbm{1}\left\{\mathbf{D}_2=\mathbf{g}_2(\mathbf{V}_1)\right\}}{\big(p_{g_2}(\mathbf{X}_1, g_1) + r(\Tilde p_{g_2}(\mathbf{X}_1, g_1) - p_{g_2}(\mathbf{X}_1, g_1))\big)\big(p_{g_1}(X_0) + r(\Tilde p_{g_1}(X_0) - p_{g_1}(X_0))\big)}
\end{align*}
Neyman orthogonality holds when the Gâteaux derivative of the score function with respect to the nuisance parameters is zero in expectation at the true values of those parameters, i.e.
\begin{align*}
    \left. \frac{\partial}{\partial r} \E \left[ \Theta^{dy}_{\mathbf{g}_2}(\mathbf{W}_2, \eta^{dy} + r(\Tilde \eta^{dy} - \eta^{dy}))- \theta^{\mathbf{g}_2}\right]\right|_{r=0} = 0\text{.}
\end{align*}
The expression is evaluated piecewise. Evaluating $A$ gives
\begin{align*}
    \left. \frac{\partial \E[A]}{\partial r} \right|_{r=0} &= \E\left[ \Tilde\nu_{\mathbf{g}_2}(X_0) - \nu_{\mathbf{g}_2}(X_0)\right]\text{.}
\end{align*}
Evaluating $B$ gives
\begin{align*}
    \left. \frac{\partial \E[B]}{\partial r} \right|_{r=0} &= \E\left[ \frac{\big(\Tilde\mu_{\mathbf{g}_2}(\mathbf{X}_1) - \mu_{\mathbf{g}_2}(\mathbf{X}_1)  \big) \mathbbm{1}\left\{D_1=g_1(V_0)\right\} }{p_{g_1}(X_0)}\right]\\
    &+ \E\left[ \frac{\mu_{\mathbf{g}_2}(\mathbf{X}_1) 
    \mathbbm{1}\left\{D_1=g_1(V_0)\right\} }{p_{g_1}(X_0)} \frac{\Tilde p_{g_1}(X_0) - p_{g_1}(X_0) }{p_{g_1}(X_0)}\right]\\
    &= \E_{X_0}\left[ \frac{p_{g_1}(X_0)}{p_{g_1}(X_0)}\E\left[\Tilde\mu_{\mathbf{g}_2}(\mathbf{X}_1) - \mu_{\mathbf{g}_2}(\mathbf{X}_1)\middle| X_0, D_1=g_1(V_0)\right]\right]\\
    &+ \E_{X_0}\left[ \frac{\Tilde p_{g_1}(X_0) - p_{g_1}(X_0) }{p_{g_1}(X_0)} \E\left[\mu_{\mathbf{g}_2}(\mathbf{X}_1)  \middle| X_0, D_1=g_1(V_0)\right]\right]\\
    &= \E_{X_0}\left[\E\left[\Tilde\mu_{\mathbf{g}_2}(\mathbf{X}_1) - \mu_{\mathbf{g}_2}(\mathbf{X}_1)\middle| X_0, D_1=g_1(V_0)\right]\right]\\
    &+ \E_{X_0}\left[ \frac{\Tilde p_{g_1}(X_0) - p_{g_1}(X_0) }{p_{g_1}(X_0)} \nu_{\mathbf{g}_2}(X_0)\right]\text{.}
\end{align*}
Evaluating $C$ gives
\begin{align*}
    \left. \frac{\partial \E[C]}{\partial r} \right|_{r=0} &= -\E\left[ \frac{\big(\Tilde\nu_{\mathbf{g}_2}(X_0) - \nu_{\mathbf{g}_2}(X_0)  \big) \mathbbm{1}\left\{D_1=g_1(V_0)\right\} }{p_{g_1}(X_0)}\right]\\
    &- \E\left[ \frac{\nu_{\mathbf{g}_2}(X_0) \mathbbm{1}\left\{D_1=g_1(V_0)\right\} }{p_{g_1}(X_0)} \frac{\Tilde p_{g_1}(X_0) - p_{g_1}(X_0) }{p_{g_1}(X_0)}\right]\\
    &= -\E_{X_0}\left[\frac{\Tilde\nu_{\mathbf{g}_2}(X_0) - \nu_{\mathbf{g}_2}(X_0) }{p_{g_1}(X_0)}\E\left[  \mathbbm{1}\left\{D_1=g_1(V_0)\right\} \middle| X_0\right]\right]\\
    &- \E_{X_0}\left[ \frac{\nu_{\mathbf{g}_2}(X_0)  }{p_{g_1}(X_0)}\frac{\Tilde p_{g_1}(X_0) - p_{g_1}(X_0) }{p_{g_1}(X_0)}\E\left[\mathbbm{1}\left\{D_1=g_1(V_0)\right\} \middle| X_0\right]\right]\\
    &= - \E\left[\Tilde\nu_{\mathbf{g}_2}(X_0) - \nu_{\mathbf{g}_2}(X_0)\right] \\
    &- \E\left[ \nu_{\mathbf{g}_2}(X_0)  \frac{\Tilde p_{g_1}(X_0) - p_{g_1}(X_0) }{p_{g_1}(X_0)}\right]\text{.}
\end{align*}
Evaluating $D$ gives
\begin{align}
    \left. \frac{\partial \E[D]}{\partial r} \right|_{r=0} 
    &= -\E\left[\frac{\big(\Tilde\mu_{\mathbf{g}_2}(\mathbf{X}_1) - \mu_{\mathbf{g}_2}(\mathbf{X}_1) \big) \mathbbm{1}\left\{\mathbf{D}_2=\mathbf{g}_2(\mathbf{V}_1)\right\}}{p_{g_2}(\mathbf{X}_1, g_1) p_{g_1}(X_0)}\right]\label{eq:D1}\\
    &-  \E\left[\frac{\big(Y - \mu_{\mathbf{g}_2}(\mathbf{X}_1) \big) \mathbbm{1}\left\{\mathbf{D}_2=\mathbf{g}_2(\mathbf{V}_1)\right\}}{p_{g_2}(\mathbf{X}_1, g_1) p_{g_1}(X_0)} \frac{ \Tilde p_{g_2}(\mathbf{X}_1, g_1) - p_{g_2}(\mathbf{X}_1, g_1)}{p_{g_2}(\mathbf{X}_1, g_1)}\right]\label{eq:D2}\\
    &-  \E\left[\frac{\big(Y - \mu_{\mathbf{g}_2}(\mathbf{X}_1) \big) \mathbbm{1}\left\{\mathbf{D}_2=\mathbf{g}_2(\mathbf{V}_1)\right\}}{p_{g_2}(\mathbf{X}_1, g_1) p_{g_1}(X_0)} \frac{\Tilde p_{g_1}(X_0) - p_{g_1}(X_0)}{p_{g_1}(X_0)}\right]\label{eq:D3}\\
    &= -\E_{X_0}\left[\E\left[\Tilde\mu_{\mathbf{g}_2}(\mathbf{X}_1) - \mu_{\mathbf{g}_2}(\mathbf{X}_1) \middle| X_0, D_1=g_1(V_0) \right]\right]\text{.}\nonumber
\end{align}
To see the result for $D$, note that (\ref{eq:D1}) can be rewritten as
\begin{align*}
    &-\E\left[\frac{\big(\Tilde\mu_{\mathbf{g}_2}(\mathbf{X}_1) - \mu_{\mathbf{g}_2}(\mathbf{X}_1) \big) \mathbbm{1}\left\{\mathbf{D}_2=\mathbf{g}_2(\mathbf{V}_1)\right\}}{p_{g_2}(\mathbf{X}_1, g_1) p_{g_1}(X_0)}\right]\\
    &= -\E_{X_0}\left[\E\left[\frac{\big(\Tilde\mu_{\mathbf{g}_2}(\mathbf{X}_1) - \mu_{\mathbf{g}_2}(\mathbf{X}_1) \big) \mathbbm{1}\left\{\mathbf{D}_2=\mathbf{g}_2(\mathbf{V}_1)\right\}}{p_{g_2}(\mathbf{X}_1, g_1) p_{g_1}(X_0)}\middle| X_0 \right]\right]\\
    &= -\E_{X_0}\left[\E\left[\frac{\big(\Tilde\mu_{\mathbf{g}_2}(\mathbf{X}_1) - \mu_{\mathbf{g}_2}(\mathbf{X}_1) \big) \mathbbm{1}\left\{{D}_2={g}_2(\mathbf{V}_1)\right\}}{p_{g_2}(\mathbf{X}_1, g_1)}\middle| X_0, D_1=g_1(V_0) \right]\right]\\
    &= -\E_{X_0}\left[\E_{X_1}\left[\E\left[\frac{\big(\Tilde\mu_{\mathbf{g}_2}(\mathbf{X}_1) - \mu_{\mathbf{g}_2}(\mathbf{X}_1) \big) \mathbbm{1}\left\{{D}_2={g}_2(\mathbf{V}_1)\right\}}{p_{g_2}(\mathbf{X}_1, g_1)}\middle| \mathbf{X}_1, {D}_1={g}_1({V}_0) \right]\middle| X_0, D_1=g_1(V_0) \right]\right]\\
    &= -\E_{X_0}\left[\E\left[\Tilde\mu_{\mathbf{g}_2}(\mathbf{X}_1) - \mu_{\mathbf{g}_2}(\mathbf{X}_1) \middle| X_0, D_1=g_1(V_0) \right]\right]\text{,}
    \intertext{that (\ref{eq:D2}) equals zero}
    &\E\left[\frac{\big(Y - \mu_{\mathbf{g}_2}(\mathbf{X}_1) \big) \mathbbm{1}\left\{\mathbf{D}_2=\mathbf{g}_2(\mathbf{V}_1)\right\}}{p_{g_2}(\mathbf{X}_1, g_1) p_{g_1}(X_0)} \frac{ \Tilde p_{g_2}(\mathbf{X}_1, g_1) - p_{g_2}(\mathbf{X}_1, g_1)}{p_{g_2}(\mathbf{X}_1, g_1)}\right]\\
    &= \E_{X_0}\left[\E\left[\frac{\big(Y - \mu_{\mathbf{g}_2}(\mathbf{X}_1) \big) \mathbbm{1}\left\{\mathbf{D}_2=\mathbf{g}_2(\mathbf{V}_1)\right\}}{p_{g_2}(\mathbf{X}_1, g_1) p_{g_1}(X_0)} \frac{ \Tilde p_{g_2}(\mathbf{X}_1, g_1) - p_{g_2}(\mathbf{X}_1, g_1)}{p_{g_2}(\mathbf{X}_1, g_1)}\middle| X_0\right]\right]\\
    &= \E_{X_0}\left[\E\left[\frac{\big(Y - \mu_{\mathbf{g}_2}(\mathbf{X}_1) \big) \mathbbm{1}\left\{{D}_2={g}_2(\mathbf{V}_1)\right\}}{p_{g_2}(\mathbf{X}_1, g_1)} \frac{ \Tilde p_{g_2}(\mathbf{X}_1, g_1) - p_{g_2}(\mathbf{X}_1, g_1)}{p_{g_2}(\mathbf{X}_1, g_1)}\middle| X_0, D_1=g_1(V_0)\right]\right]\\
    &= \E_{X_0}\left[\E_{X_1}\left[\frac{ \Tilde p_{g_2}(\mathbf{X}_1, g_1) - p_{g_2}(\mathbf{X}_1, g_1)}{p_{g_2}(\mathbf{X}_1, g_1)} \Big(\E\left[Y 
    \middle| \mathbf{X_1}, \mathbf{D}_2=\mathbf{g}_2(\mathbf{V}_1)\right]- \mu_{\mathbf{g}_2}(\mathbf{X}_1)\Big) \middle| X_0, D_1=g_1(V_0)\right]\right]\\
    &= 0\text{,}
    \intertext{and that (\ref{eq:D3}) equals zero as well}
    &\E\left[\frac{\big(Y - \mu_{\mathbf{g}_2}(\mathbf{X}_1) \big) \mathbbm{1}\left\{\mathbf{D}_2=\mathbf{g}_2(\mathbf{V}_1)\right\}}{p_{g_2}(\mathbf{X}_1, g_1) p_{g_1}(X_0)} \frac{\Tilde p_{g_1}(X_0) - p_{g_1}(X_0)}{p_{g_1}(X_0)}\right]\\
    &= \E_{X_0}\left[\frac{\Tilde p_{g_1}(X_0) - p_{g_1}(X_0)}{p_{g_1}(X_0)}\E\left[\frac{\big(Y - \mu_{\mathbf{g}_2}(\mathbf{X}_1) \big) \mathbbm{1}\left\{\mathbf{D}_2=\mathbf{g}_2(\mathbf{V}_1)\right\}}{p_{g_2}(\mathbf{X}_1, g_1) p_{g_1}(X_0)}\middle| X_0\right]\right]\\
    &= 0 \quad \text{(see Equation \ref{eq:DML_ymu}).}
\end{align*}
Putting everything together yields
\begin{align*}
    \left. \frac{\partial}{\partial r} \E \left[ \Theta^{dy}_{\mathbf{g}_2}(\mathbf{W}_2, \eta^{dy} + r(\Tilde \eta^{dy} - \eta^{dy}))- \theta^{\mathbf{g}_2}\right]\right|_{r=0}
    &\overset{A}{=} \E\left[ \Tilde\nu_{\mathbf{g}_2}(X_0) - \nu_{\mathbf{g}_2}(X_0)\right]\\
    &\overset{B}{+} \E_{X_0}\left[\E\left[\Tilde\mu_{\mathbf{g}_2}(\mathbf{X}_1) - \mu_{\mathbf{g}_2}(\mathbf{X}_1)\middle| X_0, D_1=g_1(V_0)\right]\right]\\
    &\overset{B}{+}  \E_{X_0}\left[ \frac{\Tilde p_{g_1}(X_0) - p_{g_1}(X_0) }{p_{g_1}(X_0)} \nu_{\mathbf{g}_2}(X_0)\right]\\
    &\overset{C}{-}  \E\left[ \Tilde\nu_{\mathbf{g}_2}(X_0) - \nu_{\mathbf{g}_2}(X_0)\right]\\
    &\overset{C}{-}  \E_{X_0}\left[ \frac{\Tilde p_{g_1}(X_0) - p_{g_1}(X_0) }{p_{g_1}(X_0)} \nu_{\mathbf{g}_2}(X_0)\right]\\
    &\overset{D}{-} \E_{X_0}\left[\E\left[\Tilde\mu_{\mathbf{g}_2}(\mathbf{X}_1) - \mu_{\mathbf{g}_2}(\mathbf{X}_1) \middle| X_0, D_1=g_1(V_0) \right]\right]\\
    &=0
\end{align*}

\subsection{GATE-ATE Variance} \label{ap:gate_ate_var}
Let $k \in \left\{1, ..., K\right\}$ denote one of $K$ discrete outcomes of the variables $Z_0$ and $N_k = \sum_{i=1}^N \mathbbm{1}\left\{Z_{0i}=k\right\}$ the number of observations corresponding to the realization $k$. Denote the estimate of the ATE as $\hat\theta$ and of the GATE in group $k$ as $\hat\theta(k)$. Then, compute the variance of the difference $\hat\theta(k) - \hat\theta$ as
\begin{align*}
\Var(\hat\theta(k) - \hat\theta) &= \Var(\hat\theta(k)) + \Var(\hat\theta) - 2\mathrm{Cov}(\hat\theta(k), \hat\theta)\\
    &= \Var(\hat\theta(k)) + \Var(\hat\theta) - 2N_k / N \Var(\hat\theta(k)) 
\end{align*}
since
\begin{align*}
   \mathrm{Cov}(\hat\theta(k), \hat\theta)    &=  \mathrm{Cov}\left(\hat\theta(k), \sum_{j=1}^K\frac{N_j}{N}\hat\theta(j)\right)\\
   &= \sum_{j=1}^K \frac{N_j}{N}\mathrm{Cov}(\hat\theta(k), \hat\theta(j))\\
                        &= \frac{N_k}{N} \mathrm{Cov}(\hat\theta(k), \hat\theta(k))\\
                        &= N_k / N \Var(\hat\theta(k))\text{.}
\end{align*}
The first equality follows from the definition of the ATE, the second equality follows from the properties of the covariance, the third equality follows from independence between GATEs for different groups (because observations are \textit{iid} and groups are mutually exclusive) and the last equality follows from the definition of the variance.

\newpage
\section{Comparison of DML Algorithms}\label{sec:appendix_algo}
\begin{algorithm}[ht]
    \centering
    \captionsetup{font=small, width=\linewidth}
    \caption{Overview of the DML algorithms under dynamic confounding}
    \label{alg:overview}
    \footnotesize
    \begin{tabularx}{\textwidth} {l  X  X} 
 & \textbf{(1) \citet{bodory2022evaluating}} & \textbf{(2) \citet{bradic2021high}}\\\hline
 \multicolumn{2}{>{\hsize=\dimexpr1\hsize+1\tabcolsep+\arrayrulewidth\relax}l}{\textit{Sample splitting:}}\\
 \scriptsize 1: & \multicolumn{2}{>{\hsize=\dimexpr2\hsize+2\tabcolsep+\arrayrulewidth\relax}X}{Randomly split $\mathcal{W}$ into $K$ equally sized subsamples $\mathcal{W}_k$. Define $\mathcal{W}_{-k} := \mathcal{W}\setminus\mathcal{W}_{k}$ and further split $\mathcal{W}_{-k}$ into two equal-sized sets $\mathcal{W}_{-k, 1}$ and $\mathcal{W}_{-k, 2}$. For $\mathbf{a}_2 \in \{\mathbf{g}_2, \mathbf{g}'_2\}$, define $\mathcal{W}_{-k, a_1}$ as the subsample $\mathcal{W}_{-k}$ with $D_1=a_1(V_0)$ and $\mathcal{W}_{-k, \mathbf{a}_2}$ as the subsample $\mathcal{W}_{-k}$ with $\mathbf{D}_2=({a}_1(V_0), a_2(\mathbf{V}_1))$. Similarly define $\mathcal{W}_{-k, 1, a_1}$, $\mathcal{W}_{-k, 2, a_1}$ $\mathcal{W}_{-k, 1, \mathbf{a}_2}$ and $\mathcal{W}_{-k, 2, \mathbf{a}_2}$, respectively.}\\\hline
 \multicolumn{2}{>{\hsize=\dimexpr1\hsize+1\tabcolsep+\arrayrulewidth\relax}l}{\textit{Estimation of nuisance functions:}}\\
 \scriptsize 2:  & \textbf{for} $\mathbf{a}_2 \in \{\mathbf{g}_2, \mathbf{g}'_2\}$ \textbf{do}
 & \textbf{for} $\mathbf{a}_2 \in \{\mathbf{g}_2, \mathbf{g}'_2\}$ \textbf{do} \\
\scriptsize 3:  & \leftskip 5pt \textbf{for} $k=1,..., K$ \textbf{do}  & \leftskip 5pt\textbf{for} $k=1,..., K$ \textbf{do}  \\
\scriptsize 4: & \leftskip 10pt Learn $\hat{p}_{a_1}$ on $\mathcal{W}_{-k}$ & \leftskip 10pt Learn $\hat{p}_{a_1}$ on $\mathcal{W}_{-k}$ \\
\scriptsize 5: & \leftskip 10pt Learn $\hat{p}_{a_2}$ on $\mathcal{W}_{-k, a_1}$  & \leftskip 10pt Learn $\hat{p}_{a_2}$ on $\mathcal{W}_{-k, a_1}$  \\
\scriptsize 6: & \leftskip 10pt Learn $\hat{\mu}_{\mathbf{a}_2}$ on $\mathcal{W}_{-k, \textcolor{red}{1}, \mathbf{a}_2}$ & \leftskip 10pt Learn $\hat{\mu}_{\mathbf{a}_2}$ on $\mathcal{W}_{-k, \mathbf{a}_2}$ \\
\scriptsize 7: & \leftskip 10pt Predict $\hat{\mu}_{\mathbf{a}_2}$ on $\mathcal{W}_{-k, \textcolor{red}{2}, \mathbf{a}_2}$, learn $\hat{\nu}_{\mathbf{a}_2}$ on $\mathcal{W}_{-k, \textcolor{red}{2}, \mathbf{a}_2}$ using (\ref{eq:nu_bodory}) & 
\leftskip 10pt  Learn $\hat{\mu}_{\mathbf{a}_2}$ on $\mathcal{W}_{-k, \textcolor{red}{1}, \mathbf{a}_2}$, learn $\hat{p}_{a_2}$ on $\mathcal{W}_{-k, \textcolor{red}{1}, a_1}$, predict $\hat{\mu}_{\mathbf{a}_2}$ and $\hat{p}_{a_2}$ on $\mathcal{W}_{-k, \textcolor{red}{2}, a_1}$, learn $\hat{\nu}_{\mathbf{a}_2}^{\textcolor{red}{1}}$ on $\mathcal{W}_{-k, \textcolor{red}{2}, a_1}$ using (\ref{eq:nu_bradic})\\
\scriptsize 8: & & \leftskip 10pt  Learn $\hat{\mu}_{\mathbf{a}_2}$ on $\mathcal{W}_{-k, \textcolor{red}{2}, \mathbf{a}_2}$, learn $\hat{p}_{a_2}$ on $\mathcal{W}_{-k, \textcolor{red}{2}, a_1}$, predict $\hat{\mu}_{\mathbf{a}_2}$ and $\hat{p}_{a_2}$ on $\mathcal{W}_{-k, \textcolor{red}{1}, a_1}$, learn $\hat{\nu}_{\mathbf{a}_2}^{\textcolor{red}{2}}$ on $\mathcal{W}_{-k, \textcolor{red}{1}, a_1}$ using (\ref{eq:nu_bradic}) \\
\scriptsize 9: &  \leftskip 10pt Predict $\hat{p}_{a_1}$, $\hat{p}_{a_2}$, $\hat{\mu}_{\mathbf{a}_2}$ and $\hat{\nu}_{\mathbf{a}_2}$ on $\mathcal{W}_k$ &  \leftskip 10pt Predict $\hat{p}_{a_1}$, $\hat{p}_{a_2}$, $\hat{\mu}_{\mathbf{a}_2}$, $\hat{\nu}_{\mathbf{a}_2}^{\textcolor{red}{1}}$ and $\hat{\nu}_{\mathbf{a}_2}^{\textcolor{red}{2}}$ on $\mathcal{W}_k$, obtain $\hat{\nu}_{\mathbf{a}_2}$ by computing average of $\hat{\nu}_{\mathbf{a}_2}^{\textcolor{red}{1}}$ and $\hat{\nu}_{\mathbf{a}_2}^{\textcolor{red}{2}}$ \\
\scriptsize 10: & \leftskip 5pt \textbf{end for} & \leftskip 5pt \textbf{end for} \\
\scriptsize 11: & \leftskip 5pt Compute $\hat{\Theta}^{dy}_{\mathbf{a}_2, i}$ using (\ref{eq:DML_emp}) & \leftskip 5pt Compute $\hat{\Theta}^{dy}_{\mathbf{a}_2, i}$ using (\ref{eq:DML_emp})\\
\scriptsize 12: & \textbf{end for} & \textbf{end for} \\\hline
 \multicolumn{2}{>{\hsize=\dimexpr1\hsize+1\tabcolsep+\arrayrulewidth\relax}l}{\textit{Computation of effects and standard errors:}}\\
\scriptsize 13: & \multicolumn{2}{>{\hsize=\dimexpr2\hsize+2\tabcolsep+\arrayrulewidth\relax}X}{Compute $\hat{\theta}^{\mathbf{g}_2} =  \frac{1}{N}\sum_{i=1}^{N} \hat{\Theta}^{dy}_{\mathbf{g}_2, i}$ and $\hat{\sigma}^{\mathbf{g}_2} =  \sqrt{ \frac{1}{N}\sum_{i=1}^{N} \left( \hat{\Theta}^{dy}_{\mathbf{g}_2, i} -\hat{\theta}^{\mathbf{g}_2}\right)^2}$}\\
\scriptsize 14: & \multicolumn{2}{>{\hsize=\dimexpr2\hsize+2\tabcolsep+\arrayrulewidth\relax}X}{Compute $\hat{\tau}^{\mathbf{g}_2, \mathbf{g}_2'} =  \frac{1}{N}\sum_{i=1}^{N} \hat{\Theta}^{dy}_{\mathbf{g}_2, i} - \hat{\Theta}^{dy}_{\mathbf{g}'_2, i}$ and $\hat{\sigma}^{\mathbf{g}_2, \mathbf{g}_2'} =  \sqrt{ \frac{1}{N}\sum_{i=1}^{N} \left( \hat{\Theta}^{dy}_{\mathbf{g}_2, i} - \hat{\Theta}^{dy}_{\mathbf{g}'_2, i}-\hat{\tau}^{\mathbf{g}_2, \mathbf{g}_2'}\right)^2}$}\\
\scriptsize 15: & \multicolumn{2}{>{\hsize=\dimexpr2\hsize+2\tabcolsep+\arrayrulewidth\relax}X}{Let $N_{z_0} = \sum_{i=1}^N \mathbbm{1} \{Z_{0i}=z_0\}$. Compute $\hat{\tau}^{\mathbf{g}_2, \mathbf{g}_2'}(z_0) =  \frac{1}{N_{z_0}} \sum_{i=1}^N  \mathbbm{1} \{Z_{0i}=z_0\} \left(\hat{\Theta}^{dy}_{\mathbf{g}_2, i} - \hat{\Theta}^{dy}_{\mathbf{g}'_2, i}\right)$ and $\hat{\sigma}^{\mathbf{g}_2, \mathbf{g}_2'}(z_0) =  \sqrt{\frac{1}{N_{z_0}} \sum_{i=1}^N  \mathbbm{1} \{Z_{0i}=z_0\} \left(\hat{\Theta}^{dy}_{\mathbf{g}_2, i} - \hat{\Theta}^{dy}_{\mathbf{g}'_2, i} - \hat{\tau}^{\mathbf{g}_2, \mathbf{g}_2'}(z_0)\right)^2}$} \\
\end{tabularx}
\end{algorithm}

\newpage
\section{Descriptive Statistics}\label{sec:appendix_descriptives}
\subsection{Covariate and Outcome Means by Program sequence}\label{sec:appendix_descriptives_balance}
To assess covariate and outcome differences between treatment groups standardized differences
\begin{align*}
    \Delta=\frac{\left|\bar{X}_{\mathbf{d}_2}-\bar{X}_{\mathbf{d}'_2}\right|}{\sqrt{1 / 2\left(\operatorname{Var}({X}_{\mathbf{d}_2})+\operatorname{Var}({X}_{\mathbf{d}'_2})\right)}} \cdot 100
\end{align*}
are computed, where $\bar{X}_{\mathbf{d}_2}$ and $\operatorname{Var}({X}_{\mathbf{d}_2})$ indicate the sample mean and variance of variable $X$ in the subgroup with $\mathbf{D}_2 = {\mathbf{d}_2}$. As the standardized difference is independent of sample size it is preferred over a $t$-test to compare the balance of baseline covariates across treatment groups. Imbalance is typically defined as an absolute value greater than 20 \citep{imbens2015causal, yang2012unified}.

\newpage
\newgeometry{left=2cm,bottom=2cm,top=2cm,right=2cm} 

\begin{landscape}
\setlength{\tabcolsep}{2pt} 
{\fontsize{6pt}{6}\selectfont
\begin{longtable}{l|rrrrr|rrrrr|rrrrr|rrrrr}
\caption{Pre-treatment covariate means and standardized differences (vs. first program - NP) in trimmed sample. Standardized difference $\geq20$ marked in violet.} \label{tab:balance_x0_2nd} \\*
\toprule
\textbf{} & \multicolumn{5}{c|}{\textbf{First program: JA}} & \multicolumn{5}{c|}{\textbf{First program: TC}} & \multicolumn{5}{c|}{\textbf{First program: EP}} & \multicolumn{5}{c}{\textbf{First program: WS}} \\*
\textbf{Second program:} & \textbf{NP} & \textbf{JA} & \textbf{TC} & \textbf{EP} & \textbf{WS} & \textbf{NP} & \textbf{JA} & \textbf{TC} & \textbf{EP} & \textbf{WS} & \textbf{NP} & \textbf{JA} & \textbf{TC} & \textbf{EP} & \textbf{WS} & \textbf{NP} & \textbf{JA} & \textbf{TC} & \textbf{EP} & \textbf{WS} \\*
\textbf{Number of observations in sequence:} & \textbf{26,109} & \textbf{4,759} & \textbf{3,780} & \textbf{3,785} & \textbf{4,571} & \textbf{9,840} & \textbf{1,073} & \textbf{9,865} & \textbf{1,712} & \textbf{2,370} & \textbf{6,084} & \textbf{107} & \textbf{396} & \textbf{11,274} & \textbf{1,757} & \textbf{34,485} & \textbf{2,141} & \textbf{2,057} & \textbf{2,010} & \textbf{40,279} \\*
\midrule
\endfirsthead
\caption[]{Pre-treatment covariate means and standardized differences (vs. first program - NP) in trimmed sample. Standardized difference $\geq20$ marked in violet.} \\*
\toprule
\textbf{} & \multicolumn{5}{c|}{\textbf{First program: JA}} & \multicolumn{5}{c|}{\textbf{First program: TC}} & \multicolumn{5}{c|}{\textbf{First program: EP}} & \multicolumn{5}{c}{\textbf{First program: WS}} \\*
\textbf{Second program:} & \textbf{NP} & \textbf{JA} & \textbf{TC} & \textbf{EP} & \textbf{WS} & \textbf{NP} & \textbf{JA} & \textbf{TC} & \textbf{EP} & \textbf{WS} & \textbf{NP} & \textbf{JA} & \textbf{TC} & \textbf{EP} & \textbf{WS} & \textbf{NP} & \textbf{JA} & \textbf{TC} & \textbf{EP} & \textbf{WS} \\*
\textbf{Number of observations in sequence:} & \textbf{26,109} & \textbf{4,759} & \textbf{3,780} & \textbf{3,785} & \textbf{4,571} & \textbf{9,840} & \textbf{1,073} & \textbf{9,865} & \textbf{1,712} & \textbf{2,370} & \textbf{6,084} & \textbf{107} & \textbf{396} & \textbf{11,274} & \textbf{1,757} & \textbf{34,485} & \textbf{2,141} & \textbf{2,057} & \textbf{2,010} & \textbf{40,279} \\*
\midrule
\endhead
\midrule
\multicolumn{21}{r}{Continued on next page} \\*
\midrule
\endfoot
\bottomrule
\endlastfoot
\bfseries Avg add. income in months ue 1y before ref. point & \bfseries 17.22 & \bfseries 15.67 & \bfseries 16.37 & \bfseries 17.96 & \bfseries 31.35 & \bfseries 24.42 & \bfseries 15.83 & \bfseries 21.32 & \bfseries 24.47 & \bfseries 50.21 & \bfseries 38.34 & \bfseries 13.44 & \bfseries 30.63 & \bfseries 24.46 & \bfseries 51.56 & \bfseries 99.90 & \bfseries 59.48 & \bfseries 70.63 & \bfseries 73.17 & \bfseries 114.71 \\*
\bfseries  & \bfseries  & \bfseries (1.02) & \bfseries (0.53) & \bfseries (0.44) & \bfseries (7.48) & \bfseries  & \bfseries (5.42) & \bfseries (1.68) & \bfseries (0.03) & \bfseries (10.94) & \bfseries  & \bfseries (13.74) & \bfseries (3.39) & \bfseries (6.60) & \bfseries (5.23) & \bfseries  & \bfseries (10.00) & \bfseries (7.21) & \bfseries (6.72) & \bfseries (3.17) \\
\bfseries Age at start of unemployment & \bfseries 39.20 & \bfseries 39.90 & \bfseries 40.23 & \bfseries 39.34 & \bfseries 38.93 & \bfseries 39.34 & \bfseries 40.75 & \bfseries 38.76 & \bfseries 39.51 & \bfseries 39.24 & \bfseries 38.56 & \bfseries 39.24 & \bfseries 39.14 & \bfseries 39.38 & \bfseries 38.78 & \bfseries 37.83 & \bfseries 38.79 & \bfseries 39.19 & \bfseries 38.66 & \bfseries 39.55 \\*
\bfseries  & \bfseries  & \bfseries (7.94) & \bfseries (11.72) & \bfseries (1.56) & \bfseries (3.02) & \bfseries  & \bfseries (16.70) & \bfseries (6.77) & \bfseries (2.00) & \bfseries (1.10) & \bfseries  & \bfseries (8.12) & \bfseries (6.76) & \bfseries (9.21) & \bfseries (2.52) & \bfseries  & \bfseries (10.85) & \bfseries (15.68) & \bfseries (9.35) & \bfseries (19.61) \\
\end{longtable}}
{\fontsize{6pt}{6}\selectfont
\begin{longtable}{l|rrrrr|rrrrr|rrrrr|rrrrr}
\caption{Intermediate covariate means and standardized differences (vs. first program - NP) in trimmed sample. Standardized difference $\geq20$ marked in violet.} \label{tab:balance_x1_2nd} \\*
\toprule
\textbf{} & \multicolumn{5}{c|}{\textbf{First program: JA}} & \multicolumn{5}{c|}{\textbf{First program: TC}} & \multicolumn{5}{c|}{\textbf{First program: EP}} & \multicolumn{5}{c}{\textbf{First program: WS}} \\*
\textbf{Second program:} & \textbf{NP} & \textbf{JA} & \textbf{TC} & \textbf{EP} & \textbf{WS} & \textbf{NP} & \textbf{JA} & \textbf{TC} & \textbf{EP} & \textbf{WS} & \textbf{NP} & \textbf{JA} & \textbf{TC} & \textbf{EP} & \textbf{WS} & \textbf{NP} & \textbf{JA} & \textbf{TC} & \textbf{EP} & \textbf{WS} \\*
\textbf{Number of observations in sequence:} & \textbf{26,109} & \textbf{4,759} & \textbf{3,780} & \textbf{3,785} & \textbf{4,571} & \textbf{9,840} & \textbf{1,073} & \textbf{9,865} & \textbf{1,712} & \textbf{2,370} & \textbf{6,084} & \textbf{107} & \textbf{396} & \textbf{11,274} & \textbf{1,757} & \textbf{34,485} & \textbf{2,141} & \textbf{2,057} & \textbf{2,010} & \textbf{40,279} \\*
\midrule
\endfirsthead
\caption[]{Intermediate covariate means and standardized differences (vs. first program - NP) in trimmed sample. Standardized difference $\geq20$ marked in violet.} \\*
\toprule
\textbf{} & \multicolumn{5}{c|}{\textbf{First program: JA}} & \multicolumn{5}{c|}{\textbf{First program: TC}} & \multicolumn{5}{c|}{\textbf{First program: EP}} & \multicolumn{5}{c}{\textbf{First program: WS}} \\*
\textbf{Second program:} & \textbf{NP} & \textbf{JA} & \textbf{TC} & \textbf{EP} & \textbf{WS} & \textbf{NP} & \textbf{JA} & \textbf{TC} & \textbf{EP} & \textbf{WS} & \textbf{NP} & \textbf{JA} & \textbf{TC} & \textbf{EP} & \textbf{WS} & \textbf{NP} & \textbf{JA} & \textbf{TC} & \textbf{EP} & \textbf{WS} \\*
\textbf{Number of observations in sequence:} & \textbf{26,109} & \textbf{4,759} & \textbf{3,780} & \textbf{3,785} & \textbf{4,571} & \textbf{9,840} & \textbf{1,073} & \textbf{9,865} & \textbf{1,712} & \textbf{2,370} & \textbf{6,084} & \textbf{107} & \textbf{396} & \textbf{11,274} & \textbf{1,757} & \textbf{34,485} & \textbf{2,141} & \textbf{2,057} & \textbf{2,010} & \textbf{40,279} \\*
\midrule
\endhead
\midrule
\multicolumn{21}{r}{Continued on next page} \\*
\midrule
\endfoot
\bottomrule
\endlastfoot
\bfseries Canceled appointment in period 1 & \bfseries 0.05 & \bfseries 0.01 & \bfseries 0.01 & \bfseries 0.01 & \bfseries 0.02 & \bfseries 0.04 & \bfseries 0.01 & \bfseries 0.01 & \bfseries 0.01 & \bfseries 0.01 & \bfseries 0.07 & \bfseries 0.04 & \bfseries 0.01 & \bfseries 0.01 & \bfseries 0.01 & \bfseries 0.06 & \bfseries 0.02 & \bfseries 0.01 & \bfseries 0.02 & \bfseries 0.01 \\*
\bfseries  & \bfseries  & \bfseries \color{violet} (21.18) & \bfseries \color{violet} (22.80) & \bfseries \color{violet} (21.01) & \bfseries (18.89) & \bfseries  & \bfseries \color{violet} (22.65) & \bfseries \color{violet} (21.20) & \bfseries \color{violet} (20.43) & \bfseries (19.20) & \bfseries  & \bfseries (13.22) & \bfseries \color{violet} (27.98) & \bfseries \color{violet} (28.26) & \bfseries \color{violet} (27.66) & \bfseries  & \bfseries \color{violet} (21.50) & \bfseries \color{violet} (26.22) & \bfseries \color{violet} (20.06) & \bfseries \color{violet} (25.75) \\
\bfseries No-show appointment in period 1 & \bfseries 0.05 & \bfseries 0.04 & \bfseries 0.04 & \bfseries 0.04 & \bfseries 0.04 & \bfseries 0.04 & \bfseries 0.03 & \bfseries 0.03 & \bfseries 0.04 & \bfseries 0.04 & \bfseries 0.07 & \bfseries 0.06 & \bfseries 0.04 & \bfseries 0.04 & \bfseries 0.06 & \bfseries 0.08 & \bfseries 0.07 & \bfseries 0.06 & \bfseries 0.07 & \bfseries 0.06 \\*
\bfseries  & \bfseries  & \bfseries (5.21) & \bfseries (6.58) & \bfseries (1.89) & \bfseries (2.44) & \bfseries  & \bfseries (6.20) & \bfseries (4.64) & \bfseries (1.19) & \bfseries (0.28) & \bfseries  & \bfseries (4.56) & \bfseries (11.84) & \bfseries (10.56) & \bfseries (3.48) & \bfseries  & \bfseries (5.76) & \bfseries (11.20) & \bfseries (5.36) & \bfseries (9.35) \\
\end{longtable}
}
{\fontsize{6pt}{6}\selectfont
\begin{longtable}{l|rrrrr|rrrrr|rrrrr|rrrrr}
\caption{Outcome means and standardized differences (vs. first program - NP) in trimmed sample. Standardized difference $\geq20$ marked in violet.} \label{tab:balance_yy_2nd} \\*
\toprule
\textbf{} & \multicolumn{5}{c|}{\textbf{First program: JA}} & \multicolumn{5}{c|}{\textbf{First program: TC}} & \multicolumn{5}{c|}{\textbf{First program: EP}} & \multicolumn{5}{c}{\textbf{First program: WS}} \\*
\textbf{Second program:} & \textbf{NP} & \textbf{JA} & \textbf{TC} & \textbf{EP} & \textbf{WS} & \textbf{NP} & \textbf{JA} & \textbf{TC} & \textbf{EP} & \textbf{WS} & \textbf{NP} & \textbf{JA} & \textbf{TC} & \textbf{EP} & \textbf{WS} & \textbf{NP} & \textbf{JA} & \textbf{TC} & \textbf{EP} & \textbf{WS} \\*
\textbf{Number of observations in sequence:} & \textbf{26,109} & \textbf{4,759} & \textbf{3,780} & \textbf{3,785} & \textbf{4,571} & \textbf{9,840} & \textbf{1,073} & \textbf{9,865} & \textbf{1,712} & \textbf{2,370} & \textbf{6,084} & \textbf{107} & \textbf{396} & \textbf{11,274} & \textbf{1,757} & \textbf{34,485} & \textbf{2,141} & \textbf{2,057} & \textbf{2,010} & \textbf{40,279} \\*
\midrule
\endfirsthead
\caption[]{Outcome means and standardized differences (vs. first program - NP) in trimmed sample. Standardized difference $\geq20$ marked in violet.} \\*
\toprule
\textbf{} & \multicolumn{5}{c|}{\textbf{First program: JA}} & \multicolumn{5}{c|}{\textbf{First program: TC}} & \multicolumn{5}{c|}{\textbf{First program: EP}} & \multicolumn{5}{c}{\textbf{First program: WS}} \\*
\textbf{Second program:} & \textbf{NP} & \textbf{JA} & \textbf{TC} & \textbf{EP} & \textbf{WS} & \textbf{NP} & \textbf{JA} & \textbf{TC} & \textbf{EP} & \textbf{WS} & \textbf{NP} & \textbf{JA} & \textbf{TC} & \textbf{EP} & \textbf{WS} & \textbf{NP} & \textbf{JA} & \textbf{TC} & \textbf{EP} & \textbf{WS} \\*
\textbf{Number of observations in sequence:} & \textbf{26,109} & \textbf{4,759} & \textbf{3,780} & \textbf{3,785} & \textbf{4,571} & \textbf{9,840} & \textbf{1,073} & \textbf{9,865} & \textbf{1,712} & \textbf{2,370} & \textbf{6,084} & \textbf{107} & \textbf{396} & \textbf{11,274} & \textbf{1,757} & \textbf{34,485} & \textbf{2,141} & \textbf{2,057} & \textbf{2,010} & \textbf{40,279} \\*
\midrule
\endhead
\midrule
\multicolumn{21}{r}{Continued on next page} \\*
\midrule
\endfoot
\bottomrule
\endlastfoot
\bfseries Cum. months employed in 30 months after 1st period & \bfseries 17.60 & \bfseries 13.31 & \bfseries 13.39 & \bfseries 13.21 & \bfseries 16.35 & \bfseries 17.97 & \bfseries 13.52 & \bfseries 14.11 & \bfseries 13.40 & \bfseries 16.42 & \bfseries 17.84 & \bfseries 13.03 & \bfseries 14.96 & \bfseries 13.78 & \bfseries 16.16 & \bfseries 22.20 & \bfseries 15.01 & \bfseries 14.96 & \bfseries 14.93 & \bfseries 16.66 \\*
\bfseries  & \bfseries  & \bfseries \color{violet} (44.98) & \bfseries \color{violet} (45.12) & \bfseries \color{violet} (46.78) & \bfseries (13.81) & \bfseries  & \bfseries \color{violet} (47.80) & \bfseries \color{violet} (40.63) & \bfseries \color{violet} (49.18) & \bfseries (16.87) & \bfseries  & \bfseries \color{violet} (49.41) & \bfseries \color{violet} (30.86) & \bfseries \color{violet} (42.00) & \bfseries (18.18) & \bfseries  & \bfseries \color{violet} (86.39) & \bfseries \color{violet} (87.32) & \bfseries \color{violet} (87.04) & \bfseries \color{violet} (65.03) \\
\end{longtable}
}
\end{landscape}
\normalsize

\restoregeometry
\newpage
\section{Overlap}\label{sec_app_overlap}
\setlength{\tabcolsep}{2pt}
{\fontsize{6pt}{6}\selectfont
\begin{longtable}{lrr|lrr}
\caption{Pre-treatment covariate means and standardized differences trimmed vs. retained sample.} \label{tab:balance_x0_trim} \\*
\toprule
\textbf{Variable} & \textbf{retained} & \textbf{trimmed} & \textbf{Variable } & \textbf{retained } & \textbf{trimmed } \\*
\midrule
\endfirsthead
\caption[]{Pre-treatment covariate means and standardized differences trimmed vs. retained sample.} \\*
\toprule
\textbf{Variable} & \textbf{retained} & \textbf{trimmed} & \textbf{Variable } & \textbf{retained } & \textbf{trimmed } \\*
\midrule
\endhead
\midrule
\multicolumn{6}{r}{Continued on next page} \\*
\midrule
\endfoot
\bottomrule
\endlastfoot
\bfseries Avg add. income in months ue 1y before ref. point & \bfseries 61.93 & \bfseries 135.45 & \bfseries Insured earnings before start of first program & \bfseries 4,973.60 & \bfseries 5,304.21 \\*
\bfseries  & \bfseries  & \bfseries (16.11) & \bfseries  & \bfseries  & \bfseries (14.84) \\
\bfseries Age at start of unemployment & \bfseries 38.99 & \bfseries 38.97 & \bfseries Ever insufficient jobsearch efforts 1y before ref. point & \bfseries 0.28 & \bfseries 0.26 \\*
\bfseries  & \bfseries  & \bfseries (0.23) & \bfseries  & \bfseries  & \bfseries (3.18) \\
\bfseries Age at 1st income subj. to Swiss social ins. contrib. & \bfseries 23.82 & \bfseries 23.00 & \bfseries Ever insufficient job search eff. in prev. ue spells & \bfseries 0.32 & \bfseries 0.33 \\*
\bfseries  & \bfseries  & \bfseries (11.45) & \bfseries  & \bfseries  & \bfseries (0.53) \\
\bfseries Avg monthly applications in previous ue spells & \bfseries 2.92 & \bfseries 3.46 & \bfseries Language level Italian & \bfseries 1.04 & \bfseries 1.05 \\*
\bfseries  & \bfseries  & \bfseries (13.40) & \bfseries  & \bfseries  & \bfseries (0.83) \\
\end{longtable}
}
\normalsize
\newpage

\begin{figure}[htbp!]
    \centering
    \captionsetup{font=small, width=0.95\linewidth} 
    \caption{Overlap plots of dynamic policies starting with \textit{JA} or \textit{TC} under dynamic confounding.}
    \includegraphics[width=0.49\linewidth]{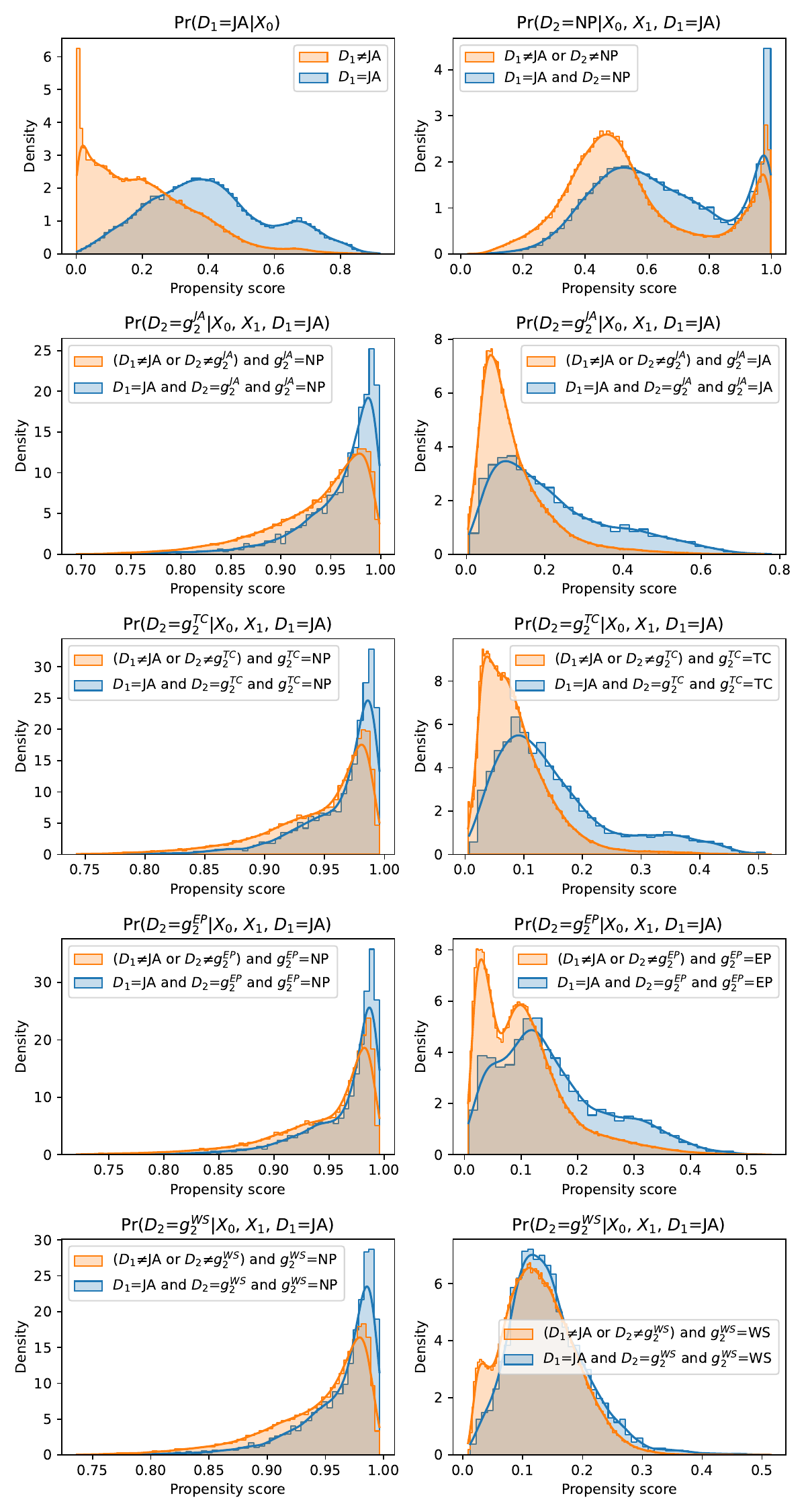}\vrule
    \includegraphics[width=0.49\linewidth]{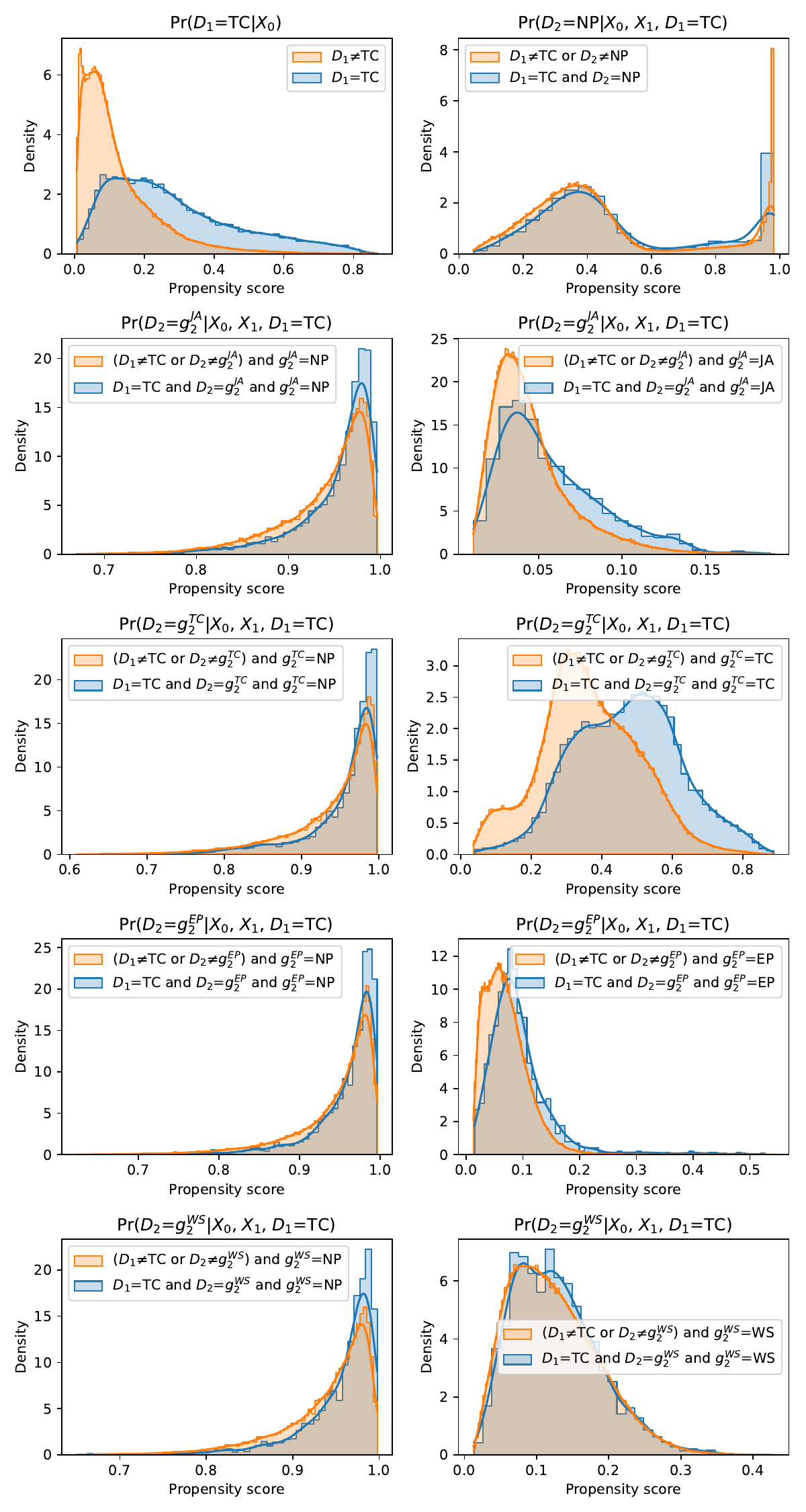}
    \caption*{\scriptsize \textit{Notes:} This figure plots propensity score densities after trimming for all dynamic policies starting with \textit{JA} or \textit{TC} considered in the main analysis. $g_2^{\text{XX}}$ refers to the dynamic policy $g_2(Y_1)$ defined in equation \ref{eq:g_application} with $d_2=\text{XX}$. For each policy, densities are plotted separately for individuals following policy $\mathbf{g}_2(Y_1)$ versus those not following the policy and separately for the different realizations of the function $g_2^{\text{XX}}$. \textit{JA}: Job-search assistance, \textit{TC}: Training course, \textit{EP}: Employment program,  \textit{WS}: Wage subsidy, \textit{NP}: No program.}
    \label{fig:app_pscores_dynpolJATC}
\end{figure}
\newpage\newpage
\begin{figure}[htbp!]
    \centering
    \captionsetup{font=small, width=0.95\linewidth} 
    \caption{Overlap plots of dynamic policies starting with \textit{EP} or \textit{WS} under dynamic confounding.}
    \includegraphics[width=0.49\linewidth]{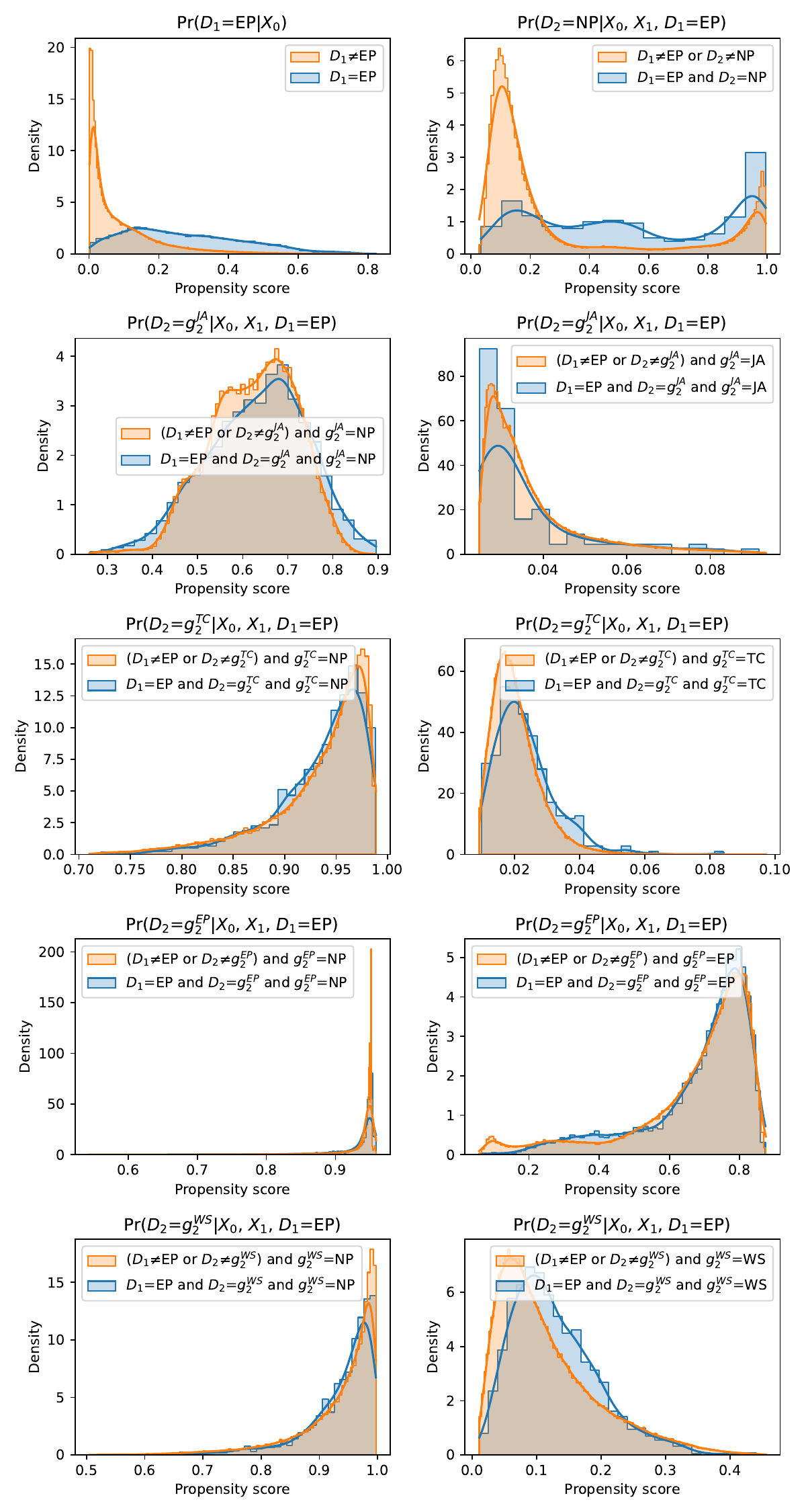}\vrule
    \includegraphics[width=0.49\linewidth]{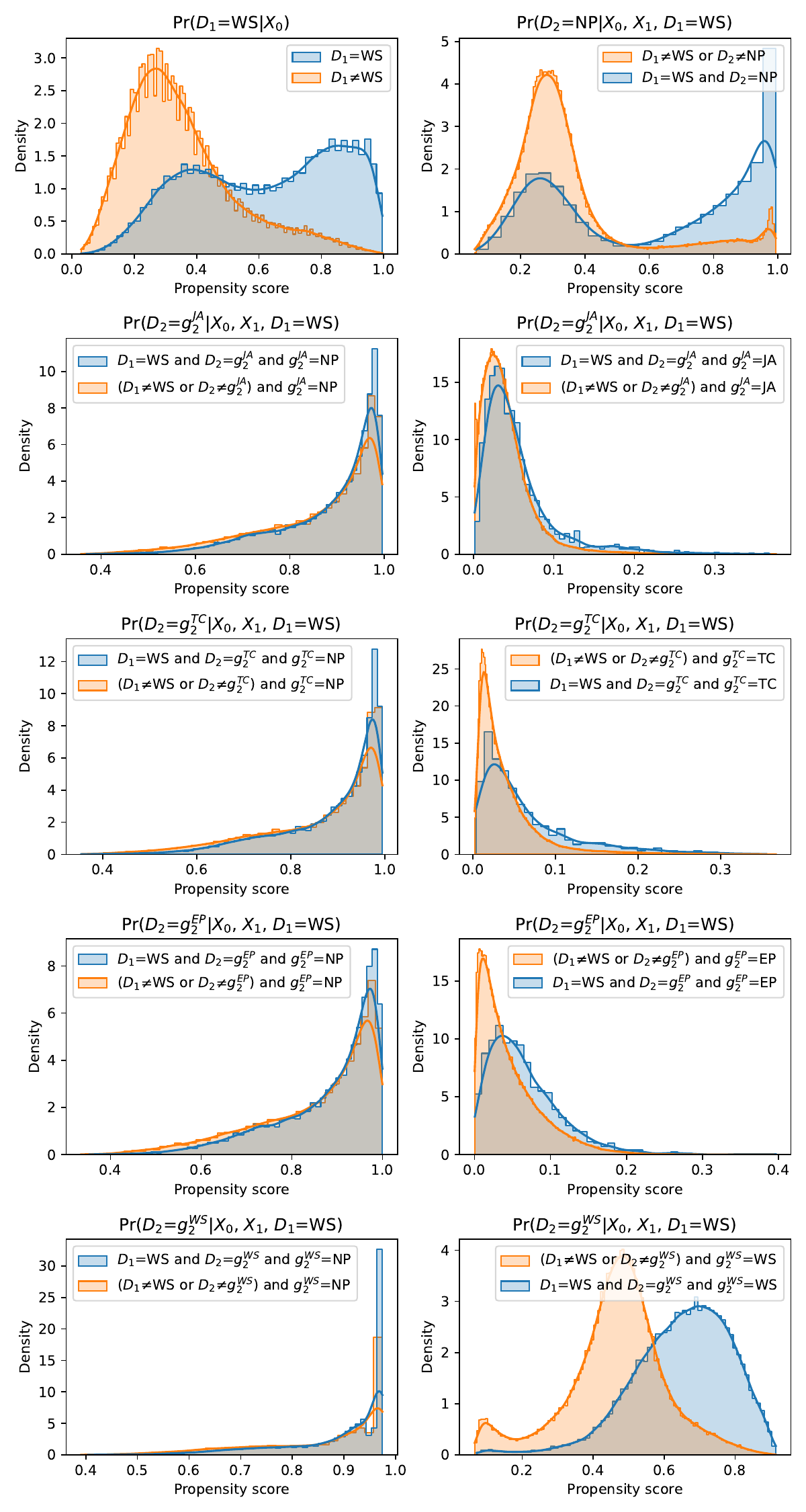}
    \caption*{\scriptsize \textit{Notes:} This figure plots propensity score densities after trimming for all dynamic policies starting with \textit{EP} or \textit{WS} considered in the main analysis. $g_2^{\text{XX}}$ refers to the dynamic policy $g_2(Y_1)$ defined in equation \ref{eq:g_application} with $d_2=\text{XX}$. For each policy, densities are plotted separately for individuals following policy $\mathbf{g}_2(Y_1)$ versus those not following the policy and separately for the different realizations of the function $g_2^{\text{XX}}$. \textit{JA}: Job-search assistance, \textit{TC}: Training course, \textit{EP}: Employment program,  \textit{WS}: Wage subsidy, \textit{NP}: No program.}
    \label{fig:app_pscores_dynpolEPWS}
\end{figure}
\newpage

\begin{figure}[htbp!]
    \centering
    \captionsetup{font=small, width=0.8\linewidth} 
    \caption{Alluvial plot of program sequences after trimming}
    \includegraphics[width=0.8\linewidth]{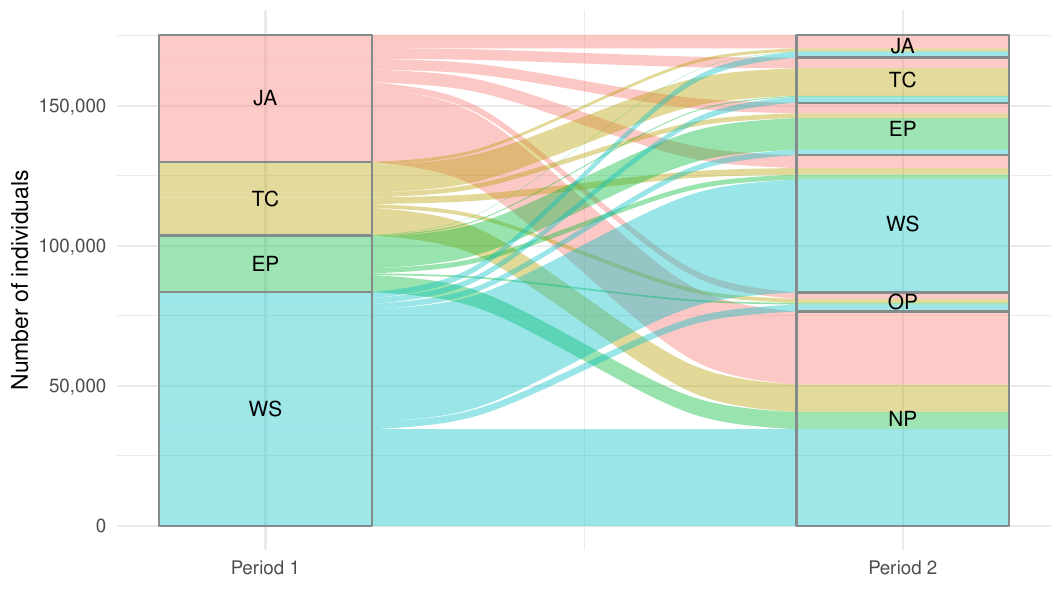}
    \caption*{\scriptsize \textit{Notes:} Program frequencies and transitions between first and second period. \textit{JA}: Job-search assistance, \textit{TC}: Training course, \textit{EP}: Employment program,  \textit{WS}: Wage subsidy, \textit{OP}: Other programs, \textit{NP}: No program.}
    \label{fig:alluvial_trim}
\end{figure}
\newpage\newpage

\section{Extended Empirical Results}
\begin{table}[htbp!]
\centering
\begin{threeparttable}
\caption{GATE-ATE by local language knowledge and previous program participation.}
\label{tab:GATEmATE_combined_em332}
\begin{tabularx}{.9\textwidth}{rXrrrrrrr}
\toprule
 &  &  & \makecell{$d_1 =$ JA\\ $d'_1 =$ TC} & \makecell{$d_1 =$ JA\\ $d'_1 =$ EP} & \makecell{$d_1 =$ JA\\ $d'_1 =$ WS} & \makecell{$d_1 =$ TC\\ $d'_1 =$ EP} & \makecell{$d_1 =$ TC\\ $d'_1 =$ WS} & \makecell{$d_1 =$ EP\\ $d'_1 =$ WS} \\
\cmidrule{4-9}&&&\multicolumn{6}{c}{$g_2(Y_1^{d_1}) = \mathbbm{1}\{Y_1^{d_1}=0\}d_1 +\mathbbm{1}\{Y_1^{d_1}=1\}\text{NP}$}\\
&&&\multicolumn{6}{c}{$g'_2(Y_1^{d'_1}) = \mathbbm{1}\{Y_1^{d'_1}=0\}d'_1 +\mathbbm{1}\{Y_1^{d'_1}=1\}\text{NP}$}\\
\midrule
\multicolumn{9}{l}{\textit{Panel A: Local language knowledge}}\\
\midrule\ & \multirow[c]{2}{*}{None to basic} &  & 1.07**\phantom{*(} & 0.90\phantom{***(} & -0.02\phantom{***(} & -0.17\phantom{***(} & -1.09***\phantom{(} & -0.92*\phantom{**(} \\
 &  &   & (0.42)\phantom{***} & (0.60)\phantom{***} & (0.35)\phantom{***} & (0.56)\phantom{***} & (0.28)\phantom{***} & (0.51)\phantom{***} \\
 & \multirow[c]{2}{*}{Intermediate} &  & 0.47\phantom{***(} & -0.50\phantom{***(} & -0.26\phantom{***(} & -0.97**\phantom{*(} & -0.73***\phantom{(} & 0.24\phantom{***(} \\
 &  &   & (0.34)\phantom{***} & (0.40)\phantom{***} & (0.26)\phantom{***} & (0.39)\phantom{***} & (0.25)\phantom{***} & (0.33)\phantom{***} \\
 & \multirow[c]{2}{*}{Good} &  & 1.01\phantom{***(} & 0.22\phantom{***(} & 0.58\phantom{***(} & -0.79\phantom{***(} & -0.43\phantom{***(} & 0.36\phantom{***(} \\
 &  &   & (0.62)\phantom{***} & (0.71)\phantom{***} & (0.54)\phantom{***} & (0.59)\phantom{***} & (0.35)\phantom{***} & (0.50)\phantom{***} \\
 & \multirow[c]{2}{*}{Fluent} &  & -0.61***\phantom{(} & -0.09\phantom{***(} & -0.03\phantom{***(} & 0.52***\phantom{(} & 0.58***\phantom{(} & 0.06\phantom{***(} \\
 &  &   & (0.18)\phantom{***} & (0.21)\phantom{***} & (0.14)\phantom{***} & (0.20)\phantom{***} & (0.13)\phantom{***} & (0.16)\phantom{***} \\
\midrule
\multicolumn{9}{l}{\textit{Panel B: Unemployment (UE) and program participation in 5 years prior to current UE spell}}\\
\midrule\ & \multirow[c]{2}{*}{UE no program} &  & -0.03\phantom{***(} & 0.17\phantom{***(} & -0.17\phantom{***(} & 0.20\phantom{***(} & -0.13\phantom{***(} & -0.34*\phantom{**(} \\
 &  &   & (0.20)\phantom{***} & (0.22)\phantom{***} & (0.14)\phantom{***} & (0.22)\phantom{***} & (0.14)\phantom{***} & (0.18)\phantom{***} \\
 & \multirow[c]{2}{*}{Not UE} &  & 0.52\phantom{***(} & -0.55\phantom{***(} & -0.27\phantom{***(} & -1.07\phantom{***(} & -0.79*\phantom{**(} & 0.28\phantom{***(} \\
 &  &   & (0.61)\phantom{***} & (0.74)\phantom{***} & (0.43)\phantom{***} & (0.76)\phantom{***} & (0.45)\phantom{***} & (0.62)\phantom{***} \\
 & \multirow[c]{2}{*}{JA} &  & -0.02\phantom{***(} & 0.12\phantom{***(} & 0.15\phantom{***(} & 0.15\phantom{***(} & 0.18\phantom{***(} & 0.03\phantom{***(} \\
 &  &   & (0.61)\phantom{***} & (0.62)\phantom{***} & (0.44)\phantom{***} & (0.64)\phantom{***} & (0.46)\phantom{***} & (0.48)\phantom{***} \\
 & \multirow[c]{2}{*}{TC} &  & 0.07\phantom{***(} & 0.35\phantom{***(} & 1.06**\phantom{*(} & 0.28\phantom{***(} & 0.99**\phantom{*(} & 0.71\phantom{***(} \\
 &  &   & (0.64)\phantom{***} & (0.61)\phantom{***} & (0.44)\phantom{***} & (0.65)\phantom{***} & (0.50)\phantom{***} & (0.46)\phantom{***} \\
 & \multirow[c]{2}{*}{EP} &  & -1.27\phantom{***(} & -1.22\phantom{***(} & -0.76\phantom{***(} & 0.05\phantom{***(} & 0.51\phantom{***(} & 0.45\phantom{***(} \\
 &  &   & (0.87)\phantom{***} & (0.83)\phantom{***} & (0.72)\phantom{***} & (0.68)\phantom{***} & (0.53)\phantom{***} & (0.47)\phantom{***} \\
 & \multirow[c]{2}{*}{WS} &  & -0.27\phantom{***(} & -0.17\phantom{***(} & 0.71**\phantom{*(} & 0.10\phantom{***(} & 0.99***\phantom{(} & 0.88**\phantom{*(} \\
 &  &   & (0.47)\phantom{***} & (0.51)\phantom{***} & (0.34)\phantom{***} & (0.51)\phantom{***} & (0.34)\phantom{***} & (0.39)\phantom{***} \\
\bottomrule
\end{tabularx}
\caption*{\scriptsize \textit{Note:} This table reports GATE-ATE by local language knowledge and previous program participation. Each column represents the comparison of two dynamic policies, where the first period program is continued in the second period if the individual remains unemployed in the first period (no program otherwise). The effects are estimated using method proposed in \citet{bodory2022evaluating}. $d_1$ and $d'_1$ represent first-period programs in the treatment and control states, respectively, while $g_2$ and $g'_2$ denote second-period policies dependent on the potential intermediate outcomes (1 if an individual exits unemployment in the first period). JA: Job-search assistance, TC: Training course, EP: Employment program, WS: Temporary wage subsidy, NP: No program. Outcome: Cumulative months in employment in the 30 months from start of the second period. For each comparison of programs, the rows show the GATE-ATE and the standard errors in parentheses for the respective category of the heterogeneity variable. *, **, *** indicate the precision of the estimate by showing whether the p-value of a two-sided significance test is below 10\%, 5\%, and 1\%, respectively.}
\end{threeparttable}
\end{table}

\begin{table}[htbp!]
\centering
\begin{threeparttable}
\caption{Average treatment effects for program duration.}
\label{tab:DATE_first+duration_em332}
\begin{tabularx}{\textwidth}{rXrrrrrrr}
\toprule
 &  &  & \makecell{$d_1 =$ JA\\ $d'_1 =$ TC} & \makecell{$d_1 =$ JA\\ $d'_1 =$ EP} & \makecell{$d_1 =$ JA\\ $d'_1 =$ WS} & \makecell{$d_1 =$ TC\\ $d'_1 =$ EP} & \makecell{$d_1 =$ TC\\ $d'_1 =$ WS} & \makecell{$d_1 =$ EP\\ $d'_1 =$ WS} \\
\midrule
\multicolumn{9}{l}{\textit{Panel A: Second period program unrestricted (single-period intervention):}}\\
\midrule\ & \multirow[c]{2}{*}{ATE (static conf.)} &  & -0.71***\phantom{(} & -0.12\phantom{***(} & -2.65***\phantom{(} & 0.59***\phantom{(} & -1.94***\phantom{(} & -2.53***\phantom{(} \\
 &  &   & (0.11)\phantom{***} & (0.16)\phantom{***} & (0.09)\phantom{***} & (0.15)\phantom{***} & (0.09)\phantom{***} & (0.14)\phantom{***} \\
 & \multirow[c]{2}{*}{$N_{d_1}$} &  & 45,358\phantom{(} & 45,358\phantom{(} & 45,358\phantom{(} & 26,233\phantom{(} & 26,233\phantom{(} & 20,242\phantom{(} \\
 &  &   & (94\%)\phantom{(} & (94\%)\phantom{(} & (94\%)\phantom{(} & (94\%)\phantom{(} & (94\%)\phantom{(} & (90\%)\phantom{(} \\
 & \multirow[c]{2}{*}{$N_{d'_1}$} &  & 26,233\phantom{(} & 20,242\phantom{(} & 83,456\phantom{(} & 20,242\phantom{(} & 83,456\phantom{(} & 83,456\phantom{(} \\
 &  &   & (94\%)\phantom{(} & (90\%)\phantom{(} & (90\%)\phantom{(} & (90\%)\phantom{(} & (90\%)\phantom{(} & (90\%)\phantom{(} \\
\midrule
\multicolumn{9}{l}{\textit{Panel B: Second period without program (static policy):} $g_2(Y_1^{d_1}) =$ NP, $g'_2(Y_1^{d'_1}) =$ NP}\\
\midrule\ & \multirow[c]{2}{*}{ATE (BHL22)} &  & -1.43***\phantom{(} & -1.47***\phantom{(} & -4.19***\phantom{(} & -0.03\phantom{***(} & -2.75***\phantom{(} & -2.72***\phantom{(} \\
 &  &   & (0.21)\phantom{***} & (0.33)\phantom{***} & (0.15)\phantom{***} & (0.34)\phantom{***} & (0.17)\phantom{***} & (0.31)\phantom{***} \\
 & \multirow[c]{2}{*}{ATE (BJZ24)} &  & -1.42***\phantom{(} & -1.48***\phantom{(} & -4.16***\phantom{(} & -0.06\phantom{***(} & -2.73***\phantom{(} & -2.68***\phantom{(} \\
 &  &   & (0.21)\phantom{***} & (0.33)\phantom{***} & (0.16)\phantom{***} & (0.34)\phantom{***} & (0.17)\phantom{***} & (0.31)\phantom{***} \\
 & \multirow[c]{2}{*}{ATE (static conf.)} &  & -1.61***\phantom{(} & -2.74***\phantom{(} & -4.67***\phantom{(} & -1.13***\phantom{(} & -3.06***\phantom{(} & -1.92***\phantom{(} \\
 &  &   & (0.23)\phantom{***} & (0.37)\phantom{***} & (0.19)\phantom{***} & (0.35)\phantom{***} & (0.15)\phantom{***} & (0.33)\phantom{***} \\
 & \multirow[c]{2}{*}{$N_{\mathbf{g}_2}$} &  & 26,109\phantom{(} & 26,109\phantom{(} & 26,109\phantom{(} & 9,840\phantom{(} & 9,840\phantom{(} & 6,084\phantom{(} \\
 &  &   & (92\%)\phantom{(} & (92\%)\phantom{(} & (92\%)\phantom{(} & (92\%)\phantom{(} & (92\%)\phantom{(} & (86\%)\phantom{(} \\
 & \multirow[c]{2}{*}{$N_{\mathbf{g}'_2}$} &  & 9,840\phantom{(} & 6,084\phantom{(} & 34,485\phantom{(} & 6,084\phantom{(} & 34,485\phantom{(} & 34,485\phantom{(} \\
 &  &   & (92\%)\phantom{(} & (86\%)\phantom{(} & (89\%)\phantom{(} & (86\%)\phantom{(} & (89\%)\phantom{(} & (89\%)\phantom{(} \\
\midrule
\multicolumn{9}{l}{\textit{Panel C: Same program for at least two periods (static policy):} $g_2(Y_1^{d_1}) = d_1$, $g'_2(Y_1^{d'_1}) = d'_1$}\\
\midrule\ & \multirow[c]{2}{*}{ATE (BHL22)} &  & -0.94***\phantom{(} & -0.38\phantom{***(} & -2.86***\phantom{(} & 0.57*\phantom{**(} & -1.91***\phantom{(} & -2.48***\phantom{(} \\
 &  &   & (0.34)\phantom{***} & (0.35)\phantom{***} & (0.26)\phantom{***} & (0.34)\phantom{***} & (0.24)\phantom{***} & (0.26)\phantom{***} \\
 & \multirow[c]{2}{*}{ATE (BJZ24)} &  & -0.90**\phantom{*(} & -0.47\phantom{***(} & -2.92***\phantom{(} & 0.44\phantom{***(} & -2.02***\phantom{(} & -2.46***\phantom{(} \\
 &  &   & (0.41)\phantom{***} & (0.41)\phantom{***} & (0.32)\phantom{***} & (0.37)\phantom{***} & (0.27)\phantom{***} & (0.28)\phantom{***} \\
 & \multirow[c]{2}{*}{ATE (static conf.)} &  & -1.37**\phantom{*(} & -0.86\phantom{***(} & -2.87***\phantom{(} & 0.51***\phantom{(} & -1.50***\phantom{(} & -2.01***\phantom{(} \\
 &  &   & (0.59)\phantom{***} & (0.59)\phantom{***} & (0.58)\phantom{***} & (0.19)\phantom{***} & (0.14)\phantom{***} & (0.15)\phantom{***} \\
 & \multirow[c]{2}{*}{$N_{\mathbf{g}_2}$} &  & 4,759\phantom{(} & 4,759\phantom{(} & 4,759\phantom{(} & 9,865\phantom{(} & 9,865\phantom{(} & 11,274\phantom{(} \\
 &  &   & (94\%)\phantom{(} & (94\%)\phantom{(} & (94\%)\phantom{(} & (94\%)\phantom{(} & (94\%)\phantom{(} & (93\%)\phantom{(} \\
 & \multirow[c]{2}{*}{$N_{\mathbf{g}'_2}$} &  & 9,865\phantom{(} & 11,274\phantom{(} & 40,279\phantom{(} & 11,274\phantom{(} & 40,279\phantom{(} & 40,279\phantom{(} \\
 &  &   & (94\%)\phantom{(} & (93\%)\phantom{(} & (90\%)\phantom{(} & (93\%)\phantom{(} & (90\%)\phantom{(} & (90\%)\phantom{(} \\
\midrule
\multicolumn{9}{l}{\textit{Panel D: Same program for at least two periods if not employed in first period (dynamic policy):}}\\
\multicolumn{9}{l}{\phantom{\textit{Panel D: }}$g_2(Y_1^{d_1}) = \mathbbm{1}\{Y_1^{d_1}=0\}d_1 +\mathbbm{1}\{Y_1^{d_1}=1\}\text{NP}$ and $g'_2(Y_1^{d'_1}) = \mathbbm{1}\{Y_1^{d'_1}=0\}d'_1 +\mathbbm{1}\{Y_1^{d'_1}=1\}\text{NP}$}\\
\midrule\ & \multirow[c]{2}{*}{ATE (BHL22)} &  & -1.24***\phantom{(} & -0.61**\phantom{*(} & -3.51***\phantom{(} & 0.63***\phantom{(} & -2.27***\phantom{(} & -2.90***\phantom{(} \\
 &  &   & (0.21)\phantom{***} & (0.24)\phantom{***} & (0.16)\phantom{***} & (0.24)\phantom{***} & (0.15)\phantom{***} & (0.19)\phantom{***} \\
 & \multirow[c]{2}{*}{ATE (BJZ24)} &  & -1.13***\phantom{(} & -0.54**\phantom{*(} & -3.40***\phantom{(} & 0.59**\phantom{*(} & -2.27***\phantom{(} & -2.86***\phantom{(} \\
 &  &   & (0.21)\phantom{***} & (0.25)\phantom{***} & (0.16)\phantom{***} & (0.24)\phantom{***} & (0.15)\phantom{***} & (0.20)\phantom{***} \\
 & \multirow[c]{2}{*}{$N_{\mathbf{g}_2}$} &  & 9,438\phantom{(} & 9,438\phantom{(} & 9,438\phantom{(} & 11,821\phantom{(} & 11,821\phantom{(} & 13,124\phantom{(} \\
 &  &   & (96\%)\phantom{(} & (96\%)\phantom{(} & (96\%)\phantom{(} & (95\%)\phantom{(} & (95\%)\phantom{(} & (93\%)\phantom{(} \\
 & \multirow[c]{2}{*}{$N_{\mathbf{g}'_2}$} &  & 11,821\phantom{(} & 13,124\phantom{(} & 55,037\phantom{(} & 13,124\phantom{(} & 55,037\phantom{(} & 55,037\phantom{(} \\
 &  &   & (95\%)\phantom{(} & (93\%)\phantom{(} & (92\%)\phantom{(} & (93\%)\phantom{(} & (92\%)\phantom{(} & (92\%)\phantom{(} \\
\bottomrule
\end{tabularx}
\caption*{\scriptsize \textit{Note:} This table reports ATEs and the number of observations for various comparisons of policies. $d_1$ and $d'_1$ represent first-period programs in the treatment and control states, respectively, while $g_2$ and $g'_2$ denote second-period policies dependent on the potential intermediate outcomes (1 if an individual exits unemployment in the first period). JA: Job-search assistance, TC: Training course, EP: Employment program, WS: Temporary wage subsidy, NP: No program. Outcome: Cumulative months in employment in the 30 months from start of the second period. Rows labeled ATE report effect sizes with standard errors in parentheses. *, **, *** indicate p-values below 10\%, 5\%, and 1\%. Rows labeled $N$ show the trimmed sample size and its proportion of the untrimmed sample (in parentheses). BHL22, BJZ24, and static conf. indicate that the estimators by \citet{bodory2022evaluating}, \citet{bradic2021high}, and \citet{chernozhukov2018double}, respectively, are used for estimation.}
\end{threeparttable}
\end{table}

\begin{table}[htbp!]
\centering
\begin{threeparttable}
\caption{Average treatment effects for program order.}
\label{tab:DATE_order_em332}
\begin{tabularx}{\textwidth}{rXrrrrrrr}
\toprule
 &  &  & \makecell{$d_1 =$ JA\\ $d'_1 =$ TC} & \makecell{$d_1 =$ JA\\ $d'_1 =$ EP} & \makecell{$d_1 =$ JA\\ $d'_1 =$ WS} & \makecell{$d_1 =$ TC\\ $d'_1 =$ EP} & \makecell{$d_1 =$ TC\\ $d'_1 =$ WS} & \makecell{$d_1 =$ EP\\ $d'_1 =$ WS} \\
\midrule
\multicolumn{9}{l}{\textit{Panel A: Program order (static policy): $g_2(Y_1^{d_1}) = d'_1$, $g'_2(Y_1^{d'_1}) = d_1$}} \\
\midrule\ & \multirow[c]{2}{*}{ATE (BHL22)} &  & -0.01\phantom{***(} & 5.30\phantom{***(} & 0.52\phantom{***(} & -0.51\phantom{***(} & 1.65***\phantom{(} & 1.89***\phantom{(} \\
 &  &   & (0.59)\phantom{***} & (5.54)\phantom{***} & (0.48)\phantom{***} & (0.62)\phantom{***} & (0.59)\phantom{***} & (0.59)\phantom{***} \\
 & \multirow[c]{2}{*}{ATE (BJZ24)} &  & 0.04\phantom{***(} & 5.31\phantom{***(} & 0.42\phantom{***(} & -0.48\phantom{***(} & 1.78***\phantom{(} & 1.77***\phantom{(} \\
 &  &   & (0.61)\phantom{***} & (4.88)\phantom{***} & (0.47)\phantom{***} & (0.69)\phantom{***} & (0.58)\phantom{***} & (0.62)\phantom{***} \\
 & \multirow[c]{2}{*}{ATE (static conf.)} &  & -0.43\phantom{***(} & -0.25\phantom{***(} & 1.03***\phantom{(} & -0.17\phantom{***(} & 1.09***\phantom{(} & 1.29***\phantom{(} \\
 &  &   & (0.31)\phantom{***} & (1.05)\phantom{***} & (0.25)\phantom{***} & (0.53)\phantom{***} & (0.28)\phantom{***} & (0.39)\phantom{***} \\
 & \multirow[c]{2}{*}{$N_{\mathbf{g}_2}$} &  & 3,780\phantom{(} & 3,785\phantom{(} & 4,571\phantom{(} & 1,712\phantom{(} & 2,370\phantom{(} & 1,757\phantom{(} \\
 &  &   & (97\%)\phantom{(} & (96\%)\phantom{(} & (96\%)\phantom{(} & (95\%)\phantom{(} & (95\%)\phantom{(} & (89\%)\phantom{(} \\
 & \multirow[c]{2}{*}{$N_{\mathbf{g}'_2}$} &  & 1,073\phantom{(} & 107\phantom{(} & 2,141\phantom{(} & 396\phantom{(} & 2,057\phantom{(} & 2,010\phantom{(} \\
 &  &   & (95\%)\phantom{(} & (94\%)\phantom{(} & (94\%)\phantom{(} & (92\%)\phantom{(} & (95\%)\phantom{(} & (89\%)\phantom{(} \\
\midrule
\multicolumn{9}{l}{\textit{Panel B: Program order (dynamic policy):}}\\
\multicolumn{9}{l}{\phantom{\textit{Panel B: }}$g_2(Y_1^{d_1}) = \mathbbm{1}\{Y_1^{d_1}=0\}d'_1 +\mathbbm{1}\{Y_1^{d_1}=1\}\text{NP}$ and $g'_2(Y_1^{d'_1}) = \mathbbm{1}\{Y_1^{d'_1}=0\}d_1 +\mathbbm{1}\{Y_1^{d'_1}=1\}\text{NP}$}\\
\midrule\ & \multirow[c]{2}{*}{ATE (BHL22)} &  & -0.58\phantom{***(} & -0.40\phantom{***(} & 0.22\phantom{***(} & -0.13\phantom{***(} & 0.19\phantom{***(} & 0.13\phantom{***(} \\
 &  &   & (0.41)\phantom{***} & (0.46)\phantom{***} & (0.31)\phantom{***} & (0.66)\phantom{***} & (0.32)\phantom{***} & (0.88)\phantom{***} \\
 & \multirow[c]{2}{*}{ATE (BJZ24)} &  & -0.69\phantom{***(} & 0.59\phantom{***(} & 0.04\phantom{***(} & -0.05\phantom{***(} & 0.23\phantom{***(} & 0.13\phantom{***(} \\
 &  &   & (0.42)\phantom{***} & (0.48)\phantom{***} & (0.31)\phantom{***} & (0.68)\phantom{***} & (0.31)\phantom{***} & (0.95)\phantom{***} \\
 & \multirow[c]{2}{*}{$N_{\mathbf{g}_2}$} &  & 8,465\phantom{(} & 8,466\phantom{(} & 9,168\phantom{(} & 3,676\phantom{(} & 4,315\phantom{(} & 3,588\phantom{(} \\
 &  &   & (98\%)\phantom{(} & (98\%)\phantom{(} & (98\%)\phantom{(} & (97\%)\phantom{(} & (97\%)\phantom{(} & (93\%)\phantom{(} \\
 & \multirow[c]{2}{*}{$N_{\mathbf{g}'_2}$} &  & 3,047\phantom{(} & 2,002\phantom{(} & 19,013\phantom{(} & 2,284\phantom{(} & 18,942\phantom{(} & 18,912\phantom{(} \\
 &  &   & (97\%)\phantom{(} & (97\%)\phantom{(} & (98\%)\phantom{(} & (96\%)\phantom{(} & (98\%)\phantom{(} & (97\%)\phantom{(} \\
\bottomrule
\end{tabularx}
\caption*{\scriptsize \textit{Note:} This table reports ATEs and the number of observations for various comparisons of policies. $d_1$ and $d'_1$ represent first-period programs in the treatment and control states, respectively, while $g_2$ and $g'_2$ denote second-period policies dependent on the potential intermediate outcomes (1 if an individual exits unemployment in the first period). JA: Job-search assistance, TC: Training course, EP: Employment program, WS: Temporary wage subsidy, NP: No program. Outcome: Cumulative months in employment in the 30 months from start of the second period. Rows labeled ATE report effect sizes with standard errors in parentheses. *, **, *** indicate p-values below 10\%, 5\%, and 1\%. Rows labeled $N$ show the trimmed sample size and its proportion of the untrimmed sample (in parentheses). BHL22, BJZ24, and static conf. indicate that the estimators by \citet{bodory2022evaluating}, \citet{bradic2021high}, and \citet{chernozhukov2018double}, respectively, are used for estimation.}
\end{threeparttable}
\end{table}

\newpage
\end{appendices}
 
\end{document}